\documentclass[conference, twocolumns, usletter]{IEEEtran}

\usepackage{amsmath,amssymb,amsthm,mathrsfs,amsfonts,dsfont,stmaryrd,wasysym,marvosym}
\usepackage{mathtools}
\usepackage{color}
\usepackage{graphicx}  
\usepackage{psfrag, xcolor}
\usepackage{enumitem}
\usepackage{subfigure,cite,epsfig,dblfloatfix,url}

\usepackage{bm,comment}

\newtheorem{Th}{Theorem}
\newtheorem{Def}{Definition}
\newtheorem{Rem}{Remark}
\newtheorem{Cor}{Corollary}
\newtheorem{Lem}{Lemma}
\newcommand{\dv}{\mathbf} 
\newcommand{\mc}{\mathcal} 
\newcommand{\bs}{\boldsymbol} 

\newcommand{\squeezeup}{\vspace{-2.5mm}}

\begin{document}
\fontencoding{OT1}\fontsize{9.4}{11}\selectfont
\title{On Learning Parametric Distributions from Quantized Samples} 


\author{%
  \IEEEauthorblockN{Septimia Sarbu \IEEEauthorrefmark{1},
										Abdellatif Zaidi \IEEEauthorrefmark{2}\IEEEauthorrefmark{1}}
	\IEEEauthorblockA{\IEEEauthorrefmark{1}%
										Huawei Technologies France, Mathematical and Algorithmic Sciences Lab,\\
										Paris Research Center, Boulogne-Billancourt, 92100, France\\
                     }
					
\IEEEauthorblockA{\IEEEauthorrefmark{2}%
										Universit\'{e} Paris-Est, Champs-sur-Marne, 77454, France\\
                    \{septimia.sarbu@huawei.com, abdellatif.zaidi@u-pem.fr\} }
}

\maketitle

\begin{abstract}
We consider the problem of learning parametric distributions from their quantized samples in a network. Specifically, $n$ agents or sensors observe independent samples of an unknown parametric distribution; and each of them uses $k$ bits to describe its observed sample to a central processor whose goal is to estimate the unknown distribution. First, we establish a generalization of the well-known van Trees inequality to general $L_p$-norms, with $p > 1$, in terms of Generalized Fisher information. Then, we develop minimax lower bounds on the estimation error for two losses: general $L_p$-norms and the related Wasserstein loss from optimal transport.
\end{abstract}

\section{Introduction and Problem Formulation}\label{Intro}

\begin{figure}[h!]
\centering
\includegraphics[width=0.6\linewidth]{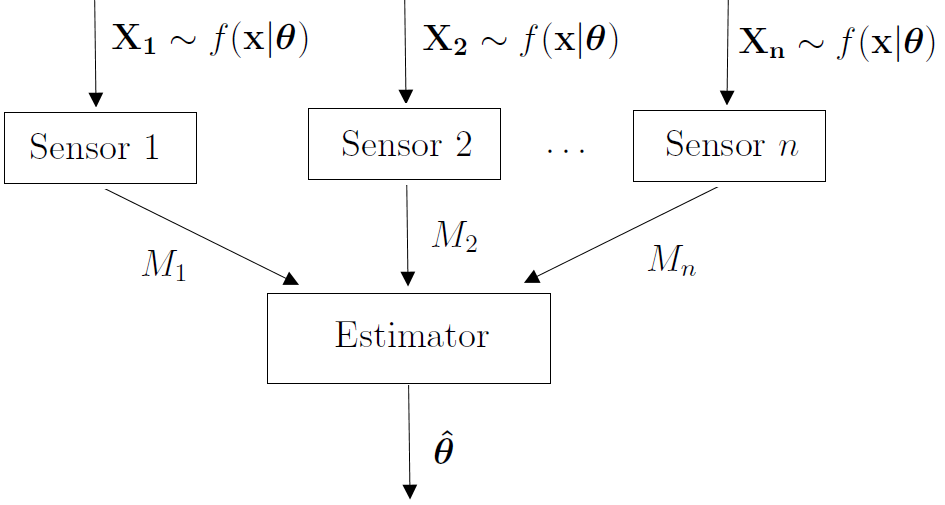}
\caption{Distribution estimation from quantized samples.} 
\squeezeup
\label{fig-distribution-estimation-from-quantized-samples}
\end{figure}

Consider the multiterminal detection system shown in Figure~\ref{fig-distribution-estimation-from-quantized-samples}. In this problem a memoryless vector source $\dv X$ has joint distribution $f(\dv x|\theta)$ that depends on an unknown (vector) parameter $\bs \theta=(\theta_1,\hdots,\theta_d) \in \mathbb{R}^d$, with $d \geq 1$. A number of agents or sensors, say $n$, observe each one independent sample of $\dv X$; and each of them uses $k \geq 1$ bits to describe its sample to a fusion center whose goal is to find a distribution $\hat{f}$ that approximates the unknown (parametric) distribution $f(\dv x|\theta)$ in a suitable sense. How well can $f(\dv x|\bs \theta)$ be approximated from the quantized samples ? This question has so far been resolved (partially) only for few special cases, among which the $L_2$ loss~\cite{HanCOLT2018,Barnes2019}. Worse, even in the extreme case in which $k$ is large (unquantized samples) little is known about this problem for general loss measures~\cite{KOPS15}.

\vspace{0.2cm}

 In this paper we study an instance of this problem under general $L_p$-norms, where $p \in \mathbb{R}$ with $p > 1$, as well as the related Wasserstein distance of order $p$. We recall that for given distributions $P$ and $Q$, the $p$-Wasserstein distance between $P$ and $Q$ is defined as~\cite{Villani2009}
\begin{equation}
W_p(P,Q)= \inf_{\nu \; \in \; \Pi(P,Q)} \: \left( \mathbb{E}_{(Z,Y) \; \sim \; \nu} \:\: \left[ d^p(Z,Y) \right]  \right)^{\frac{1}{p}}
\label{definition-Wasserstein-distance}
\end{equation}
where the random variables $Z \in \mc Z$ and $Y \in \mc Y$ have distributions $P$ and $Q$ respectively, i.e., $ Z \sim P$ and $Y \sim Q$; the set $\Pi(P,Q)$ designates the set of measures $\nu$ on $\mc Z \times \mc Y$ (called couplings) whose $Z$-marginal and $Y$-marginal coincide with $P$ and $Q$ respectively; and $d : \mc Z \times \mc Y \rightarrow \mathbb{R}_{+}$ is a given distance measure. Specifically, let
\begin{equation}
\dv X_1, \dv X_2, \hdots, \dv X_n \stackrel{\text{i.i.d.}}{\sim} f(\dv x|\bs \theta)
\end{equation}
where $\bs \theta \in \Theta \subseteq \mathbb{R}^d$. Agent $i$, $i=1,\hdots,n$, observes the sample $\dv X_i$ and sends a $k$-bit string $M_i$ to the fusion center. We assume that the agents process their observations and communicate with the fusion center simultaneously and independently of each other. The fusion center uses the tuple $\dv M^{(n)}=(M_1,\hdots,M_n)$ to find an estimate $\hat{\bs \theta} := \hat{\bs \theta}(\dv M^{(n)})$ of the unknown parameter $\bs \theta$; and then approximates the unknown source distribution as $f(\dv x|\hat{\bs \theta})$. Our goal is to design the estimator $\hat{\bs \theta}$ so as to minimize the worst case power-$p$ Wasserstein risk, i.e., to characterize
\begin{equation}
\inf_{\hat{\bs \theta}} \sup_{\bm{\theta} \in \bm{\Theta}} \: \mathbb{E} \left[  W_p^p\left(f(\mathbf{x}|\bm{\hat{\theta}}),\: f(\mathbf{x} \big |\bm{\theta})\right) \right].
\label{minimax-Wasserstein-loss}
\end{equation}
When the underlying distance in the Wasserstein risk~\eqref{minimax-Wasserstein-loss} is based on the $L_p$-norm, it is instrumental to study the following related parameter estimation problem under the $L_p$-norm,
\begin{equation}
\inf_{\hat{\bs \theta}} \sup_{\bs \theta \in \bm{\Theta}}  \: \mathbb{E} \left[\left\vert\left\vert \hat{\bs \theta} - \bs \theta \right\vert\right\vert^p_p\right]
\label{minimax-p-norm-parameter-estimation-loss}
\end{equation} 
where $\|\cdot\|_p$ designates the $L_p$ norm.

The main contributions of this paper are as follows. First, we establish a generalization of the well known van Trees inequality~\cite[p. 72]{vT68}, which is a Bayesian analog of the information inequality, to $L_p$-norms with $p >  1$, in terms of generalized Fisher information of order $p$~\cite{Boekee1977b}. This result, which holds under some mild conditions (see Section~\ref{Problem}) that are assumed to hold throughout, may be of independent interest in its own right. In particular, its proof is more direct than the traditional methods of Assouad, Fano, or Le Cam \cite{Yu1997}. Then, we develop lower bounds on the losses~\eqref{minimax-Wasserstein-loss} and~\eqref{minimax-p-norm-parameter-estimation-loss} in terms of the order $p$, the number of samples $n$, the number of quantization bits $k$ and the parameter space $d$. Some of our results generalize those of~\cite{Barnes2019}, which are established therein for the $L_2$ loss, to the case of $L_p$ loss for arbitrary $ p > 1$. Particularly interesting in these bounds is that, for some example source classes that we study, they decrease with the number of samples at least as $1/(n^{\frac{p}{2}})$; and with $k$ at least as $1/(k^\frac{p}{r})$ for some suitable value $r >0$. Key to the proofs of the results of this paper are some judicious applications of inequalities such as H\"older inequality and the Marcinkiewicz-Zygmund inequality~\cite{Ren2001}.

\subsection{Related Works}

The problem of statistical estimation in distributed settings has attracted increasing interest in recent years, in part motivated by learning applications at the wireless Edge. Most relevant to this paper are the works~\cite{Barnes2019,KOPS15}. In particular, the parameter estimation problem~\eqref{minimax-p-norm-parameter-estimation-loss} is studied in~\cite{Barnes2019} for the case $p=2$, i.e., the squared $L_2$ loss. Specifically, in~\cite{Barnes2019} the authors build upon~\cite{A11} to derive lower bounds on the risk~\eqref{minimax-p-norm-parameter-estimation-loss} that account (partially) for the loss of Fisher information (relative to the unquantized setting~\cite{KOPS15}) that is caused by quantization in the case $p=2$. In doing so, they use the standard van Trees inequality which is a Bayesian version of the well known Cram\'er-Rao inequality for the Euclidean norm $L_2$. In this paper, for the study of the problem~\eqref{minimax-p-norm-parameter-estimation-loss} for general $p>1$, after generalizing the usual van Trees inequality to general $L_p$-norms, essentially we follow the approach of~\cite{Barnes2019}. For non-parametric models of densities over $[0,1]$ that are  H\"{o}lder continuous of smoothness $s \in (0,1]$, \cite{HanISIT2018} provides upper and lower bounds on the worst case error under the $L_1$ norm. For more on this and other related works, the reader may refer to~\cite{Barnes2019,KOPS15,HanISIT2018} as well as the references mentioned therein. For related works on Wasserstein loss based learning, see, e.g.,~\cite{Frogner2015,Ambrogioni2018,Tolstikhin2018,Arjovski2017,Gulrajani2017}.

\section{Formal Problem Formulation and Definitions}\label{Problem}

Consider the model shown in Figure~\ref{fig-distribution-estimation-from-quantized-samples}. Here, there are $n$ sensors which observe each one sample of a memoryless vector source $\dv X$. We assume that the underlying distribution or density of $\dv X$ is parametrized by an unknown vector parameter $\bs \theta=(\theta_1,\hdots,\theta_d) \in \Theta \subseteq \mathbb{R}^d$ of dimension $d \geq 1$; and we write $f(\dv x) := f(\dv x|\bs \theta)$ for $\bs \theta \in \Theta$. The samples $\dv X_1, \dv X_2, \hdots, \dv X_n$ are all independent; and they are processed independently by the sensors. Sensor $i$, $i=1,\hdots,n$, encodes its sample $\dv X_i$ into a $k$-bit string $M_i \in [1,2^k]$. A (possibly stochastic) $k$-bit quantization strategy for $\dv X_i$ at Sensor $i$ can be expressed in terms of the conditional probability
\begin{equation}
p_i(m|\dv x) := p_{M_i|\dv X_i}(m|\dv x) \:\: \text{for}\:\: m \in [1,2^k]\:\:\text{and}\:\: \dv x \in \mc X.
\label{encoding-function-ith-sensor}
\end{equation}  
The sensors communicate their $k$-bit quantization messages simultaneously and independently to a fusion center whose goal is to produce an estimate of the unknown distribution $f(\dv x|\bs \theta)$ from the tuple $\dv M^{(n)}=(M_1,\hdots,M_n)$. The fusion center first finds an estimate $\hat{\bs \theta} := \hat{\bs \theta}(\dv M^{(n)})$ of the unknown parameter $\bs \theta$; and then approximates the unknown source distribution as $f(\dv x|\hat{\bs \theta})$. Let $p \in \mathbb{R}$, $p >1$, be given. Our goal is to design the estimator $\hat{\bs \theta}$ so as to minimize the worst case power-$p$ Wasserstein risk
\begin{equation}
\inf_{\hat{\bs \theta}} \sup_{\bm{\theta} \in \bm{\Theta}} \: \mathbb{E} \left[  W_p^p\left(f(\mathbf{x}|\bm{\hat{\theta}}),\: f(\mathbf{x} \big |\bm{\theta})\right) \right]
\label{minimax-Wasserstein-loss2}
\end{equation}
where the Wasserstein distance between distributions under distance $d(\cdot,\cdot)$ is defined as in~\eqref{definition-Wasserstein-distance}. As we already mentioned, when the distance $d(\cdot,\cdot)$ is the $L_p$-norm, we also consider the following parameter estimation problem under the $L_p$-norm,
\begin{equation}
\inf_{\hat{\bs \theta}} \sup_{\bs \theta \in \bm{\Theta}}  \: \mathbb{E} \left[\left\vert\left\vert \hat{\bs \theta} - \bs \theta \right\vert\right\vert^p_p\right].
\label{minimax-p-norm-parameter-estimation-loss2}
\end{equation} 

\noindent We assume that for all $i=1,\hdots,n$ there is a well defined joint probability distribution with density
\begin{equation}
f_i(\dv x, m|\bs \theta) = f(\dv x|\bs \theta) p_i(m|\dv x)
\end{equation}
and that $p_i(m|\dv x)$ is a regular conditional probability (it denotes the encoding function at the $i^{\text{th}}$ Sensor -- see~\eqref{encoding-function-ith-sensor}). For a given $\bs \theta \in \mathbb{R}^d$ and quantization strategy at the $i^{\text{th}}$ Sensor, the likelihood that the quantization message $M_i$ takes a specific value $m$ is denoted as $p_i(m|\theta)$. The vector 
\begin{align}
S_{i, \bs \theta}(m) &= \left(S_{i, \theta_1}(m), \hdots, S_{i, \theta_d}(m) \right) \nonumber\\
&= \left(\frac{\partial}{\partial \theta_1} \log p_i(m|\bs \theta), \hdots, \frac{\partial}{\partial \theta_d} \log p_i(m|\bs \theta) \right)
\label{score-function-quantization-message-ith-encoder}
\end{align}
is the score function of this likelihood. For convenience, for $\dv x \in \mc X$ we let
\begin{align}
S_{\bs \theta}(\dv x) &= \left(S_{\theta_1}(\dv x), \hdots, S_{\theta_d}(\dv x) \right) \nonumber\\
&= \left(\frac{\partial}{\partial \theta_1} \log f(\dv x|\bs \theta), \hdots, \frac{\partial}{\partial \theta_d} \log p(\dv x|\bs \theta) \right)
\label{score-function-vector-observation}
\end{align}
denote the score of the likelihood $f(\dv x|\bs \theta)$.

We make the following assumptions which we assume to hold throughout unless otherwise stated. The distributions $f(\mathbf{x}|\bm{\theta})$ and $\{p_i(m|\bs \theta)\}_{i=1}^n$ are all assumed to be continuously differentiable at every coordinate of $\bs \theta$. Also, for all $i=1,\hdots,n$ the score function $S_{i, \bs \theta}(m)$ as well as its  $p^{th}$ moment exist. Similarly, for all $i=1,\hdots,n$ the generalized Fisher information matrix of order $p$ for estimating $\bs \theta$ from $M_i$ and that for estimating it from $\mathbf{X}_i$, both defined as in Definition~\ref{definition-generalized-Fisher-information} that follows, are assumed to exist and to be continuous in $\theta_i$. 

\begin{Def}~\label{definition-generalized-Fisher-information}
Let $p \in \mathbb{R}$ with $p > 1$ be given. For a multivariate random variable $\dv X$ with  probability distribution $f(\mathbf{x}|\bm{\theta})$ that depends on an unknown vector parameter $\bm{\theta}=[\theta_1, \ldots, \theta_d] \in \mathbb{R}^d$, for all $i=1,\hdots,n$ the generalized Fisher information of order $p$ for estimating $\theta_i$ from $\dv X$ is defined as \cite{Boekee1977b,Boekee1977a}
\begin{equation}
I_{\dv X}^{(p)}(\theta_i) = \left( \mathbb{E} \left[ \left \vert \frac{\partial}{\partial \theta_i} \left[\log{ f(\dv X|\bs \theta) } \right] \right \vert^{\frac{p}{p-1}} \right] \right)^{p-1}.
\end{equation}
Also, define
\begin{equation}
\Omega^{(p)}_{\mathbf{X}}(\bm{\theta}) := \sum_{i=1}^d \left( \mathbb{E} \left[ \left \vert \frac{\partial}{\partial \theta_i} \left[\log{ f(\dv X|\bs \theta) } \right] \right \vert^{\frac{p}{p-1}} \right] \right)^{p-1}
\label{definition-trace-generalized-fisher-information-estimating-theta-from-vectorX}
\end{equation}
which can be interpreted as the trace of the generalized Fisher information matrix of order $p > 1$ for estimating $\bs \theta$ from $\dv X$. \qed
\end{Def}

It can easily be checked that for $p=2$,  the quantity $\Omega^{(2)}_{\mathbf{X}}(\bm{\theta})$ is  the trace of the standard Fisher information matrix, i.e., $\Omega^{(2)}_{\mathbf{X}}(\bm{\theta}) = \rm{Tr} \big(I_{\dv X}(\bs \theta)\big)$. As it will become clearer from the rest of this paper, throughout we will make extensive usage of the quantity $\Omega^{(p)}_{\mathbf{X}}(\bm{\theta})$ as defined by~\eqref{definition-trace-generalized-fisher-information-estimating-theta-from-vectorX}. For example, for the problem of estimating $\bs \theta$ from the quantization tuple $\dv M^{(n)}=(M_1,\hdots,M_n)$ we will use
\begin{equation}
\Omega^{(p)}_{\dv M^{(n)}}(\bm{\theta}) := \sum_{i=1}^d \left( \mathbb{E} \left[ \left \vert \frac{\partial}{\partial \theta_i} \left[\log{ p(\dv M^{(n)}|\bs \theta) } \right] \right \vert^{\frac{p}{p-1}} \right] \right)^{p-1}
\label{definition-trace-generalized-fisher-information-estimating-theta-from-quantization-tuple}
\end{equation}
where $p(\dv M^{(n)}|\bs \theta)=\prod_{i=1}^n p_i(M_i|\bs \theta)$ due to the independence of the samples and encoding functions at the sensors. Likewise, for a single quantization message $M_j$, $j=1,\hdots,n$, we use $\Omega^{(p)}_{M_j}(\bm{\theta})$ which is given by the RHS of~\eqref{definition-trace-generalized-fisher-information-estimating-theta-from-quantization-tuple} in which $p(\dv M^{(n)}|\bs \theta)$ is replaced with $p_j(M_j|\theta)$. Also, when we take a Bayesian approach and let $\mu(\bs \theta)$ be a prior on $\Theta$, we will use
\begin{equation}
\Omega^{(p)}(\mu) := \sum_{i=1}^d \left( \mathbb{E} \left[ \left \vert \frac{\partial}{\partial \theta_i} \left[\log{ \mu(\bs \theta) } \right] \right \vert^{\frac{p}{p-1}} \right] \right)^{p-1}.
\label{definition-trace-generalized-fisher-information-estimating-theta-from-prior}
\end{equation}

\section{A van Trees type inequality for $L_p$-norms}~\label{VanTreesXN1}

In this section, we take a Bayesian approach. We let the parameter space $\Theta$ to be the Cartesian product of closed intervals on the real line, i.e., $\Theta = \prod_{i=1}^d [\theta_{i, \rm{min}}, \theta_{i, \rm{max}}]$. Let $\pi$ some probability distribution on $\Theta$ with a density measure $\mu(\bs \theta)$ with respect to the Lebesgue measure (a prior on $\bs \theta$). We make the assumption that $\mu(\bs \theta)$ factorizes as $\mu(\bs \theta) = \prod_{i=1}^d \mu_i(\theta_i)$. Also, suppose that $f(\dv x|\cdot)$ and $\mu(\cdot)$ are both absolutely continuous; and that $\mu$ converges to zero at the boundaries of $\Theta$, i.e., for all $i=1,\hdots,d$
\begin{equation}
\lim_{\theta_i \to \theta_{i,\text{min}}} \mu_i(\theta_i) = \lim_{\theta_i \to \theta_{i,\text{max}}} \mu_i(\theta_i) = 0.
\label{vanishing-prior-boundaries-of-Theta}
\end{equation}

\noindent For scalar $X$ and $\theta$ (i.e, $d=1$), the usual van Trees inequality~\cite{Gill1995}, which is a Bayesian version of the well-known Cram\'{e}r-Rao inequality established for the Euclidean norm $L_2$, states that
\begin{equation}
\mathbb{E}[(\hat{\theta}(X)-\theta)^2] \geq \frac{1}{\mathbb{E}_{\theta} [I_X(\theta)] + I(\mu)}
\end{equation}
where $I_X(\theta)$ is the standard Fisher information for estimating $\theta$ from $X$ and $I(\mu)$ designates that from the prior.
 
\vspace{0.2cm}

\noindent The following theorem provides a lower bound on the average error in estimating $\bs \theta=(\theta_1,\hdots,\theta_d)$ from $\dv X$ under the $L_p$ norm, for arbitrary $p >1$. It can be seen a van Trees type inequality for $L_p$ norms. The result can also be regarded as a Bayesian version of one in~\cite{Boekee1977b}. Its proof is essentially based on a judicious application of H\"older inequality and is different from the one of~\cite{Boekee1977b}. 

\begin{Th}~\label{ThVanTreesXVTV2}
For $p > 1$, the average estimation error under the norm $L_p$ satisfies the following:
\begin{itemize}
\item[i)] If $1 < p < 2$, then we have
\begin{align}
&\mathbb{E} \left[ \left \vert \left \vert \bm{\hat{\theta}}(\mathbf{X}) - \bm{\theta} \right \vert \right \vert_p^p \right] \geq \nonumber\\
& \frac{d^p}{ \left(  \left(\mathbb{E}_{\bm{\Theta}} \left[ \left( \Omega^{(p)}_{\mathbf{X}}(\bm{\theta})\right)^{\frac{1}{p-1}} \right] \right)^{\frac{p-1}{p}} + d^{\frac{p-2}{p}}  \left(  \Omega^{(p)}(\mu) \right)^{\frac{1}{p}} \right)^{p}}
\label{lower-bound-average-error-generalized-fisher-information-p-smaller-than-two}
\end{align}
where $\Omega^{(p)}_{\mathbf{X}}(\bm{\theta})$ and $\Omega^{(p)}(\mu)$ are given by~\eqref{definition-trace-generalized-fisher-information-estimating-theta-from-vectorX} and \eqref{definition-trace-generalized-fisher-information-estimating-theta-from-prior}, respectively.
\item[ii)] If $p \geq 2$, then we have
\begin{equation}
\mathbb{E} \left[ \left \vert \left \vert \bm{\hat{\theta}}(\mathbf{X}) - \bm{\theta} \right \vert \right \vert_p^p \right] \geq \frac{d^{\left(1+\frac{p}{2}\right)}}{\left( \mathbb{E}_{\bm{\Theta}} \left[ \rm{Tr}(I_{\mathbf{X}}(\bm{\theta})) \right] + \rm{Tr}(I(\mu)) \right)^{\frac{p}{2}}}.
\label{lower-bound-average-error-generalized-fisher-information-p-larger-than-two}
\end{equation}
\end{itemize}
\end{Th}

\begin{proof}
The proof of Theorem~\ref{ThVanTreesXVTV2} is given in Section~\ref{secVI_Th1}.  
\end{proof}

\begin{Rem}
It is easy to see that for $p=2$ the result of Theorem~\ref{ThVanTreesXVTV2} is the standard van Trees inequality~\cite{vT68} (see also~\cite{Gill1995}). Also, observe that for values of $p \in \mathbb{R}$ which are such that $1<p<2$ the result involves generalized Fisher information of order $p$ for both $X$ and the prior $\mu$, whereas for $p \geq 2$ it involves standard Fisher information (i.e, of order $2$). We note that for $p \geq 2$, it is possible to derive a bound that is similar to the RHS of~\eqref{lower-bound-average-error-generalized-fisher-information-p-smaller-than-two}, i.e., one that involves generalized Fisher information of order $p$, as below
\begin{align}
&\mathbb{E}\left[ \left \vert \left \vert \bm{\hat{\theta}}(\mathbf{X}) - \bm{\theta} \right \vert \right \vert_p^p \right]  \geq  \nonumber \\
& \frac{d^{2}}{ \left( \left(\mathbb{E}_{\bm{\Theta}} \left[ \left( \Omega^{(p)}_{\mathbf{X}}(\bm{\theta}) \right)^{\frac{1}{p-1}} \right] \right)^{\frac{p-1}{p}} + \left(\Omega^{(p)}(\mu) \right)^{\frac{1}{p}} \right)^{p}}. \label{ThVanTreesXVTV-alternative-lower-bound}
\end{align}
The proof of the lower bound (\ref{ThVanTreesXVTV-alternative-lower-bound}) is given in Section \ref{secVI_Th1-alternative-lower-bound}. However, such bound does not seem to compare easily with the RHS of~\eqref{lower-bound-average-error-generalized-fisher-information-p-larger-than-two}. In addition, the RHS of ~\eqref{lower-bound-average-error-generalized-fisher-information-p-larger-than-two} turns out to be more tractable analytically for the examples that we will consider in the rest of this paper.
\end{Rem}

\noindent A more general inequality than that of Theorem~\ref{ThVanTreesXVTV2} for estimating a continuously differentiable function $\psi$ of $\theta$ is easily obtained in exactly the same way. 

\begin{Cor}~\label{ThVanTreesXVTV2Cor}
For any vector-valued function $\psi(\bm{\theta})$ which is continuously differentiable in each component $\psi_i(\bm{\theta})$, the following holds.
\begin{itemize}
\item[i)] If $1 < p < 2$, we have
\begin{align}
& \mathbb{E} \left[ \left \vert \left \vert \psi(\bm{\hat{\theta}}(\mathbf{X})) - \psi(\bm{\theta}) \right \vert \right \vert_p^p \right]  \geq \nonumber\\
& \frac{\left \vert \sum_{i=1}^d \mathbb{E}_{\bm{\Theta}} \left[ \frac{\partial \psi_i(\bm{\Theta})}{\partial \Theta_i} \right] \right \vert^{p}}{\left(  \left( \mathbb{E}_{\bm{\Theta}} \left[ \left( \Omega^{(p)}_{\mathbf{X}}(\bm{\theta}) \right)^{\frac{1}{p-1}} \right] \right)^{\frac{p-1}{p}} + d^{\frac{p-2}{p}}   \left(\Omega^{(p)}(\mu) \right)^{\frac{1}{p}}  \right)^{p}}. \nonumber
\end{align}
\item[ii)] If $p \geq 2$, we have
\begin{equation}
\mathbb{E} \left[ \left \vert \left \vert \psi(\bm{\hat{\theta}}(\mathbf{X})) - \psi(\bm{\theta}) \right \vert \right \vert_p^p \right]  \geq  \frac{d^{(1-\frac{p}{2})} \left \vert \sum_{i=1}^d \mathbb{E}_{\bm{\Theta}} \left[ \frac{\partial \psi_i(\bm{\Theta})}{\partial \Theta_i} \right] \right \vert^{p} }{\left( \mathbb{E}_{\bm{\Theta}} \left[ \rm{Tr}(I_{\mathbf{X}}(\bm{\theta})) \right] + \rm{Tr}(I(\mu)) \right)^{\frac{p}{2}}}. \nonumber
\end{equation}
\end{itemize}
\end{Cor}

\begin{proof}
The proof of Corollary~\ref{ThVanTreesXVTV2Cor} is given in Section~\ref{secVI_CorTh1}.  
\end{proof}

\section{Distributed parameter estimation from quantized samples}~\label{secIV}

\vspace{-0.3cm}

Let us now consider the minimax parameter estimation problem~\eqref{minimax-p-norm-parameter-estimation-loss2} described in Section~\ref{Problem}. Let $\mu$ be a prior on $\bs \theta$ that factorizes as in Section~\ref{VanTreesXN1} and satisfies~\eqref{vanishing-prior-boundaries-of-Theta}. Substituting $\dv X$ in Theorem~\ref{ThVanTreesXVTV2} with $\dv M^{(n)}=(M_1,\hdots,M_n)$ we obtain a lower bound on the worst case error under the $L_p$ norm. Such bound, however, does not seem to reflect the right behavior for the error decrease as a function of the number of samples $n$ (for given $p >1$ and fixed $k \geq 1$ and $d \geq 1$). A better bound, which uses the techniques of the proof of Theorem~\ref{ThVanTreesXVTV2} and combines them appropriately with Marcinkiewicz-Zygmund inequality~\cite{Ren2001}, is stated in the following theorem. 

\begin{Th}\label{ThVanTreesMVTV2}
For $p > 1$, the worst case estimation error under the norm $L_p$ satisfies the following:
\begin{itemize}
\item[i)] If $1 < p < 2$, we have
\begin{align}
&\sup_{\bm{\theta} \in \bm{\Theta}} \mathbb{E}_{\mathbf{M^{(n)}}|\bm{\Theta}} \left[ \left \vert \left \vert \bm{\hat{\theta}}(\mathbf{M^{(n)}}) - \bm{\theta} \right \vert \right \vert_p^p \left. \right \vert \bm{\Theta} \right]  \geq d^{p} \left[ d^{\frac{p-2}{p}} \left( \Omega^{(p)}(\mu) \right)^{\frac{1}{p}}  \right. \nonumber \\
&\left. +  \frac{1}{p-1} \left( \sum_{j=1}^{n} \left(\mathbb{E}_{\bm{\Theta}} \left[ \left( \Omega^{(p)}_{M_j}(\bm{\theta}) \right)^{\frac{1}{p-1}} \right] \right)^{\frac{2(p-1)}{p}} \right)^{\frac{1}{2}} \right]^{-p} \nonumber
\end{align}
where, for $j=1,\hdots,n$, $\Omega^{(p)}_{M_j}(\bm{\theta})$ is obtained using~\eqref{definition-trace-generalized-fisher-information-estimating-theta-from-quantization-tuple} and $\Omega^{(p)}(\mu)$ is given by~\eqref{definition-trace-generalized-fisher-information-estimating-theta-from-prior}.
\item[ii)] If $p \geq 2$, we have
\begin{align} 
&\sup_{\bm{\theta} \in \bm{\Theta}} \mathbb{E}_{\mathbf{M^{(n)}}|\bm{\Theta}} \left[ \left \vert \left \vert \bm{\hat{\theta}}(\mathbf{M^{(n)}}) - \bm{\theta} \right \vert \right \vert_p^p \left. \right \vert \bm{\Theta} \right]  \geq \nonumber \\
&\quad d^{\left(1+\frac{p}{2}\right)} \left( \sum_{j=1}^{n} \mathbb{E}_{\bm{\Theta}} \left[ \rm{Tr}(I_{M_j}(\bm{\theta})) \right] + \rm{Tr}(I(\mu)) \right)^{-\frac{p}{2}}. \nonumber
\end{align}
\end{itemize}
\end{Th}

\textbf{Proof:} \underline{\textit{1) Case $1 <p <2$:}} Let $q \in \mathbb{R}$ such that $\frac{1}{p}+\frac{1}{q}=1$, i.e., $q=p/(p-1)$. Also, consider the following two functions $g(\cdot)$ and $h(\cdot)$ defined, for $\dv x \in \mc X$, $\bs \theta =[\theta_1,\hdots,\theta_d] \in \Theta$ and a specific quantization messages tuple $\dv m^{(n)}=(m_1,\hdots,m_n) \in [1,2^k]^n$  as 
\begin{subequations}
\begin{align}
\label{DefG}
g(\dv m^{(n)},\bs \theta) &= \sum_{i=1}^d \frac{\partial}{\partial \theta_i} \left[ \log{\left(p(\dv m^{(n)}|\bs \theta) \mu(\bs \theta) \right)} \right]\\
h(\dv m^{(n)},\bs \theta) &=  \hat{\bs \theta}(\dv m^{(n)})- \bs \theta
\end{align}
\end{subequations}	
where in~\eqref{DefG} the quantization messages joint probability is $p(\mathbf{m^{(n)}}|\bm{\theta})=\prod_{j=1}^{n} p_j(m_j|\bm{\theta})$. For convenience,  for $i=1,\hdots,d$ we will denote the $i^{th}$ component of $h(\dv m^{(n)}, \bs \theta)$ as $h_i(\dv m^{(n)},\bs \theta)$, i.e.,
\begin{equation}
h_i(\dv m^{(n)},\bs \theta) = \hat{\theta}_i(\dv m^{(n)})- \theta_i = \left(h(\dv m^{(n)},\bs \theta)\right)_i.
\label{DefH}
\end{equation}
\noindent Using the fact that the prior measure $\mu$ converges to zero at the boundaries of $\Theta$,  it is easy to see that 
\begin{align}
\sum_{\mathbf{m^{(n)}}} \int_{\theta_i} &h_i(\mathbf{m^{(n)}},\bs \theta) \frac{\partial}{\partial \theta_i} \left[p(\mathbf{m^{(n)}}|\bm{\theta}) \mu_i(\theta_i) \right] \, \rm{d}{\theta_i}  = 1. 
\label{IntEq1MVTV2RG2}
\end{align}

\noindent By partial integration and~\eqref{IntEq1MVTV2RG2}, we get for $i=1,\hdots,d$, that
\begin{align}
&\mathbb{E}_{(\mathbf{M^{(n)}},\bm{\Theta})} \left[ h_i(\mathbf{M^{(n)}},\bs \Theta) g(\mathbf{M^{(n)}},\bm{\Theta}) \right] = d.
\end{align}
Thus, for all $i=1,\hdots,d$, we have
\begin{align}
d &\leq \mathbb{E}_{(\mathbf{M^{(n)}},\bm{\Theta})} \left[ \left \vert h_i(\mathbf{M^{(n)}},\bs \Theta) g(\mathbf{M^{(n)}},\bm{\Theta}) \right \vert \right]. 
\label{BoundLHTRG2MVTV2}
\end{align}

\noindent Applying H\"{o}lder's inequality for expectations yields
\begin{align}
&\mathbb{E}_{(\mathbf{M^{(n)}},\bm{\Theta})} \left[ \left \vert h_i(\mathbf{M^{(n)}},\Theta) g(\mathbf{M^{(n)}},\bm{\Theta}) \right \vert \right] \leq \label{OriginalHolderMVTV2RL2} \\
&\left( \mathbb{E} \left[ \left \vert h_i(\mathbf{M^{(n)}},\bs \Theta) \right \vert^p \right] \right)^{\frac{1}{p}}\left(\mathbb{E} \left[ \left \vert g(\mathbf{M^{(n)}},\bm{\Theta}) \right \vert^q \right] \right)^{\frac{1}{q}}.
\label{Holder-inequality}
\end{align}
The first element of the right-hand side produces the desired risk as 
\begin{align}
&\sup_{\bm{\theta} \in \bm{\Theta}} \mathbb{E}_{\mathbf{M^{(n)}}|\bm{\Theta}} \left[ \left \vert \left \vert \bm{\hat{\theta}}(\mathbf{M^{(n)}}) - \bm{\theta} \right \vert \right \vert_p^p \left. \right \vert \bm{\Theta} \right]  \nonumber \\
&\quad \geq \sum_{i=1}^d \mathbb{E}_{(\mathbf{M^{(n)}},\bm{\Theta})} \left[ \left \vert h_i(\mathbf{M^{(n)}},\bs \Theta) \right \vert^p \right] \label{almostDoneBound2MVTV2RL2}
\end{align}
where the inequality follows by substituting using~\eqref{DefH} and using that fact that the supremum of a function is larger than its expectation.  

\noindent We now upper bound the second expectation term of the RHS of~\eqref{Holder-inequality}. For convenience, let for $j=1,\hdots,2^k$
\begin{equation}
l(m_j,\bm{\theta})= \sum_{i=1}^d \frac{\partial}{\partial \theta_i} \left[ \log{p(m_j|\bm{\theta})} \right]
\label{intermediate-step-1}
\end{equation}
It is easy to see that for all $\bs \theta$, we have
\begin{equation}
\mathbb{E}_{M_j |\bs \Theta} \left[ l(M_j,\bs \Theta) | \bs \Theta = \bs \theta \right]  =0.
\label{proof-generalization-of-van-Trees-inequality-step8}
\end{equation}

\noindent Then, we have
\begin{align}
&\left(\mathbb{E}_{(\mathbf{M^{(n)}},\bm{\Theta})} \left[ \left \vert g(\mathbf{M^{(n)}},\bm{\Theta}) \right \vert^q \right] \right)^{\frac{1}{q}} \leq \label{ToreplacewithEqRE2RL2} \\
&\left( \mathbb{E}_{(\mathbf{M^{(n)}},\bm{\Theta})} \left[ \left \vert \sum_{j=1}^{n} l(M_j,\bm{\Theta}) \right \vert^q \right] \right)^{\frac{1}{q}} + d^{\frac{p-1}{p}} \left(\Omega^{(p)}(\mu) \right)^{\frac{1}{p}},\nonumber
\end{align}
where the inequality holds by a double application by Minkowski's inequality: first for expectations using that for all $Z$ and $T$ we have $\left(\mathbb{E} [|Z+T|^q] \right)^{\frac{1}{q}}$ $\leq \left(\mathbb{E} [|Z|^q] \right)^{\frac{1}{q}} + \left(\mathbb{E} [|T|^q] \right)^{\frac{1}{q}}$; and then that $\left(\mathbb{E} [|\sum_{i=1}^d Z_i|^q] \right)^{\frac{1}{q}} $$\leq \sum_{i=1}^d \left(\mathbb{E} [|Z_i|^q] \right)^{\frac{1}{q}}$, $\forall \, q>1$ and  $\sum_{i=1}^d u_i^\frac{1}{p} $$\leq d^{\frac{p-1}{p}} \left(\sum_{i=1}^d u_i \right)^{\frac{1}{p}}$, $\forall \, u_i>0$, $p>1$. 

\noindent Next, since the quantities $\{l(M_j,\bm{\Theta})\}_j$ are independent and satisfy that $\mathbb{E}_{M_j|\bm{\Theta}} \left[ l(M_j,\bm{\Theta}) \right]=0$ for all $j=1,\hdots,2^k$, the application of Marcinkiewicz-Zygmund inequality~\cite{Ren2001,Burkholder1988} yields
\begin{align}
& \mathbb{E}_{\mathbf{M^{(n)}}|\bm{\Theta}} \left[ \left \vert \sum_{j=1}^{n} l(M_j,\bm{\Theta}) \right \vert^q \left. \right \vert \bm{\Theta} \right] \leq \nonumber \\
&\quad \quad B_q \, \mathbb{E}_{\mathbf{M^{(n)}}|\bm{\Theta}}  \left[ \left( \sum_{j=1}^{n} l^2(M_j,\bm{\Theta})\right)^{\frac{q}{2}} \left. \right \vert \bm{\Theta} \right]
\label{intermediate-step}
\end{align}
where $B_q=(q-1)^q>0$.

\noindent Continuing from of~\eqref{intermediate-step}, we get
\begin{align}
&\left( \mathbb{E}_{(\mathbf{M^{(n)}},\bm{\Theta})} \left[ \left \vert \sum_{j=1}^{n} l(M_j,\bm{\Theta}) \right \vert^q \right] \right)^{\frac{2}{q}} \leq \nonumber \\
&\stackrel{(a)}{\leq} \frac{1}{(p-1)^2} \sum_{j=1}^{n} \left(\mathbb{E}_{(\mathbf{M^{(n)}},\bm{\Theta})}  \left[ |l(M_j,\bm{\Theta})|^q \right]\right)^{\frac{2}{q}} \nonumber \\
&\stackrel{(b)}{\leq} \frac{d^{\frac{2}{p}}}{(p-1)^2} \sum_{j=1}^{n}  \left(\mathbb{E}_{(\mathbf{M^{(n)}},\bm{\Theta})}  \left[ \sum_{i=1}^d \left \vert \frac{\partial}{\partial \theta_i} \left[\log{p(M_j|\bm{\Theta})} \right] \right \vert^q \right]\right)^{\frac{2}{q}} \nonumber \\
&\stackrel{(c)}{=}  \frac{ d^{\frac{2}{p}}}{(p-1)^2} \sum_{j=1}^{n} \left( \mathbb{E}_{\bm{\Theta}}  \left[ \left( \Omega^{(p)}_{M_j}(\bm{\theta}) \right)^{\frac{1}{p-1}} \right] \right)^{\frac{2(p-1)}{p}} , \label{CombPart2MVTV2RL2}
\end{align}
where $(a)$ follows by using the quantization messages are independent, substituting $q=p/(p-1)$ and applying Minkowski's inequality $\left(\mathbb{E} [|\sum_{i=1}^{n} Z_i|^{\frac{q}{2}}] \right)^{\frac{2}{q}} \leq \sum_{i=1}^{n} \left(\mathbb{E} [|Z_i|^{\frac{q}{2}}] \right)^{\frac{2}{q}}$ since $q=p>(p-1)>2$; $(b)$ follows by substituting using~\eqref{intermediate-step-1} and using that $\left(\sum_{i=1}^d u_i \right)^{q} \leq d^{q-1} \sum_{i=1}^d u_i^{q} $, $u_i>0$; and $(c)$ holds by ~\eqref{definition-trace-generalized-fisher-information-estimating-theta-from-quantization-tuple}. 

\noindent Finally, combining \eqref{CombPart2MVTV2RL2}, \eqref{ToreplacewithEqRE2RL2}, \eqref{BoundLHTRG2MVTV2} and \eqref{almostDoneBound2MVTV2RL2} and substituting in \eqref{OriginalHolderMVTV2RL2} yields the desired result.

\underline{\textit{2) Case $p \geq 2$:}} In this case, a direct proof can be found in a way that is essentially similar to the above (see Section~\ref{secVI_Th2_RG2} for the details). An indirect proof follows by first observing that
\begin{align}
&\mathbb{E}_{(\mathbf{M^{(n)}},\bm{\Theta})} \left[ \left \vert \left \vert \bm{\hat{\theta}}(\mathbf{M^{(n)}}) - \bm{\theta} \right \vert \right \vert_p^p \right] \geq \nonumber \\
&\quad  d^{1-\frac{p}{2}} \left( \mathbb{E}_{(\mathbf{M^{(n)}},\bm{\Theta})}\left[  \left \vert \left \vert \bm{\hat{\theta}}(\mathbf{M^{(n)}}) - \bm{\theta} \right \vert \right \vert_2^2 \right] \right)^{\frac{p}{2}} \nonumber
\end{align}
which holds due to the norms inequality $ \left \vert \left \vert \dv u \right \vert \right \vert_2 \leq d^{\frac{1}{2}-\frac{1}{p}}  \left \vert \left \vert \dv u \right \vert \right \vert_p$ for all vector $\dv u \in \mathbb{R}^d$; and then combining with the result of~\cite{Barnes2019} for the squared $L_2$ loss. 

\qed 

\vspace{0.2cm}

For some classes of sources $\dv X$ the result of Theorem~\ref{ThVanTreesMVTV2} can be used to find a more explicit lower bound. Recall that for $r \geq 1$, the $\Psi_r$ Orlicz norm of a random variable $Z$ is defined as
\begin{equation}
\|Z\|_{\Psi_r} = \inf \{K \in (0,+\infty[ \: |\: \mathbb{E}\left[\Psi_r(|Z|/K)\right] \leq 1\}
\end{equation}  
where 
\begin{equation}
\Psi_r(u) = \exp(u^r)-1.
\end{equation}
A random variable with finite $\Psi_1$ Orlicz norm is sub-exponential; and a random variable with finite $\Psi_2$ Orlicz norm is sub-Gaussian~\cite{V10}. The next theorem shows that if for some suitable $r \geq 1$ the $\Psi_r$ Orlicz norm of the projection of the score function $S_{\bs \theta}(\dv X)$ as given by~\eqref{score-function-vector-observation} onto any unit vector is bounded from the above by some constant the error decreases at least as $n^{-\frac{p}{2}}$ and at least as $k^{-\frac{p}{r}}$. For convenience, define for $p >1$ and $d \geq 1$ the following quantities,
\begin{subequations}
\begin{align}
\label{definition-constants-theorem3-constantA}
A_p &= \left( \frac{\pi}{2}  \right)^{\frac{1}{p}} \frac{2}{B} \left[\mathcal{B}\left(\frac{2p-1}{2p-2}, \frac{2p-3}{2p-2} \right) \right]^{\frac{p-1}{p}}\\
B_{p,d} & = \frac{2}{p-1} d^{\frac{2-p}{2p}}
\label{definition-constants-theorem3-constantB}
\end{align}
\label{definition-constants-theorem3}
\end{subequations}
where in~\eqref{definition-constants-theorem3-constantA} the function $\mc B(\cdot,\cdot)$ denotes the Eural integral (Beta function) given for $u>0$ and $v>0$ by
\begin{equation}
\mc B(u,v) = \int_{0}^1 t^{u-1}(1-t)^{v-1} dt.
\end{equation}

\begin{Th}~\label{ThVanTreesMVTV2Orlicz}
Suppose $\Theta=[-B,B]^d$ and let $\hat{\bs \theta} := \hat{\bs \theta}(\dv M^{(n)})$ be any estimator of $\bs \theta=[\theta_1,\hdots,\theta_d] \in [-B,B]^d$ from $\dv M^{(n)}=(M_1,\hdots,M_n)$.
\begin{itemize}
\item[i)] For $\frac{3}{2} < p < 2$: if $\:\exists\: r \geq 1/(p-1)$ and $\exists \: I_0 \geq 0$ such that for any $\bs \theta \in \Theta$ and any unit vector $\dv u \in \mathbb{R}^d$
\begin{equation}
\| \langle \dv u, S_{\bs \theta}(\dv X) \rangle \|_{\Psi_r} \leq I_0
\end{equation}
then
\begin{align}
&\sup_{\bm{\theta} \in \bm{\Theta}} \mathbb{E}_{\mathbf{M^{(n)}}|\bm{\Theta}} \left[ \left \vert \left \vert \bm{\hat{\theta}}(\mathbf{M^{(n)}}) - \bm{\theta} \right \vert \right \vert_p^p \left. \right \vert \bm{\Theta} \right]  \geq \nonumber\\
& \qquad \qquad \frac{d^p}{\left( \sqrt{nI_0} k^{\frac{1}{r}} (2^k)^{\frac{2-p}{p}} B_{p,d}   + d^{\frac{p-1}{p}} A_p \right)^{p}} \nonumber
\end{align}
where the quantities $A_p$ and $B_{p,d}$ are given by~\eqref{definition-constants-theorem3}. 
\item[ii)] For $ p \geq 2$: if $\:\exists\: r \geq 1$ and $\:\exists\: I_0 \geq 0$ such that for any $\bm{\theta} \in \bm{\Theta}$, any unit vector $\mathbf{u} \in \mathbb{R}^d$, we have $\left \vert \left \vert \langle\mathbf{u}, S_{\bm{\theta}}(\mathbf{X}) \rangle \right \vert \right \vert_{\Psi_r} \leq I_0$, then
\begin{align}
&\sup_{\bm{\theta} \in \bm{\Theta}} \mathbb{E}_{\mathbf{M^{(n)}}|\bm{\Theta}} \left[ \left \vert \left \vert \bm{\hat{\theta}}(\mathbf{M^{(n)}}) - \bm{\theta} \right \vert \right \vert_p^p \left. \right \vert \bm{\Theta} \right]  \geq \frac{d^{\left(1+\frac{p}{2}\right)}}{ \left( 4 I^2_0 k^{\frac{2}{r}} n  + \frac{d \, \pi^2}{B^2} \right)^{\frac{p}{2}}}. \nonumber
\label{ThVanTreesMVTV2Orlicz-case-p-larger-than-two}
\end{align}

\end{itemize}
\end{Th}

\begin{proof}
The proof of Theorem~\ref{ThVanTreesMVTV2Orlicz} is given in Section~\ref{secVI_Orlicz}.  
\end{proof}

\begin{Rem}
For the special case of the $L_2$ loss, setting $p=2$ in the RHS of~\eqref{ThVanTreesMVTV2Orlicz-case-p-larger-than-two} we recover the result of~\cite[Theorem 3]{Barnes2019}.  
\end{Rem}

\begin{Cor}{(Gaussian Location Model)}~\label{ThGLM} 
Let $\mathbf{X} \sim \mathcal{N}(\bm{\theta}, \sigma^2 I_d)$ with $\Theta=[-B,B]^d$.  For $p \geq 2$, we have the following: if $\pi^2 \sigma^2 d \leq n B^2 \min\{k,d\}$ then for any estimator $\bm{\hat{\theta}}(\mathbf{M^{(n)}})$ we have
\begin{align}
&\sup_{\bm{\theta} \in \bm{\Theta}} \mathbb{E}_{\mathbf{M^{(n)}}|\bm{\Theta}} \left[ \left \vert \left \vert \bm{\hat{\theta}}(\mathbf{M^{(n)}}) - \bm{\theta} \right \vert \right \vert_p^p \left. \right \vert \bm{\Theta} \right]  \geq \nonumber \\
&\qquad \qquad d^{\left( 1+\frac{p}{2} \right)} \max \left \{ \left( \frac{\sigma^2}{n \, d}\right)^{\frac{p}{2}}, \left( \frac{3\sigma^2}{32 \, n \, k} \right)^{\frac{p}{2}} \right \}.
\end{align}
\end{Cor}

\begin{proof}
The proof of Corollary~\ref{ThGLM} is given in Section~\ref{secVI_ThGLM}.  
\end{proof}

\noindent For the special case $p=2$, the result of Corollary~\ref{ThGLM} recovers that of~\cite[Corollary 5]{Barnes2019}.

\section{Estimation under the Wasserstein loss}~\label{Wass}

We now turn to the minimax risk given by~\eqref{minimax-Wasserstein-loss2} in Section~\ref{Problem}. Theorem~\ref{ThVanTreesMVTV2} of Section~\ref{secIV}, as well as its proof, are instrumental to obtaining similar bounds for the Wasserstein loss~\eqref{minimax-Wasserstein-loss2} when the underlying distance $d(\cdot,\cdot)$ is based on the $L_p$-norm. For the Gaussian location model (see Corollary~\ref{ThVanTreesWassGLM} below) this yields a lower bound on the worst-case Wasserstein loss under the $L_p$ norm which decreases at least as $n^{-\frac{p}{2}}$.

\begin{Th}~\label{ThVanTreesWass}
For any estimator $\bm{\hat{\theta}}=\bm{\hat{\theta}}(\mathbf{M^{(n)}})$, the following holds.
\begin{itemize}
\item[i)] If $1<p<2$, we have
\begin{align}
&\sup_{\bm{\theta} \in \bm{\Theta}} \mathbb{E}_{\mathbf{M^{(n)}}|\bm{\Theta}} \left[W_p^p(f(\dv x|\bm{\hat{\theta}}(\mathbf{M^{(n)}})),f(\dv x|\bm{\theta})) \left. \right \vert \bm{\Theta} \right] \geq \nonumber \\ 
& \sum_{j=1}^d \left( \left \vert \sum_{i=1}^d \mathbb{E}_{\bm{\Theta}} \left[\frac{\partial}{\partial \Theta_i} \left[ \mathbb{E}_{\mathbf{Y} \sim f(\mathbf{y}|\bm{\theta})} [Y_j] \right] \right] \right \vert^p \right) \times \nonumber \\
&\quad \left\{ d^{\frac{1}{p}} (p-1)\left[\sum_{j=1}^n \left(\mathbb{E}_{\bm{\Theta}} \left[ \left( \Omega^{(p)}_{M_j}(\bm{\Theta}) \right)^{\frac{1}{p-1}} \right] \right)^{\frac{2(p-1)}{p}}  \right]^{\frac{1}{2}}  +  \right.  \nonumber \\
&\quad \quad \left. d^{\frac{p-1}{p}} \left(\Omega^{(p)}(\mu) \right)^{\frac{1}{p}} \right \}^{-p} \nonumber
\end{align}
\item[ii)] If $p \geq 2$, we have
\begin{align}
&\sup_{\bm{\theta} \in \bm{\Theta}} \mathbb{E}_{\mathbf{M^{(n)}}|\bm{\Theta}} \left[W_p^p(f(\mathbf{x}|\bm{\hat{\theta}}(\mathbf{M^{(n)}})),f(\mathbf{x}|\bm{\theta})) \left. \right \vert \bm{\Theta} \right] \geq \nonumber \\ 
&\quad \sum_{j=1}^d \left( \left \vert \sum_{i=1}^d \mathbb{E}_{\bm{\Theta}} \left[\frac{\partial}{\partial \Theta_i} \left[ \mathbb{E}_{\mathbf{Y} \sim f(\mathbf{y}|\bm{\theta})} [Y_j] \right] \right] \right \vert^p \right) \times \nonumber \\
&\quad \quad \left( d \, \sum_{j=1}^{n} \mathbb{E}_{\bm{\Theta}} \left[\rm{Tr}(I_{M_j}(\bm{\Theta})) \right] + d \, \rm{Tr}(I(\mu)) \right)^{-\frac{p}{2}}. \nonumber
\end{align}
\end{itemize}
\end{Th}

\begin{proof}
The proof of Theorem~\ref{ThVanTreesWass} is given in Section~\ref{secVI_Wass}.  
\end{proof}

Recall for fixed $p>1$ and $d \geq 1$ the constants $A_p$ and $B_{p,d}$ as defined by~\eqref{definition-constants-theorem3}. Also, define
\begin{subequations}
\begin{align}
C_p &= (p-1) \left(\frac{\sqrt{2}}{\sigma}\right)^{\frac{1}{p}} \left[\frac{\Gamma \left(\frac{1}{2p-2}\right)}{(p-1)\sqrt{2\pi \sigma^2}}  \right]^{\frac{p-1}{p}} \\
 D_p & = (p-1) \frac{4 \sqrt{2}}{\sqrt{3} \sigma}. 
\end{align}
\end{subequations}

\begin{Cor}{(Gaussian Location Model)}~\label{ThVanTreesWassGLM}
Let $\mathbf{X} \sim \mathcal{N}(\bm{\theta}, \sigma^2 I_d)$ with $\bs \theta \in \Theta=[-B,B]^d$. For any estimator $\bm{\hat{\theta}}=\bm{\hat{\theta}}(\mathbf{M^{(n)}})$, we have the following.
\begin{itemize}
\item[i)] If $p \geq 2$, we have
\begin{align}
& \sup_{\bm{\theta} \in \bm{\Theta}} \mathbb{E}_{\mathbf{M^{(n)}}|\bm{\Theta}} \left[  W_p^p(f(\mathbf{x}|\bm{\theta}), f(\mathbf{x}|\bm{\hat{\theta}})) \left \vert \right. \bm{\Theta} \right]  \geq  \nonumber \\
&\, d^{\frac{p}{2} } \max \left \{ \left( \frac{n \, d}{\sigma^2}  + \frac{d \, \pi^2}{B^2} \right)^{-\frac{p}{2}}, \left( \frac{32 \, n \, k }{3\sigma^2}  + \frac{d \, \pi^2}{B^2} \right)^{-\frac{p}{2}} \right \}. \nonumber
\end{align}
\item[ii)] If $1 < p < 2$, we have
\begin{align}
&\sup_{\bm{\theta} \in \bm{\Theta}} \mathbb{E}_{\mathbf{M^{(n)}}|\bm{\Theta}} \left[  W_p^p(f(\mathbf{x}|\bm{\theta}), f(\mathbf{x}|\bm{\hat{\theta}})) \left \vert \right. \bm{\Theta} \right]  \geq \max \left \{  \left(C_p \times \right. \right. \nonumber \\
&\,  \left. \left. d^{\frac{2-p}{p}} \sqrt{n} + A_p \right)^{-p}, \left(D_p \, d^{\frac{4-3p}{2p}} 2^{\frac{k(2-p)}{p}} k^{\frac{1}{2}} \, \sqrt{n} + A_p \right)^{-p} \right \}. \nonumber
\end{align}
\end{itemize}
\end{Cor}

\begin{proof}
The proof of Corollary~\ref{ThVanTreesWassGLM} is given in Section~\ref{secVI_WassGLM}.  
\end{proof}

\noindent A K-subgaussian distribution, estimated with an empirical distribution smoothed by a Gaussian kernel, enjoys upper bounds on the error in the $1$-Wasserstein distance, $W_1$, of the order $n^{-\frac{1}{2}}$ and in the squared $2$-Wasserstein distance, $W_2^2$, of the order $n^{-1}$ \cite{Goldfeld2020a}. The bounds show remarkable performance improvement of this convolution over the unsmoothed empirical estimator from $n^{-\frac{1}{d}}$ to that of the order $n^{-\frac{1}{2}}$ for $W_1$ and $n^{-1}$ for the $W_2^2$. If $p=2$, we obtain a lower bound on $W_2^2$ of the order $n^{-1}$, which matches that of the upper bound in \cite{Goldfeld2020a} for the empirical estimator smoothed by a Gaussian kernel of a K-subgaussian distribution. Our technique may be useful in \cite{Goldfeld2020a}, to produce a matching lower bound, to yield optimal rates of the order $n^{-1}$.


\section{Proofs}~\label{secVI}

\subsection{Proof of Theorem \ref{ThVanTreesXVTV2}}~\label{secVI_Th1}

Let $q \in \mathbb{R}$ such that $\frac{1}{p}+\frac{1}{q}=1$, i.e., $q=p/(p-1)$. Also, consider the following two functions $g(\cdot)$ and $h(\cdot)$ defined, for $\dv x \in \mc X$ and $\bs \theta =[\theta_1,\hdots,\theta_d] \in \Theta$, as 
\begin{subequations}
\begin{align}
g(\dv x,\bs \theta) &= \sum_{i=1}^d \frac{\partial}{\partial \theta_i} \left[ \log{\left(f(\dv x|\bs \theta) \mu(\bs \theta) \right)} \right]\\
h(\dv x,\bs \theta) &=  \hat{\bs \theta}(\dv x)- \bs \theta.
\end{align}
\end{subequations}	
For convenience,  for $i=1,\hdots,d$ we will denote the $i^{th}$ component of $h(\dv x, \bs \theta)$ as $h_i(\dv x,\bs \theta)$, i.e.,
\begin{equation}
h_i(\dv x,\bs \theta) = \hat{\theta}_i(\dv x)- \theta_i = \left(h(\dv x,\bs \theta)\right)_i
\end{equation}

\noindent Using the fact that the prior measure $\mu$ converges to zero at the endpoints of $\Theta$,  it is easy to see that
\begin{equation}
\int_{\dv x} \int_{\theta_i} h_i(\dv x, \bs \theta) \frac{\partial}{\partial \theta_i} \left[f(\dv x|\bs \theta) \mu_i(\theta_i) \right] \, \rm{d}{\theta_i} \rm{d}{\dv x}  = 1. 
\label{proof-generalization-of-van-Trees-inequality-step1Th1}
\end{equation}

\noindent By partial integration and~\eqref{proof-generalization-of-van-Trees-inequality-step1Th1}, we get for $i=1,\hdots,d$, that
\begin{equation}
\mathbb{E}_{(\dv X,\bs \Theta)} \left[h_i(\dv X, \bs \Theta) g(\dv X,\bs \Theta) \right]  = d.
\label{proof-generalization-of-van-Trees-inequality-step2Th1}
\end{equation}
Thus, for all $i=1,\hdots,d$, we have
\begin{equation}
\mathbb{E}_{(\dv X,\bs \Theta)} \left[ \left \vert h_i(\dv X, \Theta) g(\dv X,\bs \Theta) \right \vert \right]  \geq d.
\label{proof-generalization-of-van-Trees-inequality-step3Th1}
\end{equation}

For convenience, let
\begin{equation}
l(\dv x,\bs \theta)= \sum_{i=1}^d \frac{\partial}{\partial \theta_i} \left[ \log{f(\dv x|\bs \theta)} \right].
\label{proof-generalization-of-van-Trees-inequality-step7Th1}
\end{equation}
It is easy to see that for all $\bs \theta$, we have
\begin{equation}
\mathbb{E}_{\dv X |\bs \Theta} \left[ l(\dv X,\bs \Theta) | \bs \Theta = \bs \theta \right]  =0
\label{proof-generalization-of-van-Trees-inequality-step8Th1}
\end{equation}
which follows by the regularity condition $\mathbb{E}_{\dv X|\bs \Theta}\left[\frac{\partial}{\partial \theta_i} \log{f(\dv x|\bs \theta)}\right] = 0$ for all $\bs \theta \in \bs \Theta$. Also, define
\begin{equation}
g(\dv x,\bs \theta) = l(\dv x,\bs \theta) + \sum_{i=1}^d \frac{\partial}{\partial \theta_i} \left[ \log{\mu(\bs \theta)} \right].
\label{proof-generalization-of-van-Trees-inequality-step9Th1}
\end{equation}

\noindent In the rest of this proof we treat separately the cases $p \geq 1$ and $1 < p <2$.

\subsubsection{Case $p \geq 2$}

\noindent In this case, the average estimation error can be lower bounded as
\begin{align}
&\mathbb{E}_{(\dv X,\bs \Theta)} \left[ \left \vert \left \vert \bs{\hat{\theta}}(\dv X) - \bs{\theta} \right \vert \right \vert_p^p \right] \nonumber \\
&\quad \stackrel{(a)}{=} \sum_{i=1}^d \mathbb{E}_{(\mathbf{X},\bs{\Theta})} \left[ \left \vert h_i(\mathbf{X},\bs \Theta) \right \vert^p \right] \\
&\quad \stackrel{(b)}{\geq } \sum_{i=1}^d \mathbb{E}_{\bs{\Theta}} \left[ \left(\mathbb{E}_{\mathbf{X}|\bs{\Theta}} \left[ \left(\left \vert h_i(\mathbf{X}, \bs \Theta) \right \vert^2 \right) | \bs \Theta = \bs \theta \right] \right)^{\frac{p}{2}} \right], 
\label{proof-generalization-of-van-Trees-inequality-step4Th1}
\end{align}
where $(a)$ follows from the definition of the $p$-norm and $(b)$ holds due to Jensen's inequality applied to the function $u \mapsto u^{\frac{p}{2}}$ which is convex for $p > 2$.

\noindent The RHS of~\eqref{proof-generalization-of-van-Trees-inequality-step4Th1} can be lower bounded as follows. First, note that we have
\begin{align}
&\mathbb{E} \left[ \left\vert h_i(\dv X,\bs \Theta) g(\dv X,\bs \Theta) \right \vert  |  \right] = \mathbb{E}_{\bs \Theta} \mathbb{E}_{\dv X | \bs \Theta} \left[ \left( \left\vert h_i(\dv X,\bs \Theta) g(\dv X,\bs \Theta) \right\vert \right)  |  \bs \Theta = \bs \theta \right] \nonumber\\
&\stackrel{(a)}{\leq} \mathbb{E}_{\bs \Theta} \left( \mathbb{E}_{\dv X | \bs \Theta} \left[ \left(\left\vert h_i(\dv X,\bs \Theta)\right\vert^2 \right) | \bs \Theta = \bs \theta \right] \right)^{\frac{1}{2}}  \nonumber \\
&\quad \times \left( \mathbb{E}_{\dv X | \bs \Theta} \left[ \left(\left\vert g(\dv X,\bs \Theta)\right\vert^2 \right) | \bs \Theta = \bs \theta \right] \right)^{\frac{1}{2}} \nonumber\\
&\stackrel{(b)}{\leq}  \left(\mathbb{E}_{\bs \Theta} \left[  \left \vert \mathbb{E}_{\dv X|\bs \Theta} \left[ \left( \left\vert h_i(\dv X,\bs \Theta) \right \vert^2\right) | \bs \Theta = \bs \theta \right] \right \vert^{\frac{p}{2}} \right] \right)^{\frac{1}{p}}  \nonumber \\
&\quad \times \left(\mathbb{E}_{\bs \Theta} \left[ \left(\mathbb{E}_{\dv X|\bs \Theta} \left[ \left( \left \vert g(\dv X,\bs \Theta) \right \vert^2 \right) | \bs \Theta = \bs \theta \right] \right)^{\frac{q}{2}} \right]\right)^{\frac{1}{q}},
\label{proof-generalization-of-van-Trees-inequality-step5Th1}
\end{align}
where $(a)$ follows by application of H\"older's inequality for every $\bs \theta \in \bs \Theta$ to the conditional expectation  $\mathbb{E}_{\dv X | \bs \Theta}[\cdot|\bs \theta]$ ; and $(b)$ follows by application of H\"older's inequality to the expectation  $\mathbb{E}_{\bs \Theta}[\cdot]$ since $p >1$, $q>1$ and are such that $\frac{1}{p}+\frac{1}{q}=1$.

\noindent Combining~\eqref{proof-generalization-of-van-Trees-inequality-step3Th1}, ~\eqref{proof-generalization-of-van-Trees-inequality-step4Th1} and~\eqref{proof-generalization-of-van-Trees-inequality-step5Th1}, we get 
\begin{align}
&\mathbb{E}_{(\dv X,\bs \Theta)} \left[ \left \vert \left \vert \bs{\hat{\theta}}(\dv X) - \bs{\theta} \right \vert \right \vert_p^p \right]  \geq \nonumber \\
&\quad \frac{d^{p+1}}{\left(\mathbb{E}_{\bs \Theta} \left[ \left(\mathbb{E}_{\dv X|\bs \Theta} \left[ \left( \left \vert g(\dv X,\bs \Theta) \right \vert^2 \right) | \bs \Theta = \bs \theta \right] \right)^{\frac{q}{2}} \right]\right)^{\frac{p}{q}}}.
\label{proof-generalization-of-van-Trees-inequality-step6Th1}
\end{align}

\noindent We now upper bound the RHS term of~\eqref{proof-generalization-of-van-Trees-inequality-step6Th1}, as follows. 

\noindent Since 
\begin{equation}
\mathbb{E}_{\dv X|\bs \Theta} \left[ l(\dv X,\bs \Theta) \left(  \sum_{i=1}^d  \frac{\partial}{\partial \theta_i} \left[\log{\mu(\bs \theta)} \right]  \right) | \bs \Theta=\bs \theta \right] = 0,
\label{proof-generalization-of-van-Trees-inequality-step10Th1}
\end{equation}
we get
\begin{align}
& \mathbb{E}_{\dv X|\bs \Theta} \left[ \left( g(\dv X,\bs \Theta) \right)^2 | \bs \Theta=\bs \theta \right]  \nonumber\\
& \quad= \mathbb{E}_{\dv X | \bs \Theta} \left[ l^2(\dv X,\bs \Theta) \: | \: \bs \Theta = \bs \theta \right] + \left( \sum_{i=1}^d  \frac{\partial}{\partial \theta_i} \left[ \log{\mu(\bs \theta)} \right] \right)^2 
\label{proof-generalization-of-van-Trees-inequality-step11Th1}
\end{align}

\noindent Thus,
\begin{align}
& \mathbb{E}_{\bs \Theta} \left[ \left(\mathbb{E}_{\dv X|\bs \Theta} \left[ \left \vert g(\dv X,\bs \Theta) \right \vert^2 | \bs \Theta=\bs \theta \right] \right)^{\frac{q}{2}} \right] \nonumber\\
&\stackrel{(a)}{\leq} \left(\mathbb{E} \left[ \left\vert g(\dv X,\bs \Theta) \right\vert^2 \right] \right)^{\frac{q}{2}} \\
&\stackrel{(b)}{\leq} \left( \mathbb{E}_{(\dv X, \bs \Theta)} \left[ l^2(\dv X,\bs \Theta) \right] + \mathbb{E}_{\bs \Theta} \left( \sum_{i=1}^d  \frac{\partial}{\partial \theta_i} \left[ \log{\mu(\bs \theta)} \right] \right)^2  \right)^{\frac{q}{2}}
\label{proof-generalization-of-van-Trees-inequality-step12Th1}
\end{align}
where $(a)$ follows using Jensen's inequality for the concave function $ u \longrightarrow u^{q/2}$ for $q = p/p-1 \leq 2$; and $(b)$ follows by substituting using~\eqref{proof-generalization-of-van-Trees-inequality-step11Th1}.

\noindent The first expectation term on the RHS of~\eqref{proof-generalization-of-van-Trees-inequality-step12Th1} is upper bounded as
\begin{align}
\mathbb{E}_{(\dv X,\bs \Theta)} \left[ l^2(\dv X,\bs \Theta) \right] &\stackrel{(a)}{=} \mathbb{E}_{(\dv X,\bs \Theta)} \left[ \left(\sum_{i=1}^d  \frac{\partial}{\partial \theta_i} \left[\log{f(\dv X|\bs \Theta)} \right] \right)^2 \right] \nonumber \\
&\stackrel{(b)}{\leq} d \, \mathbb{E}_{\bs \Theta} \left[\rm{Tr}(I_{\dv X}(\bs \theta)) \right]
\label{proof-generalization-of-van-Trees-inequality-step13Th1}
\end{align}
where $(a)$ follows by substituting using~\eqref{proof-generalization-of-van-Trees-inequality-step7Th1} and $(b)$ holds since for non-negative $\{u_i\}_{i=1}^d$ we have $\left(\sum_{i=1}^d u_i \right)^2 \leq d \, \sum_{i=1}^d u_i^2$.

\noindent Hence, we get
\begin{align}
& \left( \mathbb{E}_{\bs \Theta} \left[ \left(\mathbb{E}_{\dv X|\bs \Theta} \left[ \left \vert g(\dv X,\bs \Theta) \right \vert^2 | \bs \Theta=\bs \theta \right] \right)^{\frac{q}{2}} \right] \right)^{\frac{p}{q}} \nonumber\\
&\stackrel{(a)}{\leq} \left( \mathbb{E}_{(\dv X, \bs \Theta)} \left[ l^2(\dv X,\bs \Theta) \right] + \mathbb{E}_{\bs \Theta} \left( \sum_{i=1}^d  \frac{\partial}{\partial \theta_i} \left[ \log{\mu(\bs \theta)} \right] \right)^2  \right)^{\frac{p}{2}} \nonumber\\
&\stackrel{(b)}{\leq} \left( d \, \mathbb{E}_{\bs \Theta} \left[\rm{Tr}(I_{\dv X}(\bs \theta)) \right] + d \, \rm{Tr}(I(\mu)) \right)^{\frac{p}{2}},
\label{proof-generalization-of-van-Trees-inequality-step14Th1}
\end{align}
where $(a)$ follows by using~\eqref{proof-generalization-of-van-Trees-inequality-step12Th1} and noticing that $p/q=p-1 \geq 1$ and $(b)$ holds using~\eqref{proof-generalization-of-van-Trees-inequality-step13Th1}.

\noindent Summarizing, combining~\eqref{proof-generalization-of-van-Trees-inequality-step6Th1} and~\eqref{proof-generalization-of-van-Trees-inequality-step14Th1} we get
\begin{equation}
\mathbb{E}_{(\dv X,\bs \Theta)} \left[ \left \vert \left \vert \bs{\hat{\theta}}(\dv X) - \bs{\theta} \right \vert \right \vert_p^p \right]  \geq \frac{d^{\left(1+\frac{p}{2}\right)}}{\left(\mathbb{E}_{\bs \Theta} \left[\rm{Tr}(I_{\dv X}(\bs \theta)) \right] +  \rm{Tr}(I(\mu)) \right)^{\frac{p}{2}}}.
\label{proof-generalization-of-van-Trees-inequality-step15Th1}
\end{equation}
   
\subsubsection{Case $1 < p < 2$}

\noindent First, recall that 
\begin{equation}
\mathbb{E}_{(\mathbf{X},\bm{\Theta})} \left[ \left \vert \left \vert \bm{\hat{\theta}}(\mathbf{X}) - \bm{\theta} \right \vert \right \vert_p^p \right] = \sum_{i=1}^d \mathbb{E}_{(\mathbf{X},\bm{\Theta})} \left[ \left \vert h_i(\mathbf{X},\bm{\Theta}) \right \vert^p \right].
\label{almostDoneBound2XVTV2Th1RL2}
\end{equation}

\noindent Also, for all $i=1,\hdots,d$, an easy application of H\"{o}lder's inequality for expectations yields
\begin{align}
&\mathbb{E}_{(\mathbf{X},\bm{\Theta})} \left[ \left \vert h_i(\mathbf{X},\bm{\Theta}) g(\mathbf{X},\bm{\Theta}) \right \vert \right] \nonumber\\
&\quad \leq  \left( \mathbb{E}_{(\mathbf{X},\bm{\Theta})} \left[ \left \vert h_i(\mathbf{X},\bm{\Theta}) \right \vert^p \right] \right)^{\frac{1}{p}} \left(\mathbb{E}_{(\mathbf{X},\bm{\Theta})} \left[ \left \vert g(\mathbf{X},\bm{\Theta}) \right \vert^q \right] \right)^{\frac{1}{q}}. 
\label{OriginalHolderXVTV2Th1RL2}
\end{align}

\noindent Thus, using~\eqref{almostDoneBound2XVTV2Th1RL2} and~\eqref{OriginalHolderXVTV2Th1RL2}, we get 
\begin{equation}
\mathbb{E}_{(\dv X,\bs \Theta)} \left[ \left \vert \left \vert \bs{\hat{\theta}}(\dv X) - \bs{\theta} \right \vert \right \vert_p^p \right]  \geq \frac{d^{p+1}}{\left(\mathbb{E}_{(\mathbf{X},\bm{\Theta})} \left[ \left \vert g(\mathbf{X},\bm{\Theta}) \right \vert^q \right] \right)^{\frac{p}{q}}}.
\label{proof-generalization-of-van-Trees-inequality-step6Th1RL2}
\end{equation}

\noindent The rest of the proof in this case is devoted to upper-bounding the denominator of the RHS of~\eqref{proof-generalization-of-van-Trees-inequality-step6Th1RL2}.

\noindent Recalling~\eqref{proof-generalization-of-van-Trees-inequality-step9Th1}, we have
\begin{align}
&\left(\mathbb{E}_{(\mathbf{X},\bm{\Theta})} \left[ \left \vert g(\mathbf{X},\bm{\Theta}) \right \vert^q \right] \right)^{\frac{1}{q}} \stackrel{(a)}{\leq} \label{XToreplacewithEqRE2Th1RL2} \\
&\left( \mathbb{E}_{(\mathbf{X},\bm{\Theta})} \left[ \left \vert l(\mathbf{X},\bm{\Theta}) \right \vert^q \right] \right)^{\frac{1}{q}} + \left( \mathbb{E}_{\bm{\Theta}} \left[ \left \vert \sum_{i=1}^{d} \frac{\partial}{\partial \Theta_i} \left[ \log{\mu(\bm{\Theta})} \right] \right \vert^q \right] \right)^{\frac{1}{q}} \nonumber \\
&\stackrel{(b)}{\leq} \left( \mathbb{E}_{(\mathbf{X},\bm{\Theta})} \left[ \left \vert l(\mathbf{X},\bm{\Theta}) \right \vert^q \right] \right)^{\frac{1}{q}} + \sum_{i=1}^{d} \left( \mathbb{E}_{\bm{\Theta}} \left[ \left \vert \frac{\partial}{\partial \Theta_i} \left[ \log{\mu(\bm{\Theta})} \right] \right \vert^q \right] \right)^{\frac{1}{q}} \nonumber \\
&\stackrel{(c)}{\leq} \left( \mathbb{E}_{(\mathbf{X},\bm{\Theta})} \left[ \left \vert l(\mathbf{X},\bm{\Theta}) \right \vert^q \right] \right)^{\frac{1}{q}} \nonumber\\
& \qquad \quad + \sum_{i=1}^{d} \left( \left( \mathbb{E}_{\bm{\Theta}} \left[ \left \vert \frac{\partial}{\partial \Theta_i} \left[ \log{\mu(\bm{\Theta})} \right] \right \vert^{\frac{p}{p-1}} \right] \right)^{p-1} \right)^{\frac{1}{p}} \nonumber \\
&\stackrel{(d)}{\leq} \left( \mathbb{E}_{(\mathbf{X},\bm{\Theta})} \left[ \left \vert l(\mathbf{X},\bm{\Theta}) \right \vert^q \right] \right)^{\frac{1}{q}} + d^{\frac{p-1}{p}} \left( \Omega^{(p)}(\mu) \right)^{\frac{1}{p}}, 
\label{UBTh1RL2}
\end{align}
where: $(a)$ follows by application of the Minkowski's inequality for expectations $\left(\mathbb{E} [|Z+T|^q] \right)^{\frac{1}{q}} \leq \left(\mathbb{E} [|Z|^q] \right)^{\frac{1}{q}} + \left(\mathbb{E} [|T|^q] \right)^{\frac{1}{q}}$ for r.v.s $Z$ and $T$; $(b)$ follows by application of the Minkowski's inequality for expectations $\left(\mathbb{E} [|\sum_{i=1}^d Z_i|^q] \right)^{\frac{1}{q}} \leq \sum_{i=1}^d \left(\mathbb{E} [|Z_i|^q] \right)^{\frac{1}{q}}$ for r.v.s $(Z_1,\hdots,Z_d)$; $(c)$ holds by substituting using $q=p/p-1$ and $(d)$ holds by first using the inequality $\sum_{i=1}^d u_i^{\frac{1}{p}} \leq d^{\frac{p-1}{p}} \left(\sum_{i=1}^d u_i\right)^{\frac{1}{p}}$ for non-negative $(u_1,\hdots,u_d)$ and $p > 1$ and then substituting using~\eqref{definition-trace-generalized-fisher-information-estimating-theta-from-prior}.

\noindent Continuing from~\eqref{UBTh1RL2}, the first term of its RHS can be upper bounded as 
\begin{align}
&\left(\mathbb{E}_{(\mathbf{X},\bm{\Theta})}  \left[ |l(\mathbf{X},\bm{\Theta})|^q \right]\right)^{\frac{1}{q}} \nonumber \\
&\stackrel{(a)}{=} \left(\mathbb{E}_{(\mathbf{X},\bm{\Theta})}  \left[ \left \vert \sum_{i=1}^d   \frac{\partial}{\partial \theta_i} \left[\log{f(\mathbf{X}|\bm{\Theta})} \right] \right \vert^q \right]\right)^{\frac{1}{q}} \nonumber \\
&\stackrel{(b)}{\leq} d^{\frac{q-1}{q}} \left(\mathbb{E}_{(\mathbf{X},\bm{\Theta})}  \left[ \sum_{i=1}^d \left \vert \frac{\partial}{\partial \theta_i} \left[\log{f(\mathbf{X}|\bm{\Theta})} \right] \right \vert^q \right]\right)^{\frac{1}{q}} \nonumber \\
&\stackrel{(c)}{=} d^{\frac{1}{p}} \left(\mathbb{E}_{\bm{\Theta}}  \left[ \sum_{i=1}^d \mathbb{E}_{\mathbf{X}|\bm{\Theta}} \left[ \left \vert \frac{\partial}{\partial \theta_i} \left[\log{f(\mathbf{X}|\bm{\Theta})} \right] \right \vert^{\frac{p}{p-1}} \left \vert \right.\bm{\Theta} = \bs \theta  \right] \right] \right)^{\frac{p-1}{p}}  \nonumber \\
&\stackrel{(d)}{=} d^{\frac{1}{p}} \left(\mathbb{E}_{\bm{\Theta}}  \left[ \sum_{i=1}^d (v_i(\bs \theta))^{\frac{1}{p-1}}\right] \right)^{\frac{p-1}{p}}  \nonumber \\
&\stackrel{(e)}{\leq} d^{\frac{1}{p}} \left(\mathbb{E}_{\bm{\Theta}}  \left[ \left( \sum_{i=1}^d v_i(\bs \theta) \right)^{\frac{1}{p-1}}\right] \right)^{\frac{p-1}{p}}  \nonumber \\
&\stackrel{(f)}{=}  d^{\frac{1}{p}} \left( \mathbb{E}_{\bm{\Theta}}  \left[ \left( \Omega^{(p)}_{\mathbf{X}}(\bm{\theta}) \right)^{\frac{1}{p-1}} \right] \right)^{\frac{p-1}{p}} , \label{CombPart2XVTV2Th1RL2}
\end{align}
where: $(a)$ follows by substituting using~\eqref{proof-generalization-of-van-Trees-inequality-step7Th1}; $(b)$ holds by using the inequality $\left(\sum_{i=1}^d u_i \right)^{q} \leq d^{q-1} \sum_{i=1}^d u_i^{q} $ which holds for non-negative $(u_1,\hdots,u_d)$ and $q > 1$, $(c)$ follows by substituting using $q=\frac{p}{p-1}$; and $(d)$ holds by defining, for $i=1,\hdots,d$ and $\bs \theta \in \Theta$,
\begin{equation}
v_i(\bs \theta) = \left( \mathbb{E}_{\mathbf{X}|\bm{\Theta}} \left[ \left \vert \frac{\partial}{\partial \theta_i} \left[\log{f(\mathbf{X}|\bm{\Theta})} \right] \right \vert^{\frac{p}{p-1}} \left \vert \right.\bm{\Theta} = \bs \theta  \right] \right)^{p-1}; 
\label{DefUi2}
\end{equation}
$(e)$ holds by using the inequality $\sum_{i=1}^d u_i^{{\frac{1}{p-1}}} \leq \left(\sum_{i=1}^d u_i \right)^{\frac{1}{p-1}}$ for non-negative $(u_1,\hdots,u_d)$ and $p<2$; and $(f)$ holds by substituting using~\eqref{definition-trace-generalized-fisher-information-estimating-theta-from-vectorX}. 

\noindent Hence, combining \eqref{CombPart2XVTV2Th1RL2} and~\eqref{XToreplacewithEqRE2Th1RL2}, we get
\begin{align}
& \left(\mathbb{E}_{(\mathbf{X},\bm{\Theta})} \left[ \left \vert g(\mathbf{X},\bm{\Theta}) \right \vert^q \right] \right)^{\frac{1}{q}} \leq   \nonumber \\
&\quad d^{\frac{1}{p}} \left(\mathbb{E}_{\bm{\Theta}} \left[ \left( \Omega^{(p)}_{\mathbf{X}}(\bm{\theta}) \right)^{\frac{1}{p-1}} \right] \right)^{\frac{p-1}{p}} + d^{\frac{p-1}{p}} \left( \Omega^{(p)}(\mu) \right)^{\frac{1}{p}}. \label{XUBoundonGTh1RL2}
\end{align}

\noindent Finally, substituting in \eqref{proof-generalization-of-van-Trees-inequality-step6Th1RL2} using \eqref{XUBoundonGTh1RL2} yields the desired result,
\begin{align}
&\mathbb{E}_{(\mathbf{X},\bm{\Theta})} \left[ \left \vert \left \vert \bm{\hat{\theta}}(\mathbf{X}) - \bm{\theta} \right \vert \right \vert_p^p \right]  \geq d^{p} \left(  d^{\frac{p-2}{p}}  \left(\Omega^{(p)}(\mu) \right)^{\frac{1}{p}} + \right.  \nonumber \\
&\left. \left(\mathbb{E}_{\bm{\Theta}} \left[ \left( \Omega^{(p)}_{\mathbf{X}}(\bm{\theta}) \right)^{\frac{1}{p-1}} \right] \right)^{\frac{p-1}{p}} \right)^{-p}.
\end{align}


\subsection{Proof of Inequality (\ref{ThVanTreesXVTV-alternative-lower-bound})}~\label{secVI_Th1-alternative-lower-bound}

Let $q \in \mathbb{R}$ such that $\frac{1}{p}+\frac{1}{q}=1$, i.e., $q=p/(p-1)$. Also, consider the following two functions $g(\cdot)$ and $h(\cdot)$ defined, for $\dv x \in \mc X$ and $\bs \theta =[\theta_1,\hdots,\theta_d] \in \Theta$, as 
\begin{subequations}
\begin{align}
g(\dv x,\bs \theta) &= \sum_{i=1}^d \frac{\partial}{\partial \theta_i} \left[ \log{\left(f(\dv x|\bs \theta) \mu(\bs \theta) \right)} \right]\\
h(\dv x,\bs \theta) &=  \psi(\hat{\bs \theta}(\dv x)) - \psi(\bs \theta).
\end{align}
\end{subequations}	
For convenience,  for $i=1,\hdots,d$ we will denote the $i^{th}$ component of $h(\dv x, \bs \theta)$ as $h_i(\dv x,\bs \theta)$, i.e.,
\begin{equation}
h_i(\dv x,\bs \theta) = \psi_i(\hat{\theta}(\dv x)) - \psi_i(\theta) = \left(h(\dv x,\bs \theta)\right)_i
\end{equation}

\noindent Using the definition of the $p$-norm, the average estimation error can be lower bounded as
\begin{align}
& \mathbb{E}_{(\mathbf{X},\bm{\Theta})} \left[ \left \vert \left \vert \psi(\bm{\hat{\theta}}(\mathbf{X})) - \psi(\bm{\theta}) \right \vert \right \vert_p^p \right] = \sum_{i=1}^d \mathbb{E}_{(\mathbf{X},\bm{\Theta})} \left[ \left \vert h_i(\mathbf{X},\bm{\Theta}) \right \vert^p \right] \label{almostDoneBound2XVTV2RL2Cor2}.
\end{align}

\noindent The RHS of~\eqref{almostDoneBound2XVTV2RL2Cor2} can be lower bounded as follows. First, note that applying H\"{o}lder's inequality for expectations yields
\begin{align}
&\mathbb{E}_{(\mathbf{X},\bm{\Theta})} \left[ \left \vert h_i(\mathbf{X},\bm{\Theta}) g(\mathbf{X},\bm{\Theta}) \right \vert \right] \leq  \left( \mathbb{E}_{(\mathbf{X},\bm{\Theta})} \left[ \left \vert h_i(\mathbf{X},\bm{\Theta}) \right \vert^p \right] \right)^{\frac{1}{p}} \times \nonumber \\
&\quad  \left(\mathbb{E}_{(\mathbf{X},\bm{\Theta})} \left[ \left \vert g(\mathbf{X},\bm{\Theta}) \right \vert^q \right] \right)^{\frac{1}{q}}. \label{OriginalHolderXVTV2RL2Cor2}
\end{align}

\noindent Using the fact that the prior measure $\mu$ converges to zero at the endpoints of $\bs \Theta$ and partial integration, it is easy to see that
\begin{align}
&\int_{\theta_i} h_i(\mathbf{x},\bm{\theta}) \frac{\partial}{\partial \theta_i} \left[f(\mathbf{x}|\bm{\theta}) \mu_i(\theta_i) \right] \, \rm{d}{\theta_i}  \nonumber \\
& = \left.  h_i(\mathbf{x},\bm{\theta}) f(\mathbf{x}|\bm{\theta}) \mu_i(\theta_i) \right \vert_{\theta^{(i)}_{min}}^{\theta^{(i)}_{max}}  - \nonumber \\
&\quad \int_{\theta_i} \frac{\partial}{\partial \theta_i} \left[ h_i(\mathbf{x},\bm{\theta}) \right] f(\mathbf{x}|\bm{\theta}) \mu_i(\theta_i) \, \rm{d} \theta_i \nonumber \\
& =  - \int_{\theta_i} \frac{\partial}{\partial \theta_i} \left[ h_i(\mathbf{x},\bm{\theta}) \right] f(\mathbf{x}|\bm{\theta}) \mu_i(\theta_i) \, \rm{d} \theta_i. \label{proof-generalization-of-van-Trees-inequality-step1RL2Cor2}
\end{align}
\noindent Integration in~\eqref{proof-generalization-of-van-Trees-inequality-step1RL2Cor2}, we get for $i=1,\hdots,d$, that
\begin{align}
&\int_{\mathbf{x}} \int_{\theta_i} h_i(\mathbf{x},\bm{\theta}) \frac{\partial}{\partial \theta_i} \left[f(\mathbf{x}|\bm{\theta}) \mu_i(\theta_i) \right] \, \rm{d}{\theta_i} \, \rm{d}{\mathbf{x}}  \nonumber \\
&= - \mathbb{E}_{(\mathbf{X},\Theta_i)} \left[ \frac{\partial}{\partial \Theta_i} \left[ h_i(\mathbf{x},\bm{\Theta}) \right] \right].  \label{IntEq1MVTV2RL2PCor2}
\end{align}

Thus, with some algebraic manipulations, 
\begin{align}
&\mathbb{E}_{(\mathbf{X},\bm{\Theta})} \left[ h_i(\mathbf{X},\bm{\Theta}) g(\mathbf{X},\bm{\Theta}) \right] \nonumber \\
&=\sum_{i=1}^d \mathbb{E}_{\Theta_1} \left[ \ldots \mathbb{E}_{\Theta_d} \left[ - \mathbb{E}_{(\mathbf{X},\Theta_i}) \left[ \frac{\partial}{\partial \Theta_i} \left[ h_i(\mathbf{X},\bm{\Theta}) \right] \right] \right] \right] \nonumber \\
&= - \sum_{i=1}^d \mathbb{E}_{(\mathbf{X},\bm{\Theta})} \left[ \frac{\partial}{\partial \Theta_i} \left[ h_i(\mathbf{X},\bm{\Theta}) \right] \right]   \nonumber \\
&= \sum_{i=1}^d \mathbb{E}_{\bm{\Theta}} \left[ \frac{\partial \psi_i(\bm{\theta})}{\partial \Theta_i} \right]
\end{align}
and $|\mathbb{E}[X]| \leq \mathbb{E}[|X|]$ lower bounds the left-hand side of (\ref{OriginalHolderXVTV2RL2Cor2})
\begin{align}
&\left \vert \sum_{i=1}^d \mathbb{E}_{\bm{\Theta}} \left[ \frac{\partial \psi_i(\bm{\theta})}{\partial \Theta_i} \right]  \right \vert \leq \mathbb{E}_{(\mathbf{X},\bm{\Theta})} \left[ \left \vert h_i(\mathbf{X},\bm{\Theta}) g(\mathbf{X},\bm{\Theta}) \right \vert \right]. \label{BoundLHTRL2MVTV2PCor2}
\end{align}

\noindent Combining~\eqref{almostDoneBound2XVTV2RL2Cor2},~\eqref{OriginalHolderXVTV2RL2Cor2} and~\eqref{BoundLHTRL2MVTV2PCor2}, we get 
\begin{align}
&\mathbb{E}_{(\dv X,\bs \Theta)} \left[ \left \vert \left \vert \psi(\bs{\hat{\theta}}(\dv X)) - \psi(\bs{\theta}) \right \vert \right \vert_p^p \right]  \geq \nonumber \\
&d \, \left \vert \sum_{i=1}^d \mathbb{E}_{\bm{\Theta}} \left[ \frac{\partial \psi_i(\bm{\theta})}{\partial \Theta_i} \right]  \right \vert^p \left(\mathbb{E}_{(\mathbf{X},\bm{\Theta})} \left[ \left \vert g(\mathbf{X},\bm{\Theta}) \right \vert^q \right] \right)^{-\frac{p}{q}}.
\label{proof-generalization-of-van-Trees-inequality-step6RL2Cor2}
\end{align}

\noindent We now upper bound the second expectation of the RHS term of~\eqref{proof-generalization-of-van-Trees-inequality-step6RL2Cor2}, as follows. For convenience, let
\begin{equation}
l(\dv x,\bs \theta)= \sum_{i=1}^d \frac{\partial}{\partial \theta_i} \left[ \log{f(\dv x|\bs \theta)} \right].
\label{proof-generalization-of-van-Trees-inequality-step7RL2Cor2}
\end{equation}
It is easy to see that for all $\bs \theta$, we have
\begin{equation}
\mathbb{E}_{\dv X |\bs \Theta} \left[ l(\dv X,\bs \Theta) | \bs \Theta = \bs \theta \right]  =0
\label{proof-generalization-of-van-Trees-inequality-step8RL2Cor2}
\end{equation}
which follows by the regularity condition $\mathbb{E}_{\dv X|\bs \Theta}\left[\frac{\partial}{\partial \theta_i} \log{f(\dv x|\bs \theta)}\right] = 0$ for all $\bs \theta \in \bs \Theta$. Also,
\begin{equation}
g(\dv x,\bs \theta) = l(\dv x,\bs \theta) + \sum_{i=1}^d \frac{\partial}{\partial \theta_i} \left[ \log{\mu(\bs \theta)} \right].
\label{proof-generalization-of-van-Trees-inequality-step9RL2Cor2}
\end{equation}
Note that $l(\mathbf{x},\bm{\theta})$ is the sum of the elements of the score function associated with $\mathbf{X}$. 

\noindent From~\eqref{proof-generalization-of-van-Trees-inequality-step9RL2Cor2}, we have
\begin{align}
&\left(\mathbb{E}_{(\mathbf{X},\bm{\Theta})} \left[ \left \vert g(\mathbf{X},\bm{\Theta}) \right \vert^q \right] \right)^{\frac{1}{q}} \stackrel{(a)}{\leq} \label{XToreplacewithEqRE2Th1RL2Cor2} \\
&\left( \mathbb{E}_{(\mathbf{X},\bm{\Theta})} \left[ \left \vert l(\mathbf{X},\bm{\Theta}) \right \vert^q \right] \right)^{\frac{1}{q}} + \left( \mathbb{E}_{\bm{\Theta}} \left[ \left \vert \sum_{i=1}^{d} \frac{\partial}{\partial \Theta_i} \left[ \log{\mu(\bm{\Theta})} \right] \right \vert^q \right] \right)^{\frac{1}{q}} \nonumber \\
&\stackrel{(b)}{\leq} \left( \mathbb{E}_{(\mathbf{X},\bm{\Theta})} \left[ \left \vert l(\mathbf{X},\bm{\Theta}) \right \vert^q \right] \right)^{\frac{1}{q}} + \sum_{i=1}^{d} \left( \mathbb{E}_{\bm{\Theta}} \left[ \left \vert \frac{\partial}{\partial \Theta_i} \left[ \log{\mu(\bm{\Theta})} \right] \right \vert^q \right] \right)^{\frac{1}{q}} \nonumber \\
&\stackrel{(c)}{\leq} \left( \mathbb{E}_{(\mathbf{X},\bm{\Theta})} \left[ \left \vert l(\mathbf{X},\bm{\Theta}) \right \vert^q \right] \right)^{\frac{1}{q}} \nonumber\\
& \qquad \quad + \sum_{i=1}^{d} \left( \left( \mathbb{E}_{\bm{\Theta}} \left[ \left \vert \frac{\partial}{\partial \Theta_i} \left[ \log{\mu(\bm{\Theta})} \right] \right \vert^{\frac{p}{p-1}} \right] \right)^{p-1} \right)^{\frac{1}{p}} \nonumber \\
&\stackrel{(d)}{\leq} \left( \mathbb{E}_{(\mathbf{X},\bm{\Theta})} \left[ \left \vert l(\mathbf{X},\bm{\Theta}) \right \vert^q \right] \right)^{\frac{1}{q}} + d^{\frac{p-1}{p}} \left( \Omega^{(p)}(\mu) \right)^{\frac{1}{p}}, 
\label{UBTh1RL2Cor2}
\end{align}
where: $(a)$ follows by application of the Minkowski's inequality for expectations $\left(\mathbb{E} [|Z+T|^q] \right)^{\frac{1}{q}} \leq \left(\mathbb{E} [|Z|^q] \right)^{\frac{1}{q}} + \left(\mathbb{E} [|T|^q] \right)^{\frac{1}{q}}$ for r.v.s $Z$ and $T$; $(b)$ follows by application of the Minkowski's inequality for expectations $\left(\mathbb{E} [|\sum_{i=1}^d Z_i|^q] \right)^{\frac{1}{q}} \leq \sum_{i=1}^d \left(\mathbb{E} [|Z_i|^q] \right)^{\frac{1}{q}}$ for r.v.s $(Z_1,\hdots,Z_d)$; $(c)$ holds by substituting using $q=p/p-1$ and $(d)$ holds by first using the inequality $\sum_{i=1}^d u_i^{\frac{1}{p}} \leq d^{\frac{p-1}{p}} \left(\sum_{i=1}^d u_i\right)^{\frac{1}{p}}$ for non-negative $(u_1,\hdots,u_d)$ and $p > 1$ and then substituting using~\eqref{definition-trace-generalized-fisher-information-estimating-theta-from-prior}.

\noindent Continuing from~\eqref{UBTh1RL2Cor2}, the first term of its RHS can be upper bounded as 
\begin{align}
&\left(\mathbb{E}_{(\mathbf{X},\bm{\Theta})}  \left[ |l(\mathbf{X},\bm{\Theta})|^q \right]\right)^{\frac{1}{q}} \nonumber \\
&\stackrel{(a)}{=} \left(\mathbb{E}_{(\mathbf{X},\bm{\Theta})}  \left[ \left \vert \sum_{i=1}^d   \frac{\partial}{\partial \theta_i} \left[\log{f(\mathbf{X}|\bm{\Theta})} \right] \right \vert^q \right]\right)^{\frac{1}{q}} \nonumber \\
&\stackrel{(b)}{\leq} d^{\frac{q-1}{q}} \left(\mathbb{E}_{(\mathbf{X},\bm{\Theta})}  \left[ \sum_{i=1}^d \left \vert \frac{\partial}{\partial \theta_i} \left[\log{f(\mathbf{X}|\bm{\Theta})} \right] \right \vert^q \right]\right)^{\frac{1}{q}} \nonumber \\
&\stackrel{(c)}{=} d^{\frac{1}{p}} \left(\mathbb{E}_{\bm{\Theta}}  \left[ \sum_{i=1}^d \mathbb{E}_{\mathbf{X}|\bm{\Theta}} \left[ \left \vert \frac{\partial}{\partial \theta_i} \left[\log{f(\mathbf{X}|\bm{\Theta})} \right] \right \vert^{\frac{p}{p-1}} \left \vert \right.\bm{\Theta} = \bs \theta  \right] \right] \right)^{\frac{p-1}{p}}  \nonumber \\
&\stackrel{(d)}{=} d^{\frac{1}{p}} \left(\mathbb{E}_{\bm{\Theta}}  \left[ \sum_{i=1}^d (v_i(\bs \theta))^{\frac{1}{p-1}}\right] \right)^{\frac{p-1}{p}}  \nonumber \\
&\stackrel{(e)}{\leq} d^{\frac{p-1}{p}} \left(\mathbb{E}_{\bm{\Theta}}  \left[ \left( \sum_{i=1}^d v_i(\bs \theta) \right)^{\frac{1}{p-1}}\right] \right)^{\frac{p-1}{p}}  \nonumber \\
&\stackrel{(f)}{=}  d^{\frac{p-1}{p}} \left( \mathbb{E}_{\bm{\Theta}}  \left[ \left( \Omega^{(p)}_{\mathbf{X}}(\bm{\theta}) \right)^{\frac{1}{p-1}} \right] \right)^{\frac{p-1}{p}} , \label{CombPart2XVTV2Th1RL2Cor2}
\end{align}
where: $(a)$ follows by substituting using~\eqref{proof-generalization-of-van-Trees-inequality-step7RL2Cor2}; $(b)$ holds by using the inequality $\left(\sum_{i=1}^d u_i \right)^{q} \leq d^{q-1} \sum_{i=1}^d u_i^{q} $ which holds for non-negative $(u_1,\hdots,u_d)$ and $q > 1$, $(c)$ follows by substituting using $q=\frac{p}{p-1}$; and $(d)$ holds by defining, for $i=1,\hdots,d$ and $\bs \theta \in \Theta$,
\begin{equation}
v_i(\bs \theta) = \left( \mathbb{E}_{\mathbf{X}|\bm{\Theta}} \left[ \left \vert \frac{\partial}{\partial \theta_i} \left[\log{f(\mathbf{X}|\bm{\Theta})} \right] \right \vert^{\frac{p}{p-1}} \left \vert \right.\bm{\Theta} = \bs \theta  \right] \right)^{p-1}; 
\label{DefUi32}
\end{equation}
$(e)$ holds by using the inequality $\sum_{i=1}^d u_i^{{\frac{1}{p-1}}} \leq d^{\frac{p-2}{p-1}} \left(\sum_{i=1}^d u_i \right)^{\frac{1}{p-1}}$ for non-negative $(u_1,\hdots,u_d)$ and $p \geq 2$; and $(f)$ holds by substituting using~\eqref{definition-trace-generalized-fisher-information-estimating-theta-from-vectorX}. 

Substituting (\ref{CombPart2XVTV2Th1RL2Cor2}) in (\ref{UBTh1RL2Cor2}), we obtain
\begin{align}
& \left(\mathbb{E}_{(\mathbf{X},\bm{\Theta})} \left[ \left \vert g(\mathbf{X},\bm{\Theta}) \right \vert^q \right] \right)^{\frac{1}{q}} \leq   \nonumber \\
&\quad d^{\frac{p-1}{p}} \left(\mathbb{E}_{\bm{\Theta}} \left[ \left( \Omega^{(p)}_{\mathbf{X}}(\bm{\theta}) \right)^{\frac{1}{p-1}} \right] \right)^{\frac{p-1}{p}} + d^{\frac{p-1}{p}} \left( \Omega^{(p)}(\mu) \right)^{\frac{1}{p}}. \label{XUBoundonGRL2Cor2}
\end{align}
Substituting (\ref{XUBoundonGRL2Cor2}) in (\ref{proof-generalization-of-van-Trees-inequality-step6RL2Cor2}) produces the lower bound
\begin{align}
&\mathbb{E}_{(\mathbf{X},\bm{\Theta})} \left[ \left \vert \left \vert \psi(\bm{\hat{\theta}}(\mathbf{X})) - \psi(\bm{\theta}) \right \vert \right \vert_p^p \right]  \geq  \left \vert \sum_{i=1}^d \mathbb{E}_{\bm{\Theta}} \left[ \frac{\partial \psi_i(\bm{\theta})}{\partial \Theta_i} \right]  \right \vert^p \times  \nonumber \\
& d^{2-p} \left( \left(\Omega^{(p)}(\mu) \right)^{\frac{1}{p}} +  \left(\mathbb{E}_{\bm{\Theta}} \left[ \left( \Omega^{(p)}_{\mathbf{X}}(\bm{\theta}) \right)^{\frac{1}{p-1}} \right] \right)^{\frac{p-1}{p}} \right)^{-p}. \nonumber
\end{align}

Let $\psi(\bm{\theta}) = \bm{\theta} $, then we obtain
\begin{align}
&\mathbb{E}_{(\mathbf{X},\bm{\Theta})} \left[ \left \vert \left \vert \bm{\hat{\theta}}(\mathbf{X}) - \bm{\theta} \right \vert \right \vert_p^p \right]  \geq \nonumber \\
& d^{2} \left( \left(\Omega^{(p)}(\mu) \right)^{\frac{1}{p}} +  \left(\mathbb{E}_{\bm{\Theta}} \left[ \left( \Omega^{(p)}_{\mathbf{X}}(\bm{\theta}) \right)^{\frac{1}{p-1}} \right] \right)^{\frac{p-1}{p}} \right)^{-p}. \nonumber
\end{align}
	

\subsection{Proof of Corollary \ref{ThVanTreesXVTV2Cor}}~\label{secVI_CorTh1}

\subsubsection{Case $p \geq 2$}
Let $q \in \mathbb{R}$ such that $\frac{1}{p}+\frac{1}{q}=1$, i.e., $q=p/(p-1)$. Also, consider the following two functions $g(\cdot)$ and $h(\cdot)$ defined, for $\dv x \in \mc X$ and $\bs \theta =[\theta_1,\hdots,\theta_d] \in \Theta$, as 
\begin{subequations}
\begin{align}
g(\dv x,\bs \theta) &= \sum_{i=1}^d \frac{\partial}{\partial \theta_i} \left[ \log{\left(f(\dv x|\bs \theta) \mu(\bs \theta) \right)} \right]\\
h(\dv x,\bs \theta) &= \psi(\bs{\hat{\theta}}(\dv x)) - \psi(\bs \theta).
\end{align}
\end{subequations}	
For convenience,  for $i=1,\hdots,d$ we will denote the $i^{th}$ component of $h(\dv x, \bs \theta)$ as $h_i(\dv x,\bs \theta)$, i.e.,
\begin{equation}
h_i(\dv x,\bs \theta) = \psi_i(\bs{\hat{\theta}}(\dv x)) - \psi_i(\bs \theta) = \left(h(\dv x,\bs \theta)\right)_i
\end{equation}
The average estimation error can be lower bounded as
\begin{align}
&\mathbb{E}_{(\mathbf{X},\bm{\Theta})} \left[ \left \vert \left \vert  \psi(\bm{\hat{\theta}}(\mathbf{x})) - \psi(\bm{\theta}) \right \vert \right \vert_p^p \right] \nonumber \\
&\quad \stackrel{(a)}{=} \sum_{i=1}^d \mathbb{E}_{(\mathbf{X},\bm{\Theta})} \left[ \left \vert h_i(\mathbf{X},\bm{\Theta}) \right \vert^p \right] \label{partialFBRL2XVTV2Cor} \\
&\quad \stackrel{(b)}{\geq } \sum_{i=1}^d \mathbb{E}_{\bm{\Theta}} \left[ \left(\mathbb{E}_{\mathbf{X}|\bm{\Theta}} \left[ \left \vert h_i(\mathbf{X},\bm{\Theta}) \right \vert^2 \left \vert \right.\bm{\Theta} = \bs \theta  \right] \right)^{\frac{p}{2}} \right], \label{proof-generalization-of-van-Trees-inequality-step4Cor}
\end{align}
where $(a)$ follows from the definition of the $p$-norm and $(b)$ by replacing $\mathbb{E}_{(\mathbf{X},\bm{\Theta})}$ with $\mathbb{E}_{\bm{\Theta}}$ and $\mathbb{E}_{\mathbf{X}|\bm{\Theta}}$  and Jensen's inequality for expectations for convex functions $x \mapsto x^{\frac{p}{2}}$, for $2 < p $. 

\noindent The RHS of~\eqref{proof-generalization-of-van-Trees-inequality-step4Cor} can be lower bounded as follows. First, note that we have
\begin{align}
&\mathbb{E} \left[ \left\vert h_i(\dv X,\bs \Theta) g(\dv X,\bs \Theta) \right \vert  |  \right] = \mathbb{E}_{\bs \Theta} \mathbb{E}_{\dv X | \bs \Theta} \left[ \left( \left\vert h_i(\dv X,\bs \Theta) g(\dv X,\bs \Theta) \right\vert \right)  |  \bs \Theta = \bs \theta  \right] \nonumber\\
&\stackrel{(a)}{\leq} \mathbb{E}_{\bs \Theta} \left( \mathbb{E}_{\dv X | \bs \Theta} \left[ \left(\left\vert h_i(\dv X,\bs \Theta)\right\vert^2 \right) | \bs \Theta \right] \right)^{\frac{1}{2}}  \nonumber \\
&\quad \times \left( \mathbb{E}_{\dv X | \bs \Theta} \left[ \left(\left\vert g(\dv X,\bs \Theta)\right\vert^2 \right) | \bs \Theta = \bs \theta  \right] \right)^{\frac{1}{2}} \nonumber\\
&\stackrel{(b)}{\leq}  \left(\mathbb{E}_{\bs \Theta} \left[  \left \vert \mathbb{E}_{\dv X|\bs \Theta} \left[ \left( \left\vert h_i(\dv X,\bs \Theta) \right \vert^2\right) | \bs \Theta = \bs \theta  \right] \right \vert^{\frac{p}{2}} \right] \right)^{\frac{1}{p}}  \nonumber \\
&\quad \times \left(\mathbb{E}_{\bs \Theta} \left[ \left(\mathbb{E}_{\dv X|\bs \Theta} \left[ \left( \left \vert g(\dv X,\bs \Theta) \right \vert^2 \right) | \bs \Theta = \bs \theta  \right] \right)^{\frac{q}{2}} \right]\right)^{\frac{1}{q}},
\label{proof-generalization-of-van-Trees-inequality-step5Cor}
\end{align}
where $(a)$ follows by application of H\"older's inequality for every $\bs \theta \in \bs \Theta$ to the conditional expectation  $\mathbb{E}_{\dv X | \bs \Theta}[\cdot|\bs \theta]$ ; and $(b)$ follows by application of H\"older's inequality to the expectation  $\mathbb{E}_{\bs \Theta}[\cdot]$ since $p >1$, $q>1$ and are such that $\frac{1}{p}+\frac{1}{q}=1$.

\noindent Using the fact that the prior measure $\mu$ converges to zero at the endpoints of $\Theta$,  it is easy to see that
\begin{align}
&\int_{\theta_i} h_i(\mathbf{x},\bm{\theta}) \frac{\partial}{\partial \theta_i} \left[f(\mathbf{x}|\bm{\theta}) \mu_i(\theta_i) \right] \, \rm{d}{\theta_i}  \nonumber \\
& = \left.  h_i(\mathbf{x},\bm{\theta}) f(\mathbf{x}|\bm{\theta}) \mu_i(\theta_i) \right \vert_{\theta^{(i)}_{min}}^{\theta^{(i)}_{max}}  - \nonumber \\
&\quad \int_{\theta_i} \frac{\partial}{\partial \theta_i} \left[ h_i(\mathbf{x},\bm{\theta}) \right] f(\mathbf{x}|\bm{\theta}) \mu_i(\theta_i) \, \rm{d} \theta_i \nonumber \\
& =  - \int_{\theta_i} \frac{\partial}{\partial \theta_i} \left[ h_i(\mathbf{x},\bm{\theta}) \right] f(\mathbf{x}|\bm{\theta}) \mu_i(\theta_i) \, \rm{d} \theta_i. \label{proof-generalization-of-van-Trees-inequality-step1Cor}
\end{align}
\noindent By partial integration and~\eqref{proof-generalization-of-van-Trees-inequality-step1Cor}, we get for $i=1,\hdots,d$, that
\begin{align}
&\int_{\mathbf{x}} \int_{\theta_i} h_i(\mathbf{x},\bm{\theta}) \frac{\partial}{\partial \theta_i} \left[f(\mathbf{x}|\bm{\theta}) \mu_i(\theta_i) \right] \, \rm{d}{\theta_i} \, \rm{d}{\mathbf{x}}  \nonumber \\
&= - \mathbb{E}_{(\mathbf{X},\Theta_i)} \left[ \frac{\partial}{\partial \Theta_i} \left[ h_i(\mathbf{x},\bm{\Theta}) \right] \right].  \label{IntEq1MVTV2RG2PCor}
\end{align}

Thus, with some algebraic manipulations, 
\begin{align}
&\mathbb{E}_{(\mathbf{X},\bm{\Theta})} \left[ h_i(\mathbf{X},\bm{\Theta}) g(\mathbf{X},\bm{\Theta}) \right] \nonumber \\
&=\sum_{i=1}^d \mathbb{E}_{\Theta_1} \left[ \ldots \mathbb{E}_{\Theta_d} \left[ - \mathbb{E}_{(\mathbf{X},\Theta_i}) \left[ \frac{\partial}{\partial \Theta_i} \left[ h_i(\mathbf{X},\bm{\Theta}) \right] \right] \right] \right] \nonumber \\
&= - \sum_{i=1}^d \mathbb{E}_{(\mathbf{X},\bm{\Theta})} \left[ \frac{\partial}{\partial \Theta_i} \left[ h_i(\mathbf{X},\bm{\Theta}) \right] \right]   \nonumber \\
&= \sum_{i=1}^d \mathbb{E}_{\bm{\Theta}} \left[ \frac{\partial \psi_i(\bm{\theta})}{\partial \Theta_i} \right]
\end{align}
and $|\mathbb{E}[X]| \leq \mathbb{E}[|X|]$ lower bounds the left-hand side of (\ref{proof-generalization-of-van-Trees-inequality-step5Cor})
\begin{align}
&\left \vert \sum_{i=1}^d \mathbb{E}_{\bm{\Theta}} \left[ \frac{\partial \psi_i(\bm{\theta})}{\partial \Theta_i} \right]  \right \vert \leq \mathbb{E}_{(\mathbf{X},\bm{\Theta})} \left[ \left \vert h_i(\mathbf{X},\bm{\Theta}) g(\mathbf{X},\bm{\Theta}) \right \vert \right]. \label{BoundLHTRG2MVTV2PCor}
\end{align}

\noindent We now upper bound the second expectation of the RHS term of~\eqref{proof-generalization-of-van-Trees-inequality-step5Cor}, as follows. For convenience, let
\begin{equation}
l(\dv x,\bs \theta)= \sum_{i=1}^d \frac{\partial}{\partial \theta_i} \left[ \log{f(\dv x|\bs \theta)} \right].
\label{proof-generalization-of-van-Trees-inequality-step7Cor}
\end{equation}
It is easy to see that for all $\bs \theta$, we have
\begin{equation}
\mathbb{E}_{\dv X |\bs \Theta} \left[ l(\dv X,\bs \Theta) | \bs \Theta = \bs \theta \right]  =0
\label{proof-generalization-of-van-Trees-inequality-step8Cor}
\end{equation}
which follows by the regularity condition $\mathbb{E}_{\dv X|\bs \Theta}\left[\frac{\partial}{\partial \theta_i} \log{f(\dv x|\bs \theta)}\right] = 0$ for all $\bs \theta \in \bs \Theta$. Also,
\begin{equation}
g(\dv x,\bs \theta) = l(\dv x,\bs \theta) + \sum_{i=1}^d \frac{\partial}{\partial \theta_i} \left[ \log{\mu(\bs \theta)} \right].
\label{proof-generalization-of-van-Trees-inequality-step9Cor}
\end{equation}

\noindent Now, since 
\begin{equation}
\mathbb{E}_{\dv X|\bs \Theta} \left[ l(\dv X,\bs \Theta) \left(d \frac{\partial}{\partial \theta_i} \left[\log{\mu_i(\theta_i)} \right]  \right) | \bs \Theta=\bs \theta \right] = 0,
\label{proof-generalization-of-van-Trees-inequality-step10Cor}
\end{equation}
we get
\begin{align}
& \mathbb{E}_{\dv X|\bs \Theta} \left[ \left( g(\dv X,\bs \Theta) \right)^2 | \bs \Theta=\bs \theta \right]  \nonumber\\
& \quad= \mathbb{E}_{\dv X | \bs \Theta} \left[ l^2(\dv X,\bs \Theta) \: | \: \bs \Theta = \bs \theta \right] + \left( \sum_{i=1}^d \frac{\partial}{\partial \theta_i} \left[ \log{\mu(\bs \theta)} \right] \right)^2. 
\label{proof-generalization-of-van-Trees-inequality-step11Cor}
\end{align}
\noindent Thus,
\begin{align}
& \mathbb{E}_{\bs \Theta} \left[ \left(\mathbb{E}_{\dv X|\bs \Theta} \left[ \left \vert g(\dv X,\bs \Theta) \right \vert^2 | \bs \Theta=\bs \theta \right] \right)^{\frac{q}{2}} \right] \nonumber\\
&\stackrel{(a)}{\leq} \left(\mathbb{E} \left[ \left\vert g(\dv X,\bs \Theta) \right\vert^2 \right] \right)^{\frac{q}{2}} \\
&\stackrel{(b)}{\leq} \left( \mathbb{E}_{(\dv X, \bs \Theta)} \left[ l^2(\dv X,\bs \Theta) \right] + \mathbb{E}_{\bs \Theta} \left( \sum_{i=1}^d  \frac{\partial}{\partial \theta_i} \left[ \log{\mu(\bs \theta)} \right] \right)^2  \right)^{\frac{q}{2}}
\label{proof-generalization-of-van-Trees-inequality-step12Cor}
\end{align}
where $(a)$ follows using Jensen's inequality for the concave function $ u \longrightarrow u^{q/2}$ for $q = p/p-1 \leq 2$; and $(b)$ follows by substituting using~\eqref{proof-generalization-of-van-Trees-inequality-step11Cor}.

\noindent The first expectation term on the RHS of~\eqref{proof-generalization-of-van-Trees-inequality-step12Cor} is upper bounded as
\begin{align}
\mathbb{E}_{(\dv X,\bs \Theta)} \left[ l^2(\dv X,\bs \Theta) \right] &\stackrel{(a)}{=} \mathbb{E}_{(\dv X,\bs \Theta)} \left[ \left(\sum_{i=1}^d  \frac{\partial}{\partial \theta_i} \left[\log{f(\dv X|\bs \Theta)} \right] \right)^2 \right] \nonumber \\
&\stackrel{(b)}{\leq} d \, \mathbb{E}_{\bs \Theta} \left[\rm{Tr}(I_{\dv X}(\bs \theta)) \right]
\label{proof-generalization-of-van-Trees-inequality-step13Cor}
\end{align}
where $(a)$ follows by substituting using~\eqref{proof-generalization-of-van-Trees-inequality-step7Cor} and $(b)$ holds since for non-negative $\{u_i\}_{i=1}^d$ we have $\left(\sum_{i=1}^d u_i \right)^2 \leq d \, \sum_{i=1}^d u_i^2$. 

Hence, we get
\begin{align}
& \left( \mathbb{E}_{\bs \Theta} \left[ \left(\mathbb{E}_{\dv X|\bs \Theta} \left[ \left \vert g(\dv X,\bs \Theta) \right \vert^2 | \bs \Theta=\bs \theta \right] \right)^{\frac{q}{2}} \right] \right)^{\frac{p}{q}} \nonumber\\
&\leq \left( \mathbb{E}_{(\dv X, \bs \Theta)} \left[ l^2(\dv X,\bs \Theta) \right] + \mathbb{E}_{\bs \Theta} \left( \sum_{i=1}^d  \frac{\partial}{\partial \theta_i} \left[ \log{\mu(\bs \theta)} \right] \right)^2  \right)^{\frac{p}{2}} \nonumber\\
&\leq \left( d \, \mathbb{E}_{\bs \Theta} \left[\rm{Tr}(I_{\dv X}(\bs \theta)) \right] + d \, \rm{Tr}(I(\mu)) \right)^{\frac{p}{2}},
\label{proof-generalization-of-van-Trees-inequality-step14Cor}
\end{align}
where the last inequality follows from $\left(\sum_{i=1}^d u_i \right)^2 \leq d \, \sum_{i=1}^d u_i^2$, for non-negative $\{u_i\}_{i=1}^d$.


Summarizing, combining (\ref{proof-generalization-of-van-Trees-inequality-step4Cor}), (\ref{BoundLHTRG2MVTV2PCor}) and (\ref{proof-generalization-of-van-Trees-inequality-step14Cor}) leads us to the desired lower bound
\begin{align}
&\mathbb{E}_{(\mathbf{X},\bm{\Theta})} \left[ \left \vert \left \vert  \psi(\bm{\hat{\theta}}(\mathbf{x})) - \psi(\bm{\theta}) \right \vert \right \vert_p^p \right] \nonumber \\
&\quad \geq \sum_{i=1}^d \left[ \left \vert \sum_{i=1}^d \mathbb{E}_{\bm{\Theta}} \left[ \frac{\partial \psi_i(\bm{\theta})}{\partial \Theta_i} \right]  \right \vert^p \times \right. \nonumber \\
&\quad \quad \quad \left. \left( d \, \mathbb{E}_{\bm{\Theta}} \left[ \rm{Tr}(I_{\mathbf{X}}(\bm{\theta})) \right] + d \, \rm{Tr}(I(\mu)) \right)^{-\frac{p}{2}} \right] \nonumber \\
&\quad = d^{1-\frac{p}{2}} \,\left \vert \sum_{i=1}^d \mathbb{E}_{\bm{\Theta}} \left[ \frac{\partial \psi_i(\bm{\theta})}{\partial \Theta_i} \right]  \right \vert^p \times \nonumber \\
&\quad \quad \quad  \left( \mathbb{E}_{\bm{\Theta}} \left[ \rm{Tr}(I_{\mathbf{X}}(\bm{\theta})) \right] + \rm{Tr}(I(\mu)) \right)^{-\frac{p}{2}} \nonumber
\end{align}

\subsubsection{Case $1 < p < 2$}
Let $q \in \mathbb{R}$ such that $\frac{1}{p}+\frac{1}{q}=1$, i.e., $q=p/(p-1)$. Also, consider the following two functions $g(\cdot)$ and $h(\cdot)$ defined, for $\dv x \in \mc X$ and $\bs \theta =[\theta_1,\hdots,\theta_d] \in \Theta$, as 
\begin{subequations}
\begin{align}
g(\dv x,\bs \theta) &= \sum_{i=1}^d \frac{\partial}{\partial \theta_i} \left[ \log{\left(f(\dv x|\bs \theta) \mu(\bs \theta) \right)} \right]\\
h(\dv x,\bs \theta) &=  \psi(\hat{\bs \theta}(\dv x)) - \psi(\bs \theta).
\end{align}
\end{subequations}	
For convenience,  for $i=1,\hdots,d$ we will denote the $i^{th}$ component of $h(\dv x, \bs \theta)$ as $h_i(\dv x,\bs \theta)$, i.e.,
\begin{equation}
h_i(\dv x,\bs \theta) = \psi_i(\hat{\theta}(\dv x)) - \psi_i(\theta) = \left(h(\dv x,\bs \theta)\right)_i
\end{equation}



\noindent Using the definition of the $p$-norm, the average estimation error can be lower bounded as
\begin{align}
& \mathbb{E}_{(\mathbf{X},\bm{\Theta})} \left[ \left \vert \left \vert \psi(\bm{\hat{\theta}}(\mathbf{X})) - \psi(\bm{\theta}) \right \vert \right \vert_p^p \right] = \sum_{i=1}^d \mathbb{E}_{(\mathbf{X},\bm{\Theta})} \left[ \left \vert h_i(\mathbf{X},\bm{\Theta}) \right \vert^p \right] \label{almostDoneBound2XVTV2RL2Cor}.
\end{align}

\noindent The RHS of~\eqref{almostDoneBound2XVTV2RL2Cor} can be lower bounded as follows. First, note that applying H\"{o}lder's inequality for expectations yields
\begin{align}
&\mathbb{E}_{(\mathbf{X},\bm{\Theta})} \left[ \left \vert h_i(\mathbf{X},\bm{\Theta}) g(\mathbf{X},\bm{\Theta}) \right \vert \right] \leq  \left( \mathbb{E}_{(\mathbf{X},\bm{\Theta})} \left[ \left \vert h_i(\mathbf{X},\bm{\Theta}) \right \vert^p \right] \right)^{\frac{1}{p}} \times \nonumber \\
&\quad  \left(\mathbb{E}_{(\mathbf{X},\bm{\Theta})} \left[ \left \vert g(\mathbf{X},\bm{\Theta}) \right \vert^q \right] \right)^{\frac{1}{q}}. \label{OriginalHolderXVTV2RL2Cor}
\end{align}

\noindent Using the fact that the prior measure $\mu$ converges to zero at the endpoints of $\bs \Theta$ and partial integration, it is easy to see that
\begin{align}
&\int_{\theta_i} h_i(\mathbf{x},\bm{\theta}) \frac{\partial}{\partial \theta_i} \left[f(\mathbf{x}|\bm{\theta}) \mu_i(\theta_i) \right] \, \rm{d}{\theta_i}  \nonumber \\
& = \left.  h_i(\mathbf{x},\bm{\theta}) f(\mathbf{x}|\bm{\theta}) \mu_i(\theta_i) \right \vert_{\theta^{(i)}_{min}}^{\theta^{(i)}_{max}}  - \nonumber \\
&\quad \int_{\theta_i} \frac{\partial}{\partial \theta_i} \left[ h_i(\mathbf{x},\bm{\theta}) \right] f(\mathbf{x}|\bm{\theta}) \mu_i(\theta_i) \, \rm{d} \theta_i \nonumber \\
& =  - \int_{\theta_i} \frac{\partial}{\partial \theta_i} \left[ h_i(\mathbf{x},\bm{\theta}) \right] f(\mathbf{x}|\bm{\theta}) \mu_i(\theta_i) \, \rm{d} \theta_i. \label{proof-generalization-of-van-Trees-inequality-step1RL2Cor}
\end{align}
\noindent Integration in~\eqref{proof-generalization-of-van-Trees-inequality-step1RL2Cor}, we get for $i=1,\hdots,d$, that
\begin{align}
&\int_{\mathbf{x}} \int_{\theta_i} h_i(\mathbf{x},\bm{\theta}) \frac{\partial}{\partial \theta_i} \left[f(\mathbf{x}|\bm{\theta}) \mu_i(\theta_i) \right] \, \rm{d}{\theta_i} \, \rm{d}{\mathbf{x}}  \nonumber \\
&= - \mathbb{E}_{(\mathbf{X},\Theta_i)} \left[ \frac{\partial}{\partial \Theta_i} \left[ h_i(\mathbf{x},\bm{\Theta}) \right] \right].  \label{IntEq1MVTV2RL2PCor}
\end{align}

Thus, with some algebraic manipulations, 
\begin{align}
&\mathbb{E}_{(\mathbf{X},\bm{\Theta})} \left[ h_i(\mathbf{X},\bm{\Theta}) g(\mathbf{X},\bm{\Theta}) \right] \nonumber \\
&=\sum_{i=1}^d \mathbb{E}_{\Theta_1} \left[ \ldots \mathbb{E}_{\Theta_d} \left[ - \mathbb{E}_{(\mathbf{X},\Theta_i}) \left[ \frac{\partial}{\partial \Theta_i} \left[ h_i(\mathbf{X},\bm{\Theta}) \right] \right] \right] \right] \nonumber \\
&= - \sum_{i=1}^d \mathbb{E}_{(\mathbf{X},\bm{\Theta})} \left[ \frac{\partial}{\partial \Theta_i} \left[ h_i(\mathbf{X},\bm{\Theta}) \right] \right]   \nonumber \\
&= \sum_{i=1}^d \mathbb{E}_{\bm{\Theta}} \left[ \frac{\partial \psi_i(\bm{\theta})}{\partial \Theta_i} \right]
\end{align}
and $|\mathbb{E}[X]| \leq \mathbb{E}[|X|]$ lower bounds the left-hand side of (\ref{OriginalHolderXVTV2RL2Cor})
\begin{align}
&\left \vert \sum_{i=1}^d \mathbb{E}_{\bm{\Theta}} \left[ \frac{\partial \psi_i(\bm{\theta})}{\partial \Theta_i} \right]  \right \vert \leq \mathbb{E}_{(\mathbf{X},\bm{\Theta})} \left[ \left \vert h_i(\mathbf{X},\bm{\Theta}) g(\mathbf{X},\bm{\Theta}) \right \vert \right]. \label{BoundLHTRL2MVTV2PCor}
\end{align}

\noindent Combining~\eqref{almostDoneBound2XVTV2RL2Cor},~\eqref{OriginalHolderXVTV2RL2Cor} and~\eqref{BoundLHTRL2MVTV2PCor}, we get 
\begin{align}
&\mathbb{E}_{(\dv X,\bs \Theta)} \left[ \left \vert \left \vert \psi(\bs{\hat{\theta}}(\dv X)) - \psi(\bs{\theta}) \right \vert \right \vert_p^p \right]  \geq \nonumber \\
&d \, \left \vert \sum_{i=1}^d \mathbb{E}_{\bm{\Theta}} \left[ \frac{\partial \psi_i(\bm{\theta})}{\partial \Theta_i} \right]  \right \vert^p \left(\mathbb{E}_{(\mathbf{X},\bm{\Theta})} \left[ \left \vert g(\mathbf{X},\bm{\Theta}) \right \vert^q \right] \right)^{-\frac{p}{q}}.
\label{proof-generalization-of-van-Trees-inequality-step6RL2Cor}
\end{align}

\noindent We now upper bound the second expectation of the RHS term of~\eqref{proof-generalization-of-van-Trees-inequality-step6RL2Cor}, as follows. For convenience, let
\begin{equation}
l(\dv x,\bs \theta)= \sum_{i=1}^d \frac{\partial}{\partial \theta_i} \left[ \log{f(\dv x|\bs \theta)} \right].
\label{proof-generalization-of-van-Trees-inequality-step7RL2Cor}
\end{equation}
It is easy to see that for all $\bs \theta$, we have
\begin{equation}
\mathbb{E}_{\dv X |\bs \Theta} \left[ l(\dv X,\bs \Theta) | \bs \Theta = \bs \theta \right]  =0
\label{proof-generalization-of-van-Trees-inequality-step8RL2Cor}
\end{equation}
which follows by the regularity condition $\mathbb{E}_{\dv X|\bs \Theta}\left[\frac{\partial}{\partial \theta_i} \log{f(\dv x|\bs \theta)}\right] = 0$ for all $\bs \theta \in \bs \Theta$. Also,
\begin{equation}
g(\dv x,\bs \theta) = l(\dv x,\bs \theta) + \sum_{i=1}^d \frac{\partial}{\partial \theta_i} \left[ \log{\mu(\bs \theta)} \right].
\label{proof-generalization-of-van-Trees-inequality-step9RL2Cor}
\end{equation}
Note that $l(\mathbf{x},\bm{\theta})$ is the sum of the elements of the score function associated with $\mathbf{X}$. 

\noindent From~\eqref{proof-generalization-of-van-Trees-inequality-step9RL2Cor}, we have
\begin{align}
&\left(\mathbb{E}_{(\mathbf{X},\bm{\Theta})} \left[ \left \vert g(\mathbf{X},\bm{\Theta}) \right \vert^q \right] \right)^{\frac{1}{q}} \stackrel{(a)}{\leq} \label{XToreplacewithEqRE2Th1RL2Cor} \\
&\left( \mathbb{E}_{(\mathbf{X},\bm{\Theta})} \left[ \left \vert l(\mathbf{X},\bm{\Theta}) \right \vert^q \right] \right)^{\frac{1}{q}} + \left( \mathbb{E}_{\bm{\Theta}} \left[ \left \vert \sum_{i=1}^{d} \frac{\partial}{\partial \Theta_i} \left[ \log{\mu(\bm{\Theta})} \right] \right \vert^q \right] \right)^{\frac{1}{q}} \nonumber \\
&\stackrel{(b)}{\leq} \left( \mathbb{E}_{(\mathbf{X},\bm{\Theta})} \left[ \left \vert l(\mathbf{X},\bm{\Theta}) \right \vert^q \right] \right)^{\frac{1}{q}} + \sum_{i=1}^{d} \left( \mathbb{E}_{\bm{\Theta}} \left[ \left \vert \frac{\partial}{\partial \Theta_i} \left[ \log{\mu(\bm{\Theta})} \right] \right \vert^q \right] \right)^{\frac{1}{q}} \nonumber \\
&\stackrel{(c)}{\leq} \left( \mathbb{E}_{(\mathbf{X},\bm{\Theta})} \left[ \left \vert l(\mathbf{X},\bm{\Theta}) \right \vert^q \right] \right)^{\frac{1}{q}} \nonumber\\
& \qquad \quad + \sum_{i=1}^{d} \left( \left( \mathbb{E}_{\bm{\Theta}} \left[ \left \vert \frac{\partial}{\partial \Theta_i} \left[ \log{\mu(\bm{\Theta})} \right] \right \vert^{\frac{p}{p-1}} \right] \right)^{p-1} \right)^{\frac{1}{p}} \nonumber \\
&\stackrel{(d)}{\leq} \left( \mathbb{E}_{(\mathbf{X},\bm{\Theta})} \left[ \left \vert l(\mathbf{X},\bm{\Theta}) \right \vert^q \right] \right)^{\frac{1}{q}} + d^{\frac{p-1}{p}} \left( \Omega^{(p)}(\mu) \right)^{\frac{1}{p}}, 
\label{UBTh1RL2Cor}
\end{align}
where: $(a)$ follows by application of the Minkowski's inequality for expectations $\left(\mathbb{E} [|Z+T|^q] \right)^{\frac{1}{q}} \leq \left(\mathbb{E} [|Z|^q] \right)^{\frac{1}{q}} + \left(\mathbb{E} [|T|^q] \right)^{\frac{1}{q}}$ for r.v.s $Z$ and $T$; $(b)$ follows by application of the Minkowski's inequality for expectations $\left(\mathbb{E} [|\sum_{i=1}^d Z_i|^q] \right)^{\frac{1}{q}} \leq \sum_{i=1}^d \left(\mathbb{E} [|Z_i|^q] \right)^{\frac{1}{q}}$ for r.v.s $(Z_1,\hdots,Z_d)$; $(c)$ holds by substituting using $q=p/p-1$ and $(d)$ holds by first using the inequality $\sum_{i=1}^d u_i^{\frac{1}{p}} \leq d^{\frac{p-1}{p}} \left(\sum_{i=1}^d u_i\right)^{\frac{1}{p}}$ for non-negative $(u_1,\hdots,u_d)$ and $p > 1$ and then substituting using~\eqref{definition-trace-generalized-fisher-information-estimating-theta-from-prior}.

\noindent Continuing from~\eqref{UBTh1RL2Cor}, the first term of its RHS can be upper bounded as 
\begin{align}
&\left(\mathbb{E}_{(\mathbf{X},\bm{\Theta})}  \left[ |l(\mathbf{X},\bm{\Theta})|^q \right]\right)^{\frac{1}{q}} \nonumber \\
&\stackrel{(a)}{=} \left(\mathbb{E}_{(\mathbf{X},\bm{\Theta})}  \left[ \left \vert \sum_{i=1}^d   \frac{\partial}{\partial \theta_i} \left[\log{f(\mathbf{X}|\bm{\Theta})} \right] \right \vert^q \right]\right)^{\frac{1}{q}} \nonumber \\
&\stackrel{(b)}{\leq} d^{\frac{q-1}{q}} \left(\mathbb{E}_{(\mathbf{X},\bm{\Theta})}  \left[ \sum_{i=1}^d \left \vert \frac{\partial}{\partial \theta_i} \left[\log{f(\mathbf{X}|\bm{\Theta})} \right] \right \vert^q \right]\right)^{\frac{1}{q}} \nonumber \\
&\stackrel{(c)}{=} d^{\frac{1}{p}} \left(\mathbb{E}_{\bm{\Theta}}  \left[ \sum_{i=1}^d \mathbb{E}_{\mathbf{X}|\bm{\Theta}} \left[ \left \vert \frac{\partial}{\partial \theta_i} \left[\log{f(\mathbf{X}|\bm{\Theta})} \right] \right \vert^{\frac{p}{p-1}} \left \vert \right.\bm{\Theta} = \bs \theta  \right] \right] \right)^{\frac{p-1}{p}}  \nonumber \\
&\stackrel{(d)}{=} d^{\frac{1}{p}} \left(\mathbb{E}_{\bm{\Theta}}  \left[ \sum_{i=1}^d (v_i(\bs \theta))^{\frac{1}{p-1}}\right] \right)^{\frac{p-1}{p}}  \nonumber \\
&\stackrel{(e)}{\leq} d^{\frac{1}{p}} \left(\mathbb{E}_{\bm{\Theta}}  \left[ \left( \sum_{i=1}^d v_i(\bs \theta) \right)^{\frac{1}{p-1}}\right] \right)^{\frac{p-1}{p}}  \nonumber \\
&\stackrel{(f)}{=}  d^{\frac{1}{p}} \left( \mathbb{E}_{\bm{\Theta}}  \left[ \left( \Omega^{(p)}_{\mathbf{X}}(\bm{\theta}) \right)^{\frac{1}{p-1}} \right] \right)^{\frac{p-1}{p}} , \label{CombPart2XVTV2Th1RL2Cor}
\end{align}
where: $(a)$ follows by substituting using~\eqref{proof-generalization-of-van-Trees-inequality-step7RL2Cor}; $(b)$ holds by using the inequality $\left(\sum_{i=1}^d u_i \right)^{q} \leq d^{q-1} \sum_{i=1}^d u_i^{q} $ which holds for non-negative $(u_1,\hdots,u_d)$ and $q > 1$, $(c)$ follows by substituting using $q=\frac{p}{p-1}$; and $(d)$ holds by defining, for $i=1,\hdots,d$ and $\bs \theta \in \Theta$,
\begin{equation}
v_i(\bs \theta) = \left( \mathbb{E}_{\mathbf{X}|\bm{\Theta}} \left[ \left \vert \frac{\partial}{\partial \theta_i} \left[\log{f(\mathbf{X}|\bm{\Theta})} \right] \right \vert^{\frac{p}{p-1}} \left \vert \right.\bm{\Theta} = \bs \theta  \right] \right)^{p-1}; 
\label{DefUi3}
\end{equation}
$(e)$ holds by using the inequality $\sum_{i=1}^d u_i^{{\frac{1}{p-1}}} \leq \left(\sum_{i=1}^d u_i \right)^{\frac{1}{p-1}}$ for non-negative $(u_1,\hdots,u_d)$ and $p<2$; and $(f)$ holds by substituting using~\eqref{definition-trace-generalized-fisher-information-estimating-theta-from-vectorX}. 

Substituting (\ref{CombPart2XVTV2Th1RL2Cor}) in (\ref{UBTh1RL2Cor}), we obtain
\begin{align}
& \left(\mathbb{E}_{(\mathbf{X},\bm{\Theta})} \left[ \left \vert g(\mathbf{X},\bm{\Theta}) \right \vert^q \right] \right)^{\frac{1}{q}} \leq   \nonumber \\
&\quad d^{\frac{1}{p}} \left(\mathbb{E}_{\bm{\Theta}} \left[ \left( \Omega^{(p)}_{\mathbf{X}}(\bm{\theta}) \right)^{\frac{1}{p-1}} \right] \right)^{\frac{p-1}{p}} + d^{\frac{p-1}{p}} \left( \Omega^{(p)}(\mu) \right)^{\frac{1}{p}}. \label{XUBoundonGRL2Cor}
\end{align}
Substituting (\ref{XUBoundonGRL2Cor}) in (\ref{proof-generalization-of-van-Trees-inequality-step6RL2Cor}) produces the desired lower bound
\begin{align}
&\mathbb{E}_{(\mathbf{X},\bm{\Theta})} \left[ \left \vert \left \vert \psi(\bm{\hat{\theta}}(\mathbf{X})) - \psi(\bm{\theta}) \right \vert \right \vert_p^p \right]  \geq  \left \vert \sum_{i=1}^d \mathbb{E}_{\bm{\Theta}} \left[ \frac{\partial \psi_i(\bm{\theta})}{\partial \Theta_i} \right]  \right \vert^p \times  \nonumber \\
&\left(  d^{\frac{p-2}{p}}  \left(\Omega^{(p)}(\mu) \right)^{\frac{1}{p}} +  \left(\mathbb{E}_{\bm{\Theta}} \left[ \left( \Omega^{(p)}_{\mathbf{X}}(\bm{\theta}) \right)^{\frac{1}{p-1}} \right] \right)^{\frac{p-1}{p}} \right)^{-p}. \nonumber
\end{align}

\subsection{Proof of Theorem~\ref{ThVanTreesMVTV2}}

\subsubsection{Case  $p \geq 2$}~\label{secVI_Th2_RG2}

Let $q \in \mathbb{R}$ such that $\frac{1}{p}+\frac{1}{q}=1$, i.e., $q=p/(p-1)$. Also, consider the following two functions $g(\cdot)$ and $h(\cdot)$ defined, for $\dv x \in \mc X$, $\bs \theta =[\theta_1,\hdots,\theta_d] \in \Theta$ and a specific quantization messages tuple $\dv m^{(n)}=(m_1,\hdots,m_n) \in [1,2^k]^n$  as 
\begin{subequations}
\begin{align}
\label{DefGTh2RG2}
g(\dv m^{(n)},\bs \theta) &= \sum_{i=1}^d \frac{\partial}{\partial \theta_i} \left[ \log{\left(p(\dv m^{(n)}|\bs \theta) \mu(\bs \theta) \right)} \right]\\
h(\dv m^{(n)},\bs \theta) &=  \hat{\bs \theta}(\dv m^{(n)})- \bs \theta
\end{align}
\end{subequations}	
where in~\eqref{DefGTh2RG2} the quantization messages joint probability is $p(\mathbf{m^{(n)}}|\bm{\theta})=\prod_{j=1}^{n} p_j(m_j|\bm{\theta})$. For convenience,  for $i=1,\hdots,d$ we will denote the $i^{th}$ component of $h(\dv m^{(n)}, \bs \theta)$ as $h_i(\dv m^{(n)},\bs \theta)$, i.e.,
\begin{equation}
h_i(\dv m^{(n)},\bs \theta) = \hat{\theta}_i(\dv m^{(n)})- \theta_i = \left(h(\dv m^{(n)},\bs \theta)\right)_i.
\label{DefHTh2RG2}
\end{equation}
\noindent Using the fact that the prior measure $\mu$ converges to zero at the boundaries of $\Theta$,  it is easy to see that 
\begin{align}
\sum_{\mathbf{m^{(n)}}} \int_{\theta_i} &h_i(\mathbf{m^{(n)}},\bs \theta) \frac{\partial}{\partial \theta_i} \left[p(\mathbf{m^{(n)}}|\bm{\theta}) \mu_i(\theta_i) \right] \, \rm{d}{\theta_i}  = 1. 
\label{IntEq1MVTV2Th2RG2}
\end{align}

\noindent By partial integration and~\eqref{IntEq1MVTV2Th2RG2}, we get for $i=1,\hdots,d$, that
\begin{align}
&\mathbb{E}_{(\mathbf{M^{(n)}},\bm{\Theta})} \left[ h_i(\mathbf{M^{(n)}},\bs \Theta) g(\mathbf{M^{(n)}},\bm{\Theta}) \right] = d.
\end{align}
Thus, for all $i=1,\hdots,d$, we have
\begin{align}
d &\leq \mathbb{E}_{(\mathbf{M^{(n)}},\bm{\Theta})} \left[ \left \vert h_i(\mathbf{M^{(n)}},\bs \Theta) g(\mathbf{M^{(n)}},\bm{\Theta}) \right \vert \right]. 
\label{BoundLHTRG2MVTV2Th2RG2}
\end{align}

\noindent Note that we have
\begin{align}
&\mathbb{E} \left[ \left\vert h_i(\dv M^{(n)},\bs \Theta) g(\dv M^{(n)},\bs \Theta) \right \vert  \right] \label{OriginalHolderMVTV2Th2RG2}
 \\
&= \mathbb{E}_{\bs \Theta} \mathbb{E}_{\dv X | \bs \Theta} \left[ \left( \left\vert h_i(\dv M^{(n)},\bs \Theta) g(\dv M^{(n)},\bs \Theta) \right\vert \right)  |  \bs \Theta = \bs \theta \right] \nonumber\\
&\stackrel{(a)}{\leq} \mathbb{E}_{\bs \Theta} \left( \mathbb{E}_{\dv M^{(n)} | \bs \Theta} \left[ \left(\left\vert h_i(\dv M^{(n)},\bs \Theta)\right\vert^2 \right) | \bs \Theta = \bs \theta \right] \right)^{\frac{1}{2}}  \nonumber \\
&\quad \times \left( \mathbb{E}_{\dv M^{(n)} | \bs \Theta} \left[ \left(\left\vert g(\dv M^{(n)},\bs \Theta)\right\vert^2 \right) | \bs \Theta = \bs \theta \right] \right)^{\frac{1}{2}} \nonumber\\
&\stackrel{(b)}{\leq}  \left(\mathbb{E}_{\bs \Theta} \left[  \left \vert \mathbb{E}_{\dv M^{(n)}|\bs \Theta} \left[ \left( \left\vert h_i(\dv M^{(n)},\bs \Theta) \right \vert^2\right) | \bs \Theta = \bs \theta \right] \right \vert^{\frac{p}{2}} \right] \right)^{\frac{1}{p}}  \nonumber \\
&\quad \times \left(\mathbb{E}_{\bs \Theta} \left[ \left(\mathbb{E}_{\dv M^{(n)}|\bs \Theta} \left[ \left( \left \vert g(\dv M^{(n)},\bs \Theta) \right \vert^2 \right) | \bs \Theta = \bs \theta \right] \right)^{\frac{q}{2}} \right]\right)^{\frac{1}{q}}, \nonumber
\end{align}
where $(a)$ follows by application of H\"older's inequality for every $\bs \theta \in \bs \Theta$ to the conditional expectation  $\mathbb{E}_{\dv M^{(n)} | \bs \Theta}[\cdot|\bs \theta]$ ; and $(b)$ follows by application of H\"older's inequality to the expectation  $\mathbb{E}_{\bs \Theta}[\cdot]$ since $p >1$, $q>1$ and are such that $\frac{1}{p}+\frac{1}{q}=1$.

The first element of the right-hand side of (\ref{OriginalHolderMVTV2Th2RG2}(b)) produces the desired risk
\begin{align}
&\sup_{\bm{\theta} \in \bm{\Theta}} \mathbb{E}_{\mathbf{M^{(n)}}|\bm{\Theta}} \left[ \left \vert \left \vert \bm{\hat{\theta}}(\mathbf{M^{(n)}}) - \bm{\theta} \right \vert \right \vert_p^p \left \vert \right. \bm{\Theta} \right]  \nonumber \\
&\quad \stackrel{(a)}{\geq}  \mathbb{E}_{(\mathbf{M^{(n)}},\bm{\Theta})} \left[ \left \vert \left \vert \bm{\hat{\theta}}(\mathbf{M^{(n)}}) - \bm{\theta} \right \vert \right \vert_p^p \right] \nonumber \\
&\quad \stackrel{(b)}{=} \sum_{i=1}^d \mathbb{E}_{(\mathbf{M^{(n)}},\bm{\Theta})} \left[ \left \vert h_i(\mathbf{M^{(n)}},\bm{\Theta}) \right \vert^p \right] \label{partialFBRL2MVTV2PTh2RG2} \\
&\quad \stackrel{(c)}{\geq } \sum_{i=1}^d \mathbb{E}_{\bm{\Theta}} \left[ \left(\mathbb{E}_{\mathbf{M^{(n)}}|\bm{\Theta}} \left[ \left \vert h_i(\mathbf{M^{(n)}},\bm{\Theta}) \right \vert^2 \left \vert \right. \bm{\Theta} \right] \right)^{\frac{p}{2}} \right], \label{almostDoneBound2RG2MVTV2PTh2RG2}
\end{align}
where in $(a)$ the supremum upper bounds the expectation, $(b)$ follows by the definition of the $p$-norm and $(c)$ by replacing $\mathbb{E}_{(\mathbf{M^{(n)}},\bm{\Theta})}$ with $\mathbb{E}_{\bm{\Theta}}$ and $\mathbb{E}_{\mathbf{M^{(n)}}|\bm{\Theta}}$  and Jensen's inequality for expectations for convex functions $x \mapsto x^{\frac{p}{2}}$, for $2 < p $.

Combining (\ref{BoundLHTRG2MVTV2Th2RG2}), (\ref{OriginalHolderMVTV2Th2RG2}(b)) and (\ref{almostDoneBound2RG2MVTV2PTh2RG2}), we get
\begin{align}
&\sup_{\bm{\theta} \in \bm{\Theta}} \mathbb{E}_{\mathbf{M^{(n)}}|\bm{\Theta}} \left[ \left \vert \left \vert \bm{\hat{\theta}}(\mathbf{M^{(n)}}) - \bm{\theta} \right \vert \right \vert_p^p \left \vert \right. \bm{\Theta} \right] \geq \label{laststepTh2RG2} \\
&d^{p+1} \left(\mathbb{E}_{\bs \Theta} \left[ \left(\mathbb{E}_{\dv M^{(n)}|\bs \Theta} \left[ \left( \left \vert g(\dv M^{(n)},\bs \Theta) \right \vert^2 \right) | \bs \Theta = \bs \theta \right] \right)^{\frac{q}{2}} \right]\right)^{-\frac{p}{q}}. \nonumber
\end{align}

We now move on to the last step of the proof, to upper bound the expectation of the RHS of (\ref{laststepTh2RG2}). For convenience, let
\begin{align}
&l(m_j,\bm{\theta})= \sum_{i=1}^d \frac{\partial}{\partial \theta_i} \left[ \log{p(\mathbf{m_j}|\bm{\theta})} \right] \text{ with } \mathbb{E}_{M_j|\bm{\Theta}} \left[ l(M_j,\bm{\Theta}) \right]=0. \nonumber \\ 
&\text{ Then, } g(\mathbf{m^{(n)}},\bm{\theta})=\sum_{j=1}^{n} l(m_j,\bm{\theta}) + \sum_{i=1}^d  \frac{\partial}{\partial \theta_i} \left[ \log{\mu(\bm{\theta})} \right].
\end{align}
Note that $l(m_j,\bm{\theta})$ is the sum of the elements of the score function associated with $M_j$. We expand the square and cancel the product of the two elements, due to the property that $\mathbb{E}_{M_j|\bm{\Theta}} \left[ l(M_j,\bm{\Theta}) \right]=0$, to arrive at the trace of the Fisher information matrix of $M_j$ and that of the prior, respectively, as follows 
\begin{align}
&\left(\mathbb{E}_{\mathbf{M^{(n)}}|\bm{\Theta}} \left[ \left \vert g(\mathbf{M^{(n)}},\bm{\Theta}) \right \vert^2 \left \vert \right. \bm{\Theta} \right] \right)^{\frac{1}{2}} = \label{ToreplacewithEqRE2RG2PTh2RG2} \\
&\left[ \mathbb{E}_{\mathbf{M^{(n)}}|\bm{\Theta}} \left[ \left( \sum_{j=1}^{n} l(M_j,\bm{\Theta}) \right)^2 \left \vert \right. \bm{\Theta} \right] + \left( \sum_{i=1}^d  \frac{\partial}{\partial \Theta_i} \left[ \log{\mu(\bm{\Theta})} \right] \right)^2 \right]^{\frac{1}{2}}, \nonumber 
\end{align}
which holds by 
\[\mathbb{E}_{\mathbf{M^{(n)}}|\bm{\Theta}} \left[ \left( \sum_{j=1}^{n} l(M_j,\bm{\Theta}) \right) \left(\sum_{i=1}^d  \frac{\partial}{\partial \theta_i} \left[\log{\mu(\bm{\theta})} \right]  \right) \left \vert \right. \bm{\Theta} \right]=0. \] Further, by Jensen's inequality for expectations for concave functions $x \mapsto x^{\frac{q}{2}}$, for $q < 2 $, we have
\begin{align}
& \left(\mathbb{E}_{\bm{\Theta}} \left[ \left(\mathbb{E}_{\mathbf{M^{(n)}}|\bm{\Theta}} \left[ \left \vert g(\mathbf{M^{(n)}},\bm{\Theta}) \right \vert^2 \left \vert \right. \bm{\Theta} \right] \right)^{\frac{q}{2}} \right] \right)^{\frac{1}{q}}  \nonumber \\
&\leq \left(\mathbb{E}_{(\mathbf{M^{(n)}},\bm{\Theta})} \left[ \left( \sum_{j=1}^{n} l(M_j,\bm{\Theta}) \right)^2 \right] + \right. \nonumber \\
&\left. \quad + \mathbb{E}_{\bm{\Theta}} \left[ \left(\sum_{i=1}^d  \frac{\partial}{\partial \Theta_i} \left[ \log{\mu(\bm{\Theta})} \right] \right)^2 \right] \right)^{\frac{1}{2}} \nonumber \\ 
&\leq \left(\mathbb{E}_{(\mathbf{M^{(n)}},\bm{\Theta})} \left[ \left( \sum_{j=1}^{n} l(M_j,\bm{\Theta}) \right)^2 \right] + \right. \nonumber \\
&\left. \quad + d \, \sum_{i=1}^d   \mathbb{E}_{\bm{\Theta}} \left[ \left(\frac{\partial}{\partial \Theta_i} \left[ \log{\mu(\bm{\Theta})} \right]  \right)^2 \right] \right)^{\frac{1}{2}} \nonumber \\
&= \left(\mathbb{E}_{(\mathbf{M^{(n)}},\bm{\Theta})} \left[ \left( \sum_{j=1}^{n} l(M_j,\bm{\Theta}) \right)^2 \right]  + d \, \rm{Tr}(I(\mu)) \right)^{\frac{1}{2}}.  \label{partialReztosubstituteHereRG2PTh2RG2}
\end{align}

For $l(M_j,\bm{\Theta})$ with $\mathbb{E}_{M_j|\bm{\Theta}} \left[ l(M_j,\bm{\Theta}) \right]=0$ and independent, by the Marcinkiewicz-Zygmund inequality in the form of $(2)$ of \cite{Ren2001}, there exists a constant $B_2=1$ \cite{Burkholder1988}, such that
\begin{align}
&\mathbb{E}_{\mathbf{M^{(n)}}|\bm{\Theta}} \left[ \left(\sum_{j=1}^{n} l(M_j,\bm{\Theta}) \right)^2 \left \vert \right. \bm{\Theta} \right] = \nonumber \\
&\quad B_2 \mathbb{E}_{\mathbf{M^{(n)}}|\bm{\Theta}}  \left[ \sum_{j=1}^{n} l^2(M_j,\bm{\Theta}) \left \vert \right. \bm{\Theta} \right] \nonumber \\
&\quad \stackrel{(a)}{=} \sum_{j=1}^{n} \, \mathbb{E}_{\mathbf{M^{(n)}}|\bm{\Theta}} \left[ \left(\sum_{i=1}^d  \frac{\partial}{\partial \theta_i} \left[\log{p(M_j|\bm{\Theta})} \right] \right)^2 \left \vert \right. \bm{\Theta} \right] , \nonumber \\
&\quad \stackrel{(b)}{\leq} \, d \, \sum_{j=1}^{n} \rm{Tr}(I_{M_j}(\bm{\theta})), \label{CombPart2RG2MVTV2PTh2RG2}
\end{align}
where, by the independence of $M_j$, the expectation of each element of the summation is identical and this leads to the term $n$ in $(a)$, which also follows from the definition of $l(M_j,\bm{\Theta})$, $(b)$ is given by the inequality $\left(\sum_{i=1}^d x_i \right)^2 \leq d \, \sum_{i=1}^d x_i^2$, $x_i>0$, required in order to pass the summation inside the expectation and obtain the trace of the Fisher information matrix for $M_j$. 

Substituting (\ref{CombPart2RG2MVTV2PTh2RG2}) in (\ref{partialReztosubstituteHereRG2PTh2RG2}), we obtain
\begin{align}
& \left(\mathbb{E}_{\bm{\Theta}} \left[ \left(\mathbb{E}_{\mathbf{M^{(n)}}|\bm{\Theta}} \left[ \left \vert g(\mathbf{M^{(n)}},\bm{\Theta}) \right \vert^2 \left \vert \right. \bm{\Theta}\right] \right)^{\frac{q}{2}} \right] \right)^{\frac{1}{q}} \leq \nonumber \\
&\left( d \, \mathbb{E}_{\bm{\Theta}} \left[ \sum_{j=1}^{n} \rm{Tr}(I_{M_j}(\bm{\Theta})) \right] + d \, \rm{Tr}(I(\mu)) \right)^{\frac{1}{2}}. \label{UBoundonGRG2PTh2RG2}
\end{align}

Substituting (\ref{UBoundonGRG2PTh2RG2}) in (\ref{laststepTh2RG2}), produces the desired lower bound
\begin{align}
&\sup_{\bm{\theta} \in \bm{\Theta}} \mathbb{E}_{\mathbf{M^{(n)}}|\bm{\Theta}} \left[ \left \vert \left \vert \bm{\hat{\theta}}(\mathbf{M^{(n)}}) - \bm{\theta} \right \vert \right \vert_p^p \left \vert \right. \bm{\Theta}\right] \geq \nonumber \\
&d^{\left( 1+\frac{p}{2} \right)} \left( \sum_{j=1}^{n} \mathbb{E}_{\bm{\Theta}} \left[\rm{Tr}(I_{M_j}(\bm{\Theta})) \right] + \rm{Tr}(I(\mu)) \right)^{-\frac{p}{2}}.
\end{align}

\begin{Rem} For a random variable with bounded support, $\theta_i \in [-B,B]$, the prior distribution $\mu_i(\theta_i)$ that minimizes the Fisher information is the raised cosine distribution. That is, for $\theta_i \in [-B,B]$
\begin{align}
&\mu_i(\theta_i)=\frac{1}{B} \cos^2\left(\frac{\pi \theta_i}{2B} \right), I(\mu_i)=\frac{\pi^2}{B^2}, \Omega^{(p)}(\mu) =  \nonumber \\
& \sum_{i=1}^d I^{(p)}(\mu_i) = d \left( \mathbb{E}_{\Theta_i} \left[ \left \vert \frac{\partial}{\partial \Theta_i} \left[\log{ \mu_i(\Theta_i) } \right] \right \vert^{\frac{p}{p-1}} \right] \right)^{p-1} \nonumber \\
&= \frac{\pi d}{2} \left(\frac{2}{B} \right)^p \left[\mathcal{B}\left(\frac{2p-1}{2p-2}, \frac{2p-3}{2p-2} \right) \right]^{p-1}, \label{priorFisher}
\end{align} 
where $I(\mu_i)$ represents the Fisher information associated to the prior. The condition $p>1.5$ is required to ensure the existence of the Beta function $\mathcal{B}(\cdot)$.
\label{RemarkDefPrior}
\end{Rem}

\subsection{Proof of Theorem~\ref{ThVanTreesMVTV2Orlicz}}~\label{secVI_Orlicz}

\subsubsection{Case $p \geq 2$}

By Theorem \ref{ThVanTreesMVTV2}, if $p \geq 2$, we have that
\begin{align} 
&\sup_{\bm{\theta} \in \bm{\Theta}} \mathbb{E}_{\mathbf{M^{(n)}}|\bm{\Theta}} \left[ \left \vert \left \vert \bm{\hat{\theta}}(\mathbf{M^{(n)}}) - \bm{\theta} \right \vert \right \vert_p^p \left. \right \vert \bm{\Theta} \right]  \geq \nonumber \\
&\quad d^{\left(1+\frac{p}{2}\right)} \left( \sum_{j=1}^{n} \mathbb{E}_{\bm{\Theta}} \left[ \rm{Tr}(I_{M_j}(\bm{\theta})) \right] + \rm{Tr}(I(\mu)) \right)^{-\frac{p}{2}}. \nonumber
\end{align}

We need to compute an upper bound on $\rm{Tr}(I_{M_j}(\bm{\theta}))$. If $p \geq 2$, then, Theorem \ref{ExtensionTh2RG2} gives us that, for some $r \geq 1$, the upper bound holds
\begin{align}
\mathrm{Tr}(I_{M_j}(\bm{\theta})) &\leq \min{\{\mathrm{Tr}(I_{\mathbf{X}}(\bm{\theta}), 4 I_0 k^{\frac{2}{r}}\}}. \nonumber
\end{align}
Then, also by Remark \ref{RemarkDefPrior}, we obtain
\begin{align}
&\sup_{\bm{\theta} \in \bm{\Theta}} \mathbb{E}_{\mathbf{M^{(n)}}|\bm{\Theta}} \left[ \left \vert \left \vert \bm{\hat{\theta}}(\mathbf{M^{(n)}}) - \bm{\theta} \right \vert \right \vert_p^p \left \vert \right. \bm{\Theta} \right]  \geq \nonumber \\
&\quad \quad d^{\left( 1+\frac{p}{2} \right)} \left( 4 I_0 k^{\frac{2}{r}} n  + \frac{d \, \pi^2}{B^2} \right)^{-\frac{p}{2}}.
\end{align}

\subsubsection{Case $1 < p < 2$}

By Theorem \ref{ThVanTreesMVTV2}, if $ 1 < p < 2$, we have that
\begin{align}
&\sup_{\bm{\theta} \in \bm{\Theta}} \mathbb{E}_{\mathbf{M^{(n)}}|\bm{\Theta}} \left[ \left \vert \left \vert \bm{\hat{\theta}}(\mathbf{M^{(n)}}) - \bm{\theta} \right \vert \right \vert_p^p \left. \right \vert \bm{\Theta} \right]  \geq d^{p} \left[ d^{\frac{p-2}{p}} \left( \Omega^{(p)}(\mu) \right)^{\frac{1}{p}}  \right. \nonumber \\
&\left. +  \frac{1}{p-1} \left( \sum_{j=1}^{n} \left(\mathbb{E}_{\bm{\Theta}} \left[ \left( \Omega^{(p)}_{M_j}(\bm{\theta}) \right)^{\frac{1}{p-1}} \right] \right)^{\frac{2(p-1)}{p}} \right)^{\frac{1}{2}} \right]^{-p} \nonumber
\end{align}

We need to compute an upper bound on $\Omega_{M_j}^{(p)}(\bm{\theta}))$. If $1 < p < 2$, then, Theorem \ref{ExtensionTh2RL2} gives us that, for some $r \geq \frac{p}{2(p-1)}$, the upper bound holds
\begin{align}
\Omega_{M_j}^{(p)}(\bm{\theta}) &\leq \min{\{\Omega_{\mathbf{X}}^{(p)}(\bm{\theta}), d^{\frac{2-p}{2}} \, I_0^{\frac{p}{2}} \, (2^k)^{2-p} \, 2^p \, k^{\frac{p}{r}}\}}. \nonumber
\end{align}
Then, also by Remark \ref{RemarkDefPrior} $\forall p > 1.5$, which is required for the Beta function $\mathcal{B}(\cdot)$ to exist, we obtain
\begin{align}
&I^{(p)}(\mu_i) = \frac{\pi}{2} \left(\frac{2}{B} \right)^p \left[\mathcal{B}\left(\frac{2p-1}{2p-2}, \frac{2p-3}{2p-2} \right) \right]^{p-1} \text{ and } \nonumber \\
&\nonumber \\
&\sup_{\bm{\theta} \in \bm{\Theta}} \mathbb{E}_{\mathbf{M^{(n)}}|\bm{\Theta}} \left[ \left \vert \left \vert \bm{\hat{\theta}}(\mathbf{M^{(n)}}) - \bm{\theta} \right \vert \right \vert_p^p \left \vert \right. \bm{\Theta} \right]  \geq d^p  \left(\frac{2\sqrt{n}}{p-1} d^{\frac{2-p}{2p}} \,I_0^{\frac{1}{2}} \, k^{\frac{1}{r}} \times  \right. \nonumber \\
& \left. \, (2^k)^{\frac{2-p}{p}}  + d^{\frac{p-1}{p}}  \left( \frac{\pi}{2}  \right)^{\frac{1}{p}} \frac{2}{B} \left[\mathcal{B}\left(\frac{2p-1}{2p-2}, \frac{2p-3}{2p-2} \right) \right]^{\frac{p-1}{p}} \right)^{-p}. \nonumber
\end{align}

\subsection{Proof of Corollary \ref{ThGLM}}~\label{secVI_ThGLM}

Using the expression of the score function,
\begin{align}
&S_{\bm{\theta}}(\mathbf{X})=\frac{\partial}{\partial \bm{\theta}} \log{f(\mathbf{x}|\bm{\theta})}=\frac{1}{\sigma^2}(\mathbf{x}-\bm{\theta}), \nonumber
\end{align}
we compute
\begin{align}
&\Omega_{\mathbf{X}}^{(p)}(\bm{\theta}) =\sum_{i=1}^{d} \left( \mathbb{E}_{\mathbf{X}|\bm{\Theta}} \left[ \left \vert S_{\theta_i}(\mathbf{X}) \right \vert^{\frac{p}{p-1}} \left \vert \right.\bm{\Theta}  \right]\right)^{p-1} \nonumber \\
&=\left(\frac{1}{\sigma^2}\right)^p \sum_{i=1}^{d} \left( \mathbb{E}_{X_{i}|\bm{\Theta}} \left[ \left \vert X_{i}-\theta_i \right \vert^{\frac{p}{p-1}} \right]\right)^{p-1} \nonumber \\
& =\left(\frac{1}{\sigma^2}\right)^p \left(\frac{1}{\sqrt{2\pi \sigma^2}}\right)^{p-1} \times \nonumber \\
&\quad \times \sum_{i=1}^{d} \left( \int_{x_{i}} \left \vert x_{i}-\theta_i \right \vert^{\frac{p}{p-1}} \exp{\left[-\frac{(x_{i}-\theta)^2}{2 \sigma^2} \right]} \, \rm{d}x_{i} \right)^{p-1}. \nonumber
\end{align}
Using the gamma function, we compute the above integral as
\begin{align}
&\int_{x_{i}} |x_{i}-\theta_i|^{\frac{p}{p-1}} \exp{\left[-\frac{(x_{i}-\theta_i)^2}{2 \sigma^2} \right]} \, \rm{d}x_{i} = \nonumber \\
&\quad = \frac{2^{\frac{1}{2p-2}} \, \sigma^{\frac{2p-1}{p-1}}}{p-1} \Gamma\left(\frac{1}{2p-2} \right). \label{ComputeIntGamma2GLM}
\end{align}
Then, we obtain further 
\begin{align}
\Omega_{\mathbf{X}}^{(p)}(\bm{\theta})&= \frac{d \sqrt{2}}{\sigma} \left[\frac{\Gamma \left(\frac{1}{2p-2}\right)}{(p-1)\sqrt{2\pi \sigma^2}}  \right]^{p-1}. 
\end{align}

\subsubsection{Case $p \geq 2$}

By Theorem \ref{ThVanTreesMVTV2}, if $p \geq 2$, we have that
\begin{align} 
&\sup_{\bm{\theta} \in \bm{\Theta}} \mathbb{E}_{\mathbf{M^{(n)}}|\bm{\Theta}} \left[ \left \vert \left \vert \bm{\hat{\theta}}(\mathbf{M^{(n)}}) - \bm{\theta} \right \vert \right \vert_p^p \left. \right \vert \bm{\Theta} \right]  \geq \nonumber \\
&\quad d^{\left(1+\frac{p}{2}\right)} \left( \sum_{j=1}^{n} \mathbb{E}_{\bm{\Theta}} \left[ \rm{Tr}(I_{M_j}(\bm{\theta})) \right] + \rm{Tr}(I(\mu)) \right)^{-\frac{p}{2}}. \nonumber
\end{align}

We need to compute an upper bound on $\rm{Tr}(I_{M_j}(\bm{\theta}))$. If $p \geq 2$, then, Theorem \ref{ExtensionTh2RG2} gives us that, for some $r \geq 1$, the upper bound holds
\begin{align}
\mathrm{Tr}(I_{M_j}(\bm{\theta})) &\leq \min{\{\mathrm{Tr}(I_{\mathbf{X}}(\bm{\theta}), 4 I_0 k^{\frac{2}{r}}\}}. \nonumber
\end{align}
We move on to compute the value of $I_0$. That is, using the same approach as in Corollary $1$ of \cite{Barnes2019}, for $r=2\geq 1$, we obtain $I_0=\frac{8}{3\sigma^2}$. Then, by Remark \ref{RemarkDefPrior}, we get
\begin{align}
&\sup_{\bm{\theta} \in \bm{\Theta}} \mathbb{E}_{\mathbf{M^{(n)}}|\bm{\Theta}} \left[ \left \vert \left \vert \bm{\hat{\theta}}(\mathbf{M^{(n)}}) - \bm{\theta} \right \vert \right \vert_p^p \left \vert \right. \bm{\Theta} \right]  \geq \nonumber \\
&d^{\left( 1+\frac{p}{2} \right)} \max \left \{ \left( \frac{n \, d}{\sigma^2}  + \frac{d \, \pi^2}{B^2} \right)^{-\frac{p}{2}}, \left( \frac{32 \, n \, k}{3\sigma^2}  + \frac{d \, \pi^2}{B^2} \right)^{-\frac{p}{2}} \right \}.
\end{align}
For $\pi^2 \sigma^2 d \leq n B^2 \min\{k,d\}$, we can ignore the prior to obtain
\begin{align}
&\sup_{\bm{\theta} \in \bm{\Theta}} \mathbb{E}_{\mathbf{M^{(n)}}|\bm{\Theta}} \left[ \left \vert \left \vert \bm{\hat{\theta}}(\mathbf{M^{(n)}}) - \bm{\theta} \right \vert \right \vert_p^p \left \vert \right. \bm{\Theta} \right]  \geq \nonumber \\
&\quad \quad d^{\left( 1+\frac{p}{2} \right)} \max \left \{ \left( \frac{\sigma^2}{n \, d}\right)^{\frac{p}{2}}, \left( \frac{3\sigma^2}{32 \, n \, k} \right)^{\frac{p}{2}} \right \}.
\end{align}

\subsubsection{Case $1 < p < 2$}

By Theorem \ref{ThVanTreesMVTV2}, if $ 1 < p < 2$, we have that
\begin{align}
&\sup_{\bm{\theta} \in \bm{\Theta}} \mathbb{E}_{\mathbf{M^{(n)}}|\bm{\Theta}} \left[ \left \vert \left \vert \bm{\hat{\theta}}(\mathbf{M^{(n)}}) - \bm{\theta} \right \vert \right \vert_p^p \left. \right \vert \bm{\Theta} \right]  \geq d^{p} \left[ d^{\frac{p-2}{p}} \left( \Omega^{(p)}(\mu) \right)^{\frac{1}{p}}  \right. \nonumber \\
&\left. +  \frac{1}{p-1} \left( \sum_{j=1}^{n} \left(\mathbb{E}_{\bm{\Theta}} \left[ \left( \Omega^{(p)}_{M_j}(\bm{\theta}) \right)^{\frac{1}{p-1}} \right] \right)^{\frac{2(p-1)}{p}} \right)^{\frac{1}{2}} \right]^{-p} \nonumber
\end{align}

We need to compute an upper bound on $\Omega_{M_j}^{(p)}(\bm{\theta})$. If $1 < p < 2$, then, Theorem \ref{ExtensionTh2RL2} gives us that, for some $r \geq \frac{p}{2(p-1)}$, the upper bound holds
\begin{align}
\Omega_{M_j}^{(p)}(\bm{\theta}) &\leq \min{\{\Omega_{\mathbf{X}}^{(p)}(\bm{\theta}), d^{\frac{2-p}{2}} \, I_0^{\frac{p}{2}} \, (2^k)^{2-p} \, 2^p \, k^{\frac{p}{r}}\}}. \nonumber
\end{align}
We move on to compute the value of $I_0$. That is, using the same approach as in Corollary $1$ of \cite{Barnes2019}, for $r=2 \geq \frac{p}{2(p-1)}$ for $p > 1$, we obtain $I_0=\frac{8}{3\sigma^2}$. Then, we obtain
\begin{align}
&\Omega_{M_j}^{(p)}(\bm{\theta}) \leq \label{LowerBoundonTrMRL2W} \\
&\min \left \{ \frac{d \sqrt{2}}{\sigma} \left[\frac{\Gamma \left(\frac{1}{2p-2}\right)}{(p-1)\sqrt{2\pi \sigma^2}}  \right]^{p-1}, 2^{p+k(2-p)} d^{\frac{2-p}{2}} \left(\frac{8k}{3\sigma^2} \right)^{\frac{p}{2}} \right \}. \nonumber
\end{align}

Then, also by Remark \ref{RemarkDefPrior} $\forall p > 1.5$, which is required for the Beta function $\mathcal{B}(\cdot)$ to exist, we obtain
\begin{align}
&\Omega^{(p)}(\mu) = \frac{\pi \, d}{2} \left(\frac{2}{B} \right)^p \left[\mathcal{B}\left(\frac{2p-1}{2p-2}, \frac{2p-3}{2p-2} \right) \right]^{p-1} \text{ and } \nonumber \\
&\nonumber \\
&\sup_{\bm{\theta} \in \bm{\Theta}} \mathbb{E}_{\mathbf{M^{(n)}}|\bm{\Theta}} \left[ \left \vert \left \vert \bm{\hat{\theta}}(\mathbf{M^{(n)}}) - \bm{\theta} \right \vert \right \vert_p^p \left \vert \right. \bm{\Theta} \right]  \geq d^{p} \times \nonumber \\
&\max \left \{  \left(\frac{\sqrt{n}}{p-1} \left(\frac{d \sqrt{2}}{\sigma}\right)^{\frac{1}{p}} \left[\frac{\Gamma \left(\frac{1}{2p-2}\right)}{(p-1)\sqrt{2\pi \sigma^2}}  \right]^{\frac{p-1}{p}} + \right. \right. \nonumber \\
&\quad \left. \left. + \, d^{\frac{p-1}{p}}  \left( \frac{\pi}{2}  \right)^{\frac{1}{p}} \frac{2}{B} \left[\mathcal{B}\left(\frac{2p-1}{2p-2}, \frac{2p-3}{2p-2} \right) \right]^{\frac{p-1}{p}} \right)^{-p}, \right. \nonumber \\
& \left. \left(\frac{\sqrt{n}}{p-1} 2^{\frac{p+k(2-p)}{p}} d^{\frac{2-p}{2p}} \left(\frac{8k}{3\sigma^2} \right)^{\frac{1}{2}} + \right. \right. \nonumber \\
&\quad \left. \left. + \, d^{\frac{p-1}{p}}  \left( \frac{\pi}{2}  \right)^{\frac{1}{p}} \frac{2}{B} \left[\mathcal{B}\left(\frac{2p-1}{2p-2}, \frac{2p-3}{2p-2} \right) \right]^{\frac{p-1}{p}} \right)^{-p} \right \}. \nonumber
\end{align}


\subsection{Proof of Theorem \ref{ThVanTreesWass}}~\label{secVI_Wass}

From the definition of the Wasserstein distance, we have
\begin{align}
& \mathbb{E}_{(\mathbf{M^{(n)}},\bm{\Theta})} \left[ W_p^p(f(\mathbf{x}|\bm{\hat{\theta}}(\mathbf{M^{(n)}})),f(\mathbf{x}|\bm{\theta}))  \right]  \nonumber \\
&\quad =\mathbb{E}_{(\mathbf{M^{(n)}},\bm{\Theta})} \left[ \mathbb{E}_{(\mathbf{Z},\mathbf{Y}) \sim \mu^*} \left[ d^p(\mathbf{Z},\mathbf{Y}) \right] \right]  \nonumber \\
&\quad = \mathbb{E}_{(\mathbf{M^{(n)}},\bm{\Theta})} \left[ \int_{\mathbf{z}} \int_{\mathbf{y}} d^p(\mathbf{z},\mathbf{y}) \mu^*(\mathbf{z},\mathbf{y}) \, \rm{d}\mathbf{z} \, \rm{d}\mathbf{y} \right]
\end{align}
If $d(\mathbf{z},\mathbf{y})=||\mathbf{z}-\mathbf{y}||_p$, then
\begin{align}
& \mathbb{E}_{(\mathbf{M^{(n)}},\bm{\Theta})} \left[ W_p^p(f(\mathbf{x}|\bm{\hat{\theta}}(\mathbf{M^{(n)}})),f(\mathbf{x}|\bm{\theta}))  \right]  \label{NewWassIneqP} \\
&\, =\mathbb{E}_{(\mathbf{M^{(n)}},\bm{\Theta})} \left[ \mathbb{E}_{(\mathbf{Z},\mathbf{Y}) \sim \mu^*} \left[ d^p(\mathbf{Z},\mathbf{Y}) \right] \right]  \nonumber \\
&\, = \mathbb{E}_{(\mathbf{M^{(n)}},\bm{\Theta})} \left[ \int_{\mathbf{z}} \int_{\mathbf{y}} \left(\sum_{i=1}^d |z_i-y_i|^p \right) \mu^*(\mathbf{z},\mathbf{y}) \, \rm{d}\mathbf{z} \, \rm{d}\mathbf{y} \right] \nonumber \\
&\, = \mathbb{E}_{(\mathbf{M^{(n)}},\bm{\Theta})} \left[  \mathbb{E}_{(\mathbf{Z},\mathbf{Y}) \sim \mu^*} \left[ \sum_{i=1}^d |Z_i-Y_i|^p \right]\right] \nonumber \\
&\, = \mathbb{E}_{(\mathbf{M^{(n)}},\bm{\Theta})} \left[ \sum_{i=1}^d \mathbb{E}_{(\mathbf{Z},\mathbf{Y}) \sim \mu^*} \left[ |Z_i-Y_i|^p \right] \right] \nonumber \\
&\, \stackrel{(a)}{\geq} \mathbb{E}_{(\mathbf{M^{(n)}},\bm{\Theta})} \left[ \sum_{i=1}^d \left(  \mathbb{E}_{(\mathbf{Z},\mathbf{Y}) \sim \mu^*} \left[ |Z_i-Y_i| \right] \right)^p \right] \nonumber \\
&\, \stackrel{(b)}{\geq}  \mathbb{E}_{(\mathbf{M^{(n)}},\bm{\Theta})} \left[ \sum_{i=1}^d \left \vert  \mathbb{E}_{(\mathbf{Z},\mathbf{Y}) \sim \mu^*} \left[ Z_i-Y_i \right] \right \vert^p \right] \nonumber \\
&\, = \mathbb{E}_{(\mathbf{M^{(n)}},\bm{\Theta})} \left[ \sum_{i=1}^d \left \vert  \mathbb{E}_{\mathbf{Z} \sim f(\mathbf{z}|\bm{\hat{\theta}}(\mathbf{M^{(n)}}))} [ Z_i] - \mathbb{E}_{\mathbf{Y} \sim f(\mathbf{y}|\bm{\Theta})} [Y_i] \right \vert^p \right], \nonumber
\end{align}
where $(a)$ follows from Jensen's inequality for convex functions of expectations, $\mathbb{E}[|X|^p] \geq \left( \mathbb{E}[|X|] \right)^p$, $p>1$ and $(b)$ is given by $\mathbb{E}[|X|] \geq \left \vert \mathbb{E}[X] \right \vert$.

\subsubsection{Case $1 < p < 2$}

Let $q \in \mathbb{R}$ such that $\frac{1}{p}+\frac{1}{q}=1$, i.e., $q=p/(p-1)$. Also, consider the following two functions $g(\cdot)$ and $h(\cdot)$ defined, for $\dv x \in \mc X$, $\bs \theta =[\theta_1,\hdots,\theta_d] \in \Theta$ and a specific quantization messages tuple $\dv m^{(n)}=(m_1,\hdots,m_n) \in [1,2^k]^n$  as 
\begin{subequations}
\begin{align}
\label{DefGWass}
g(\dv m^{(n)},\bs \theta) &= \sum_{i=1}^d \frac{\partial}{\partial \theta_i} \left[ \log{\left(p(\dv m^{(n)}|\bs \theta) \mu(\bs \theta) \right)} \right]\\
h(\dv m^{(n)},\bs \theta) &=  \mathbb{E}_{\mathbf{Z} \sim f(\mathbf{z}|\bm{\hat{\theta}}(\mathbf{M^{(n)}}))} [ \mathbf{Z}] - \mathbb{E}_{\mathbf{Y} \sim f(\mathbf{y}|\bm{\Theta})} [\mathbf{Y}]
\end{align}
\end{subequations}	
where in~\eqref{DefGWass} the quantization messages joint probability is $p(\mathbf{m^{(n)}}|\bm{\theta})=\prod_{j=1}^{n} p_j(m_j|\bm{\theta})$. For convenience,  for $i=1,\hdots,d$ we will denote the $i^{th}$ component of $h(\dv m^{(n)}, \bs \theta)$ as $h_i(\dv m^{(n)},\bs \theta)$, i.e.,
\begin{align}
h_i(\dv m^{(n)},\bs \theta) &= \mathbb{E}_{\mathbf{Z} \sim f(\mathbf{z}|\bm{\hat{\theta}}(\mathbf{M^{(n)}}))} [ Z_i] - \mathbb{E}_{\mathbf{Y} \sim f(\mathbf{y}|\bm{\Theta})} [Y_i] \nonumber \\
&= \left(h(\dv m^{(n)},\bs \theta)\right)_i.
\label{DefHWass}
\end{align}

\noindent Applying H\"{o}lder's inequality for expectations yields
\begin{align}
&\mathbb{E}_{(\mathbf{M^{(n)}},\bm{\Theta})} \left[ \left \vert h_i(\mathbf{M^{(n)}},\bs \Theta) g(\mathbf{M^{(n)}},\bm{\Theta}) \right \vert \right] \leq \nonumber \\
&\left( \mathbb{E} \left[ \left \vert h_i(\mathbf{M^{(n)}},\bs \Theta) \right \vert^p \right] \right)^{\frac{1}{p}}\left(\mathbb{E} \left[ \left \vert g(\mathbf{M^{(n)}},\bm{\Theta}) \right \vert^q \right] \right)^{\frac{1}{q}}.
\label{Holder-inequalityWass}
\end{align}
The first element of the right-hand side produces the desired risk as 
\begin{align}
&\sup_{\bm{\theta} \in \bm{\Theta}} \mathbb{E}_{\mathbf{M^{(n)}}|\bm{\Theta}} \left[ W_p^p(f(\mathbf{x}|\bm{\hat{\theta}}(\mathbf{M^{(n)}})),f(\mathbf{x}|\bm{\theta}))  \right]  \nonumber \\
&\quad \geq \sum_{i=1}^d \mathbb{E}_{(\mathbf{M^{(n)}},\bm{\Theta})} \left[ \left \vert h_i(\mathbf{M^{(n)}},\bs \Theta) \right \vert^p \right] \label{almostDoneBound2MVTV2RL2Wass}
\end{align}
where the inequality follows by substituting using~\eqref{DefHWass} and \ref{NewWassIneqP} and the fact that the supremum of a function is larger than its expectation. In the following, in order to avoid confusion in the indeces, we will use the notation $h_j(\dot)$.



\noindent Using the fact that the prior measure $\mu$ converges to zero at the endpoints of $\bs \Theta$ and partial integration, it is easy to see that
\begin{align}
&\int_{\theta_i} h_j(\mathbf{m^{(n)}},\bm{\theta}) \frac{\partial}{\partial \theta_i} \left[p(\dv m^{(n)}|\bs \theta)  \mu_i(\theta_i) \right] \, \rm{d}{\theta_i}  \nonumber \\
& = \left.  h_j(\mathbf{m^{(n)}},\bm{\theta}) p(\dv m^{(n)}|\bs \theta)  \mu_i(\theta_i) \right \vert_{\theta^{(i)}_{min}}^{\theta^{(i)}_{max}}  - \nonumber \\
&\quad \int_{\theta_i} \frac{\partial}{\partial \theta_i} \left[ h_i(\mathbf{m^{(n)}},\bm{\theta}) \right] p(\dv m^{(n)}|\bs \theta)  \mu_i(\theta_i) \, \rm{d} \theta_i \nonumber \\
& =  - \int_{\theta_i} \frac{\partial}{\partial \theta_i} \left[ h_i(\mathbf{m^{(n)}},\bm{\theta}) \right] p(\dv m^{(n)}|\bs \theta)  \mu_i(\theta_i) \, \rm{d} \theta_i. \label{proof-generalization-of-van-Trees-inequality-step1RL2Wass}
\end{align}
\noindent Summing over all messages in~\eqref{proof-generalization-of-van-Trees-inequality-step1RL2Wass}, we get for $i=1,\hdots,d$, that
\begin{align}
&\sum_{\mathbf{m^{(n)}}} \int_{\theta_i} h_j(\mathbf{m^{(n)}},\bm{\theta}) \frac{\partial}{\partial \theta_i} \left[p(\dv m^{(n)}|\bs \theta)  \mu_i(\theta_i) \right] \, \rm{d}{\theta_i}  \nonumber \\
&= - \mathbb{E}_{(\mathbf{M^{(n)}},\Theta_i)} \left[ \frac{\partial}{\partial \Theta_i} \left[ h_j(\mathbf{m^{(n)}},\bm{\Theta}) \right] \right].  \label{IntEq1MVTV2RL2PWass}
\end{align}

Thus, with some algebraic manipulations, 
\begin{align}
&\mathbb{E}_{(\mathbf{M^{(n)}},\bm{\Theta})} \left[ h_j(\mathbf{M^{(n)}},\bm{\Theta}) g(\mathbf{M^{(n)}},\bm{\Theta}) \right] \nonumber \\
&=\sum_{i=1}^d \mathbb{E}_{\Theta_1} \left[ \ldots \mathbb{E}_{\Theta_d} \left[ - \mathbb{E}_{(\mathbf{M^{(n)}},\Theta_i}) \left[ \frac{\partial}{\partial \Theta_i} \left[ h_j(\mathbf{M^{(n)}},\bm{\Theta}) \right] \right] \right] \right] \nonumber \\
&= - \sum_{i=1}^d \mathbb{E}_{(\mathbf{M^{(n)}},\bm{\Theta})} \left[ \frac{\partial}{\partial \Theta_i} \left[ h_j(\mathbf{M^{(n)}},\bm{\Theta}) \right] \right]   \nonumber \\
&= \sum_{i=1}^d \mathbb{E}_{\bm{\Theta}} \left[ \frac{\partial }{\partial \Theta_i} \mathbb{E}_{\mathbf{Y} \sim f(\mathbf{y}|\bm{\Theta})} [Y_j] \right] \nonumber
\end{align}
and $|\mathbb{E}[X]| \leq \mathbb{E}[|X|]$ lower bounds the left-hand side of (\ref{Holder-inequalityWass}) as
\begin{align}
&\left \vert \sum_{i=1}^d \mathbb{E}_{\bm{\Theta}} \left[ \frac{\partial }{\partial \Theta_i} \left[ \mathbb{E}_{\mathbf{Y} \sim f(\mathbf{y}|\bm{\Theta})} [Y_j] \right] \right]  \right \vert \leq \nonumber \\
&\quad \quad \quad \mathbb{E}_{(\mathbf{M^{(n)}},\bm{\Theta})} \left[ \left \vert h_j(\mathbf{X},\bm{\Theta}) g(\mathbf{X},\bm{\Theta}) \right \vert \right]. \label{BoundLHTRL2MVTV2PWass}
\end{align}
\noindent Combining~\eqref{NewWassIneqP},~\eqref{Holder-inequalityWass},~\eqref{almostDoneBound2MVTV2RL2Wass} and~\eqref{BoundLHTRL2MVTV2PWass}, we get 
\begin{align}
& \mathbb{E}_{(\mathbf{M^{(n)}},\bm{\Theta})} \left[ W_p^p(f(\mathbf{x}|\bm{\hat{\theta}}(\mathbf{M^{(n)}})),f(\mathbf{x}|\bm{\theta}))  \right]  \geq  \nonumber \\
& \quad \left( \sum_{j=1}^d \left \vert \sum_{i=1}^d \mathbb{E}_{\bm{\Theta}} \left[ \frac{\partial }{\partial \Theta_i} \left[ \mathbb{E}_{\mathbf{Y} \sim f(\mathbf{y}|\bm{\Theta})} [Y_j] \right] \right] \right \vert^p \right) \times \nonumber \\
& \quad  \left(\mathbb{E}_{(\mathbf{M^{(n)}},\bm{\Theta})} \left[ \left \vert g(\mathbf{M^{(n)}},\bm{\Theta}) \right \vert^q \right] \right)^{-\frac{p}{q}}.
\label{proof-generalization-of-van-Trees-inequality-step6RL2Wass}
\end{align}

\noindent We now upper bound the second expectation term of the RHS of~\eqref{proof-generalization-of-van-Trees-inequality-step6RL2Wass}. For convenience, let for $j=1,\hdots,2^k$
\begin{equation}
l(m_j,\bm{\theta})= \sum_{i=1}^d \frac{\partial}{\partial \theta_i} \left[ \log{p(m_j|\bm{\theta})} \right]
\label{intermediate-step-1Wass}
\end{equation}
It is easy to see that for all $\bs \theta$, we have
\begin{equation}
\mathbb{E}_{M_j |\bs \Theta} \left[ l(M_j,\bs \Theta) | \bs \Theta = \bs \theta \right]  =0.
\label{proof-generalization-of-van-Trees-inequality-step8Wass}
\end{equation}

\noindent Then, we have
\begin{align}
&\left(\mathbb{E}_{(\mathbf{M^{(n)}},\bm{\Theta})} \left[ \left \vert g(\mathbf{M^{(n)}},\bm{\Theta}) \right \vert^q \right] \right)^{\frac{1}{q}} \leq \label{ToreplacewithEqRE2RL2Wass} \\
&\left( \mathbb{E}_{(\mathbf{M^{(n)}},\bm{\Theta})} \left[ \left \vert \sum_{j=1}^{n} l(M_j,\bm{\Theta}) \right \vert^q \right] \right)^{\frac{1}{q}} + d^{\frac{p-1}{p}} \left(\Omega^{(p)}(\mu) \right)^{\frac{1}{p}},\nonumber
\end{align}
where the inequality holds by a double application by Minkowski's inequality: first for expectations using that for all $Z$ and $T$ we have $\left(\mathbb{E} [|Z+T|^q] \right)^{\frac{1}{q}}$ $\leq \left(\mathbb{E} [|Z|^q] \right)^{\frac{1}{q}} + \left(\mathbb{E} [|T|^q] \right)^{\frac{1}{q}}$; and then that $\left(\mathbb{E} [|\sum_{i=1}^d Z_i|^q] \right)^{\frac{1}{q}} $$\leq \sum_{i=1}^d \left(\mathbb{E} [|Z_i|^q] \right)^{\frac{1}{q}}$, $\forall \, q>1$ and  $\sum_{i=1}^d u_i^\frac{1}{p} $$\leq d^{\frac{p-1}{p}} \left(\sum_{i=1}^d u_i \right)^{\frac{1}{p}}$, $\forall \, u_i>0$, $p>1$. 

\noindent Next, since the quantities $\{l(M_j,\bm{\Theta})\}_j$ are independent and satisfy that $\mathbb{E}_{M_j|\bm{\Theta}} \left[ l(M_j,\bm{\Theta}) \right]=0$ for all $j=1,\hdots,2^k$, the application of Marcinkiewicz-Zygmund inequality~\cite{Ren2001,Burkholder1988} yields
\begin{align}
& \mathbb{E}_{\mathbf{M^{(n)}}|\bm{\Theta}} \left[ \left \vert \sum_{j=1}^{n} l(M_j,\bm{\Theta}) \right \vert^q \left. \right \vert \bm{\Theta} \right] \leq \nonumber \\
&\quad \quad B_q \, \mathbb{E}_{\mathbf{M^{(n)}}|\bm{\Theta}}  \left[ \left( \sum_{j=1}^{n} l^2(M_j,\bm{\Theta})\right)^{\frac{q}{2}} \left. \right \vert \bm{\Theta} \right]
\label{intermediate-stepWass}
\end{align}
where $B_q=1/(q-1)^q>0$.

\noindent Continuing from of~\eqref{intermediate-stepWass}, we get
\begin{align}
&\left( \mathbb{E}_{(\mathbf{M^{(n)}},\bm{\Theta})} \left[ \left \vert \sum_{j=1}^{n} l(M_j,\bm{\Theta}) \right \vert^q \right] \right)^{\frac{2}{q}} \leq \nonumber \\
&\stackrel{(a)}{\leq} (p-1)^2 \sum_{j=1}^{n} \left(\mathbb{E}_{(\mathbf{M^{(n)}},\bm{\Theta})}  \left[ |l(M_j,\bm{\Theta})|^q \right]\right)^{\frac{2}{q}} \nonumber \\
&\stackrel{(b)}{\leq} d^{\frac{2}{p}}(p-1)^2 \sum_{j=1}^{n}  \left(\mathbb{E}_{(\mathbf{M^{(n)}},\bm{\Theta})}  \left[ \sum_{i=1}^d \left \vert \frac{\partial}{\partial \theta_i} \left[\log{p(M_j|\bm{\Theta})} \right] \right \vert^q \right]\right)^{\frac{2}{q}} \nonumber \\
&\stackrel{(c)}{=} d^{\frac{2}{p}}(p-1)^2 \sum_{j=1}^{n} \left( \mathbb{E}_{\bm{\Theta}}  \left[ \left( \Omega^{(p)}_{M_j}(\bm{\theta}) \right)^{\frac{1}{p-1}} \right] \right)^{\frac{2(p-1)}{p}} , \label{CombPart2MVTV2RL2Wass}
\end{align}
where $(a)$ follows by using the quantization messages are independent, substituting $q=p/(p-1)$ and applying Minkowski's inequality $\left(\mathbb{E} [|\sum_{i=1}^{n} Z_i|^{\frac{q}{2}}] \right)^{\frac{2}{q}} \leq \sum_{i=1}^{n} \left(\mathbb{E} [|Z_i|^{\frac{q}{2}}] \right)^{\frac{2}{q}}$ since $q=p>(p-1)>2$; $(b)$ follows by substituting using~\eqref{intermediate-step-1Wass} and using that $\left(\sum_{i=1}^d u_i \right)^{q} \leq d^{q-1} \sum_{i=1}^d u_i^{q} $, $u_i>0$; and $(c)$ holds by ~\eqref{definition-trace-generalized-fisher-information-estimating-theta-from-quantization-tuple}.

Finally, combining \eqref{CombPart2MVTV2RL2Wass} with \eqref{ToreplacewithEqRE2RL2Wass} and substituting in \eqref{proof-generalization-of-van-Trees-inequality-step6RL2Wass} yields the desired result
\begin{align}
&\sup_{\bm{\theta} \in \bm{\Theta}} \mathbb{E}_{\mathbf{M^{(n)}}|\bm{\Theta}} \left[W_p^p(f(\dv x|\bm{\hat{\theta}}(\mathbf{M^{(n)}})),f(\dv x|\bm{\theta})) \left. \right \vert \bm{\Theta} \right] \geq \nonumber \\ 
& \left( \sum_{j=1}^d \left \vert \sum_{i=1}^d \mathbb{E}_{\bm{\Theta}} \left[\frac{\partial}{\partial \Theta_i} \left[ \mathbb{E}_{\mathbf{Y} \sim f(\mathbf{y}|\bm{\theta})} [Y_i] \right] \right] \right \vert^p \right) \times \nonumber \\
&\quad \left\{ d^{\frac{1}{p}} (p-1) \left[\sum_{j=1}^n \left(\mathbb{E}_{\bm{\Theta}} \left[ \left( \Omega^{(p)}_{M_j}(\bm{\Theta}) \right)^{\frac{1}{p-1}} \right] \right)^{\frac{2(p-1)}{p}}  \right]^{\frac{1}{2}}  +  \right.  \nonumber \\
&\quad \quad \left. d^{\frac{p-1}{p}} \left(\Omega^{(p)}(\mu) \right)^{\frac{1}{p}} \right \}^{-p} \nonumber
\end{align}

\subsubsection{Case $p \geq 2$}
Let $q \in \mathbb{R}$ such that $\frac{1}{p}+\frac{1}{q}=1$, i.e., $q=p/(p-1)$. Also, consider the following two functions $g(\cdot)$ and $h(\cdot)$ defined, for $\dv x \in \mc X$, $\bs \theta =[\theta_1,\hdots,\theta_d] \in \Theta$ and a specific quantization messages tuple $\dv m^{(n)}=(m_1,\hdots,m_n) \in [1,2^k]^n$  as 
\begin{subequations}
\begin{align}
\label{DefGWassRG2}
g(\dv m^{(n)},\bs \theta) &= \sum_{i=1}^d \frac{\partial}{\partial \theta_i} \left[ \log{\left(p(\dv m^{(n)}|\bs \theta) \mu(\bs \theta) \right)} \right]\\
h(\dv m^{(n)},\bs \theta) &=  \mathbb{E}_{\mathbf{Z} \sim f(\mathbf{z}|\bm{\hat{\theta}}(\mathbf{M^{(n)}}))} [ \mathbf{Z}] - \mathbb{E}_{\mathbf{Y} \sim f(\mathbf{y}|\bm{\Theta})} [\mathbf{Y}]
\end{align}
\end{subequations}	
where in~\eqref{DefGWassRG2} the quantization messages joint probability is $p(\mathbf{m^{(n)}}|\bm{\theta})=\prod_{j=1}^{n} p_j(m_j|\bm{\theta})$. For convenience,  for $i=1,\hdots,d$ we will denote the $i^{th}$ component of $h(\dv m^{(n)}, \bs \theta)$ as $h_i(\dv m^{(n)},\bs \theta)$, i.e.,
\begin{align}
h_i(\dv m^{(n)},\bs \theta) &= \mathbb{E}_{\mathbf{Z} \sim f(\mathbf{z}|\bm{\hat{\theta}}(\mathbf{M^{(n)}}))} [ Z_i] - \mathbb{E}_{\mathbf{Y} \sim f(\mathbf{y}|\bm{\Theta})} [Y_i] \nonumber \\
&= \left(h(\dv m^{(n)},\bs \theta)\right)_i.
\label{DefHWassRG2}
\end{align}

\noindent Note that we have
\begin{align}
&\mathbb{E} \left[ \left\vert h_i(\dv M^{(n)},\bs \Theta) g(\dv M^{(n)},\bs \Theta) \right \vert  \right] \label{OriginalHolderMVTV2WassRG2}
 \\
&= \mathbb{E}_{\bs \Theta} \mathbb{E}_{\dv X | \bs \Theta} \left[ \left( \left\vert h_i(\dv M^{(n)},\bs \Theta) g(\dv M^{(n)},\bs \Theta) \right\vert \right)  |  \bs \Theta = \bs \theta \right] \nonumber\\
&\stackrel{(a)}{\leq} \mathbb{E}_{\bs \Theta} \left( \mathbb{E}_{\dv M^{(n)} | \bs \Theta} \left[ \left(\left\vert h_i(\dv M^{(n)},\bs \Theta)\right\vert^2 \right) | \bs \Theta = \bs \theta \right] \right)^{\frac{1}{2}}  \nonumber \\
&\quad \times \left( \mathbb{E}_{\dv M^{(n)} | \bs \Theta} \left[ \left(\left\vert g(\dv M^{(n)},\bs \Theta)\right\vert^2 \right) | \bs \Theta = \bs \theta \right] \right)^{\frac{1}{2}} \nonumber\\
&\stackrel{(b)}{\leq}  \left(\mathbb{E}_{\bs \Theta} \left[  \left \vert \mathbb{E}_{\dv M^{(n)}|\bs \Theta} \left[ \left( \left\vert h_i(\dv M^{(n)},\bs \Theta) \right \vert^2\right) | \bs \Theta = \bs \theta \right] \right \vert^{\frac{p}{2}} \right] \right)^{\frac{1}{p}}  \nonumber \\
&\quad \times \left(\mathbb{E}_{\bs \Theta} \left[ \left(\mathbb{E}_{\dv M^{(n)}|\bs \Theta} \left[ \left( \left \vert g(\dv M^{(n)},\bs \Theta) \right \vert^2 \right) | \bs \Theta = \bs \theta \right] \right)^{\frac{q}{2}} \right]\right)^{\frac{1}{q}}, \nonumber
\end{align}
where $(a)$ follows by application of H\"older's inequality for every $\bs \theta \in \bs \Theta$ to the conditional expectation  $\mathbb{E}_{\dv M^{(n)} | \bs \Theta}[\cdot|\bs \theta]$ ; and $(b)$ follows by application of H\"older's inequality to the expectation  $\mathbb{E}_{\bs \Theta}[\cdot]$ since $p >1$, $q>1$ and are such that $\frac{1}{p}+\frac{1}{q}=1$.


The first element of the right-hand side of (\ref{OriginalHolderMVTV2WassRG2}(b)) produces the desired risk
\begin{align}
&\sup_{\bm{\theta} \in \bm{\Theta}} \mathbb{E}_{\mathbf{M^{(n)}}|\bm{\Theta}} \left[ W_p^p(f(\mathbf{x}|\bm{\hat{\theta}}(\mathbf{M^{(n)}})),f(\mathbf{x}|\bm{\theta})) \left. \right \vert \bm{\Theta} = \bs \theta \right]  \nonumber \\ 
&\quad \stackrel{(a)}{\geq} \sum_{i=1}^d \mathbb{E}_{(\mathbf{M^{(n)}},\bm{\Theta})} \left[ \left \vert h_i(\mathbf{M^{(n)}},\bm{\Theta}) \right \vert^p \right] \label{partialFBRL2MVTV2PWassRG2} \\
&\quad \stackrel{(b)}{\geq } \sum_{i=1}^d \mathbb{E}_{\bm{\Theta}} \left[ \left(\mathbb{E}_{\mathbf{M^{(n)}}|\bm{\Theta}} \left[ \left \vert h_i(\mathbf{M^{(n)}},\bm{\Theta}) \right \vert^2 \left \vert \right. \bm{\Theta} = \bs \theta \right] \right)^{\frac{p}{2}} \right], \label{almostDoneBound2RG2MVTV2PWassRG2}
\end{align}
where $(a)$ follows from (\ref{NewWassIneqP}) and the fact that the supremum upper bounds the expectation and $(b)$ by replacing $\mathbb{E}_{(\mathbf{M^{(n)}},\bm{\Theta})}$ with $\mathbb{E}_{\bm{\Theta}}$ and $\mathbb{E}_{\mathbf{M^{(n)}}|\bm{\Theta}}$  and Jensen's inequality for expectations for convex functions $x \mapsto x^{\frac{p}{2}}$, for $2 < p $. In order to avoid confusion in the indeces, we will use the notation $h_j(\dot)$.

Using the fact that the prior measure $\mu$ converges to zero at the endpoints of $\bs \Theta$ and partial integration, it is easy to see that
\begin{align}
&\int_{\theta_i} h_j(\mathbf{m^{(n)}},\bm{\theta}) \frac{\partial}{\partial \theta_i} \left[p(\dv m^{(n)}|\bs \theta)  \mu_i(\theta_i) \right] \, \rm{d}{\theta_i}  \nonumber \\
& = \left.  h_j(\mathbf{m^{(n)}},\bm{\theta}) p(\dv m^{(n)}|\bs \theta)  \mu_i(\theta_i) \right \vert_{\theta^{(i)}_{min}}^{\theta^{(i)}_{max}}  - \nonumber \\
&\quad \int_{\theta_i} \frac{\partial}{\partial \theta_i} \left[ h_j(\mathbf{m^{(n)}},\bm{\theta}) \right] p(\dv m^{(n)}|\bs \theta)  \mu_i(\theta_i) \, \rm{d} \theta_i \nonumber \\
& =  - \int_{\theta_i} \frac{\partial}{\partial \theta_i} \left[ h_j(\mathbf{m^{(n)}},\bm{\theta}) \right] p(\dv m^{(n)}|\bs \theta)  \mu_i(\theta_i) \, \rm{d} \theta_i. \label{proof-generalization-of-van-Trees-inequality-step1RL2WassRG2}
\end{align}
\noindent Summing over all messages in~\eqref{proof-generalization-of-van-Trees-inequality-step1RL2WassRG2}, we get for $i=1,\hdots,d$, that
\begin{align}
&\sum_{\mathbf{m^{(n)}}} \int_{\theta_i} h_j(\mathbf{m^{(n)}},\bm{\theta}) \frac{\partial}{\partial \theta_i} \left[p(\dv m^{(n)}|\bs \theta)  \mu_i(\theta_i) \right] \, \rm{d}{\theta_i}  \nonumber \\
&= - \mathbb{E}_{(\mathbf{M^{(n)}},\Theta_i)} \left[ \frac{\partial}{\partial \Theta_i} \left[ h_j(\mathbf{m^{(n)}},\bm{\Theta}) \right] \right].  \label{IntEq1MVTV2RL2PWassRG2}
\end{align}

Thus, with some algebraic manipulations, 
\begin{align}
&\mathbb{E}_{(\mathbf{M^{(n)}},\bm{\Theta})} \left[ h_j(\mathbf{M^{(n)}},\bm{\Theta}) g(\mathbf{M^{(n)}},\bm{\Theta}) \right] \nonumber \\
&=\sum_{i=1}^d \mathbb{E}_{\Theta_1} \left[ \ldots \mathbb{E}_{\Theta_d} \left[ - \mathbb{E}_{(\mathbf{M^{(n)}},\Theta_i}) \left[ \frac{\partial}{\partial \Theta_i} \left[ h_j(\mathbf{M^{(n)}},\bm{\Theta}) \right] \right] \right] \right] \nonumber \\
&= - \sum_{i=1}^d \mathbb{E}_{(\mathbf{M^{(n)}},\bm{\Theta})} \left[ \frac{\partial}{\partial \Theta_i} \left[ h_j(\mathbf{M^{(n)}},\bm{\Theta}) \right] \right]   \nonumber \\
&= \sum_{i=1}^d \mathbb{E}_{\bm{\Theta}} \left[ \frac{\partial }{\partial \Theta_i} \mathbb{E}_{\mathbf{Y} \sim f(\mathbf{y}|\bm{\Theta})} [Y_j] \right]
\end{align}
and $|\mathbb{E}[X]| \leq \mathbb{E}[|X|]$ lower bounds the left-hand side of (\ref{OriginalHolderMVTV2WassRG2}(b)) as
\begin{align}
&\left \vert \sum_{i=1}^d \mathbb{E}_{\bm{\Theta}} \left[ \frac{\partial }{\partial \Theta_i} \left[ \mathbb{E}_{\mathbf{Y} \sim f(\mathbf{y}|\bm{\Theta})} [Y_j] \right] \right]  \right \vert \leq \nonumber \\
&\quad \quad \quad \mathbb{E}_{(\mathbf{M^{(n)}},\bm{\Theta})} \left[ \left \vert h_j(\mathbf{X},\bm{\Theta}) g(\mathbf{X},\bm{\Theta}) \right \vert \right]. \label{BoundLHTRL2MVTV2PWassRG2}
\end{align}

Combining (\ref{OriginalHolderMVTV2WassRG2}(b)), (\ref{almostDoneBound2RG2MVTV2PWassRG2}) and (\ref{BoundLHTRL2MVTV2PWassRG2}), we get
\begin{align}
&\sup_{\bm{\theta} \in \bm{\Theta}} \mathbb{E}_{\mathbf{M^{(n)}}|\bm{\Theta}} \left[ W_p^p(f(\mathbf{x}|\bm{\hat{\theta}}(\mathbf{M^{(n)}})),f(\mathbf{x}|\bm{\theta})) \left. \right \vert \bm{\Theta} = \bs \theta  \right] \geq \label{laststepWassRG2} \\
&\quad \sum_{j=1}^d \left( \left \vert \sum_{i=1}^d \mathbb{E}_{\bm{\Theta}} \left[ \frac{\partial }{\partial \Theta_i} \left[ \mathbb{E}_{\mathbf{Y} \sim f(\mathbf{y}|\bm{\Theta})} [Y_j] \right] \right]  \right \vert^p \right) \times \nonumber \\
&\quad \quad \left(\mathbb{E}_{\bs \Theta} \left[ \left(\mathbb{E}_{\dv M^{(n)}|\bs \Theta} \left[ \left( \left \vert g(\dv M^{(n)},\bs \Theta) \right \vert^2 \right) | \bs \Theta = \bs \theta \right] \right)^{\frac{q}{2}} \right]\right)^{-\frac{p}{q}}. \nonumber
\end{align}

We now move on to the last step of the proof, to upper bound the second expectation of the RHS of (\ref{laststepWassRG2}). For convenience, let
\begin{align}
&l(m_j,\bm{\theta})= \sum_{i=1}^d \frac{\partial}{\partial \theta_i} \left[ \log{p(\mathbf{m_j}|\bm{\theta})} \right] \text{ with } \mathbb{E}_{M_j|\bm{\Theta}} \left[ l(M_j,\bm{\Theta}) \right]=0. \nonumber \\ 
&\text{ Then, } g(\mathbf{m^{(n)}},\bm{\theta})=\sum_{j=1}^{n} l(m_j,\bm{\theta}) + \sum_{i=1}^d  \frac{\partial}{\partial \theta_i} \left[ \log{\mu(\bm{\theta})} \right].
\end{align}
Note that $l(m_j,\bm{\theta})$ is the sum of the elements of the score function associated with $M_j$. We expand the square and cancel the product of the two elements, due to the property that $\mathbb{E}_{M_j|\bm{\Theta}} \left[ l(M_j,\bm{\Theta}) \right]=0$, to arrive at the trace of the Fisher information matrix of $M_j$ and that of the prior, respectively, as follows 
\begin{align}
&\left(\mathbb{E}_{\mathbf{M^{(n)}}|\bm{\Theta}} \left[ \left \vert g(\mathbf{M^{(n)}},\bm{\Theta}) \right \vert^2 \left \vert \right. \bm{\Theta} \right] \right)^{\frac{1}{2}} = \label{ToreplacewithEqRE2RG2PWassRG2} \\
&\left[ \mathbb{E}_{\mathbf{M^{(n)}}|\bm{\Theta}} \left[ \left( \sum_{j=1}^{n} l(M_j,\bm{\Theta}) \right)^2 \left \vert \right. \bm{\Theta} \right] + \left( \sum_{i=1}^d  \frac{\partial}{\partial \Theta_i} \left[ \log{\mu(\bm{\Theta})} \right] \right)^2 \right]^{\frac{1}{2}}, \nonumber 
\end{align}
which holds by 
\[\mathbb{E}_{\mathbf{M^{(n)}}|\bm{\Theta}} \left[ \left( \sum_{j=1}^{n} l(M_j,\bm{\Theta}) \right) \left(\sum_{i=1}^d  \frac{\partial}{\partial \theta_i} \left[\log{\mu(\bm{\theta})} \right]  \right) \left \vert \right. \bm{\Theta} \right]=0. \] Further, by Jensen's inequality for expectations for concave functions $x \mapsto x^{\frac{q}{2}}$, for $q < 2 $, we have
\begin{align}
& \left(\mathbb{E}_{\bm{\Theta}} \left[ \left(\mathbb{E}_{\mathbf{M^{(n)}}|\bm{\Theta}} \left[ \left \vert g(\mathbf{M^{(n)}},\bm{\Theta}) \right \vert^2 \left \vert \right. \bm{\Theta} \right] \right)^{\frac{q}{2}} \right] \right)^{\frac{1}{q}}  \nonumber \\
&\leq \left(\mathbb{E}_{(\mathbf{M^{(n)}},\bm{\Theta})} \left[ \left( \sum_{j=1}^{n} l(M_j,\bm{\Theta}) \right)^2 \right] + \right. \nonumber \\
&\left. \quad + \mathbb{E}_{\bm{\Theta}} \left[ \left(\sum_{i=1}^d  \frac{\partial}{\partial \Theta_i} \left[ \log{\mu(\bm{\Theta})} \right] \right)^2 \right] \right)^{\frac{1}{2}} \nonumber \\ 
&\leq \left(\mathbb{E}_{(\mathbf{M^{(n)}},\bm{\Theta})} \left[ \left( \sum_{j=1}^{n} l(M_j,\bm{\Theta}) \right)^2 \right] + \right. \nonumber \\
&\left. \quad + d \, \sum_{i=1}^d   \mathbb{E}_{\bm{\Theta}} \left[ \left(\frac{\partial}{\partial \Theta_i} \left[ \log{\mu(\bm{\Theta})} \right]  \right)^2 \right] \right)^{\frac{1}{2}} \nonumber \\
&= \left(\mathbb{E}_{(\mathbf{M^{(n)}},\bm{\Theta})} \left[ \left( \sum_{j=1}^{n} l(M_j,\bm{\Theta}) \right)^2 \right]  + d \, \rm{Tr}(I(\mu)) \right)^{\frac{1}{2}}.  \label{partialReztosubstituteHereRG2PWassRG2}
\end{align}

For $l(M_j,\bm{\Theta})$ with $\mathbb{E}_{M_j|\bm{\Theta}} \left[ l(M_j,\bm{\Theta}) \right]=0$ and independent, by the Marcinkiewicz-Zygmund inequality in the form of $(2)$ of \cite{Ren2001}, there exists a constant $B_2=1$ \cite{Burkholder1988}, such that
\begin{align}
&\mathbb{E}_{\mathbf{M^{(n)}}|\bm{\Theta}} \left[ \left(\sum_{j=1}^{n} l(M_j,\bm{\Theta}) \right)^2 \left \vert \right. \bm{\Theta} \right] = \nonumber \\
&\quad B_2 \mathbb{E}_{\mathbf{M^{(n)}}|\bm{\Theta}}  \left[ \sum_{j=1}^{n} l^2(M_j,\bm{\Theta}) \left \vert \right. \bm{\Theta} \right] \nonumber \\
&\quad \stackrel{(a)}{=} \sum_{j=1}^{n} \, \mathbb{E}_{\mathbf{M^{(n)}}|\bm{\Theta}} \left[ \left(\sum_{i=1}^d  \frac{\partial}{\partial \theta_i} \left[\log{p(M_j|\bm{\Theta})} \right] \right)^2 \left \vert \right. \bm{\Theta} \right] , \nonumber \\
&\quad \stackrel{(b)}{\leq} \, d \, \sum_{j=1}^{n} \rm{Tr}(I_{M_j}(\bm{\theta})), \label{CombPart2RG2MVTV2PWassRG2}
\end{align}
where, by the independence of $M_j$, the expectation of each element of the summation is identical and this leads to the term $n$ in $(a)$, which also follows from the definition of $l(M_j,\bm{\Theta})$, $(b)$ is given by the inequality $\left(\sum_{i=1}^d x_i \right)^2 \leq d \, \sum_{i=1}^d x_i^2$, $x_i>0$, required in order to pass the summation inside the expectation and obtain the trace of the Fisher information matrix for $M_j$. 

Substituting (\ref{CombPart2RG2MVTV2PWassRG2}) in (\ref{partialReztosubstituteHereRG2PWassRG2}), we obtain
\begin{align}
& \left(\mathbb{E}_{\bm{\Theta}} \left[ \left(\mathbb{E}_{\mathbf{M^{(n)}}|\bm{\Theta}} \left[ \left \vert g(\mathbf{M^{(n)}},\bm{\Theta}) \right \vert^2 \left \vert \right. \bm{\Theta}\right] \right)^{\frac{q}{2}} \right] \right)^{\frac{1}{q}} \leq \nonumber \\
&\left( d \, \mathbb{E}_{\bm{\Theta}} \left[ \sum_{j=1}^{n} \rm{Tr}(I_{M_j}(\bm{\Theta})) \right] + d \, \rm{Tr}(I(\mu)) \right)^{\frac{1}{2}}. \label{UBoundonGRG2PWassRG2}
\end{align}

Substituting (\ref{UBoundonGRG2PWassRG2}) in (\ref{laststepWassRG2}), produces the desired lower bound
\begin{align}
&\sup_{\bm{\theta} \in \bm{\Theta}} \mathbb{E}_{\mathbf{M^{(n)}}|\bm{\Theta}} \left[W_p^p(f(\mathbf{x}|\bm{\hat{\theta}}(\mathbf{M^{(n)}})),f(\mathbf{x}|\bm{\theta})) \left. \right \vert \bm{\Theta} \right] \geq \nonumber \\ 
&\quad \sum_{j=1}^d \left( \left \vert \sum_{i=1}^d \mathbb{E}_{\bm{\Theta}} \left[\frac{\partial}{\partial \Theta_i} \left[ \mathbb{E}_{\mathbf{Y} \sim f(\mathbf{y}|\bm{\theta})} [Y_j] \right] \right] \right \vert^p \right) \times \nonumber \\
&\quad \quad \left( d \, \sum_{j=1}^{n} \mathbb{E}_{\bm{\Theta}} \left[\rm{Tr}(I_{M_j}(\bm{\Theta})) \right] + d \, \rm{Tr}(I(\mu)) \right)^{-\frac{p}{2}}. \nonumber
\end{align}

\subsection{Proof of Corollary \ref{ThVanTreesWassGLM}}~\label{secVI_WassGLM}

Using the expression of the score function,
\begin{align}
&S_{\bm{\theta}}(\mathbf{X})=\frac{\partial}{\partial \bm{\theta}} \log{f(\mathbf{x}|\bm{\theta})}=\frac{1}{\sigma^2}(\mathbf{x}-\bm{\theta}), \nonumber
\end{align}
we can compute 
\begin{align}
&\Omega_{\mathbf{X}}^{(p)}(\bm{\theta}) =\sum_{i=1}^{d} \left( \mathbb{E}_{\mathbf{X}|\bm{\Theta}} \left[ \left \vert S_{\theta_i}(\mathbf{X}) \right \vert^{\frac{p}{p-1}} \left \vert \right.\bm{\Theta} = \bs \theta  \right]\right)^{p-1} \nonumber \\
&=\left(\frac{1}{\sigma^2}\right)^p \sum_{i=1}^{d} \left( \mathbb{E}_{X_{i}|\bm{\Theta}} \left[ \left \vert X_{i}-\theta_i \right \vert^{\frac{p}{p-1}} \right]\right)^{p-1} \nonumber \\
& =\left(\frac{1}{\sigma^2}\right)^p \left(\frac{1}{\sqrt{2\pi \sigma^2}}\right)^{p-1} \times \nonumber \\
&\quad \times \sum_{i=1}^{d} \left( \int_{x_{i}} \left \vert x_{i}-\theta_i \right \vert^{\frac{p}{p-1}} \exp{\left[-\frac{(x_{i}-\theta)^2}{2 \sigma^2} \right]} \, \rm{d}x_{i} \right)^{p-1}. \nonumber
\end{align}
Using the gamma function, the above integral becomes
\begin{align}
&\int_{x_{i}} |x_{i}-\theta_i|^{\frac{p}{p-1}} \exp{\left[-\frac{(x_{i}-\theta_i)^2}{2 \sigma^2} \right]} \, \rm{d}x_{i} = \nonumber \\
&\quad = \frac{2^{\frac{1}{2p-2}} \, \sigma^{\frac{2p-1}{p-1}}}{p-1} \Gamma\left(\frac{1}{2p-2} \right). \label{ComputeIntGamma2GLMW}
\end{align}
Then, we obtain further 
\begin{align}
\Omega_{\mathbf{X}}^{(p)}(\bm{\theta}) &= \frac{d \sqrt{2}}{\sigma} \left[\frac{\Gamma \left(\frac{1}{2p-2}\right)}{(p-1)\sqrt{2\pi \sigma^2}}  \right]^{p-1}. 
\end{align}

\subsubsection{Case $p \geq 2$}

From Theorem \ref{ThVanTreesWass}, if $p \geq 2$, we know that
\begin{align}
&\sup_{\bm{\theta} \in \bm{\Theta}} \mathbb{E}_{\mathbf{M^{(n)}}|\bm{\Theta}} \left[W_p^p(f(\mathbf{x}|\bm{\hat{\theta}}(\mathbf{M^{(n)}})),f(\mathbf{x}|\bm{\theta})) \left. \right \vert \bm{\Theta} \right] \geq \label{ThWpart1W} \\ 
&\quad \sum_{j=1}^d \left( \left \vert \sum_{i=1}^d \mathbb{E}_{\bm{\Theta}} \left[\frac{\partial}{\partial \Theta_i} \left[ \mathbb{E}_{\mathbf{Y} \sim f(\mathbf{y}|\bm{\theta})} [Y_j] \right] \right] \right \vert^p \right) \times \nonumber \\
&\quad \quad \left( d \, \sum_{j=1}^{n} \mathbb{E}_{\bm{\Theta}} \left[\rm{Tr}(I_{M_j}(\bm{\Theta})) \right] + d \, \rm{Tr}(I(\mu)) \right)^{-\frac{p}{2}}.
\end{align}

For the Gaussian location model, 
\begin{align}
&\mathbb{E}_{\bm{\Theta}} \left[\frac{\partial}{\partial \Theta_i} \left[ \mathbb{E}_{\mathbf{Y} \sim f(\mathbf{y}|\bm{\theta})} [Y_i] \right] \right] = \mathbb{E}_{\bm{\Theta}} \left[\frac{\partial}{\partial \Theta_i} \left[ \Theta_i \right] \right]=1. \label{ThWpart2W}
\end{align}

We need to compute an upper bound on $\rm{Tr}(I_{M}(\bm{\theta}))$. If $p \geq 2$, then, Theorem \ref{ExtensionTh2RG2} gives us that, for some $r \geq 1$, the upper bound holds
\begin{align}
\mathrm{Tr}(I_{M}(\bm{\theta})) &\leq \min{\{\mathrm{Tr}(I_{\mathbf{X}}(\bm{\theta}), 4 I_0 k^{\frac{2}{r}}\}}. \label{last1W}
\end{align}
We move on to compute the value of $I_0$. That is, using the same approach as in Corollary $1$ of \cite{Barnes2019}, for $r=2\geq 1$, we obtain $I_0=\frac{8}{3\sigma^2}$. Then, combining (\ref{ThWpart1W}), (\ref{ThWpart2W}), (\ref{last1W}) and by Remark \ref{RemarkDefPrior}, we get
\begin{align}
&\sup_{\bm{\theta} \in \bm{\Theta}} \mathbb{E}_{\mathbf{M^{(n)}}|\bm{\Theta}} \left[  W_p^p(f(\mathbf{x}|\bm{\theta}), f(\mathbf{x}|\bm{\hat{\theta}})) \left \vert \right. \bm{\Theta} \right]  \geq d^{\frac{p}{2}} \times \nonumber \\
& \max \left \{ \left( \frac{n \, d}{\sigma^2}  + \frac{d \, \pi^2}{B^2} \right)^{-\frac{p}{2}}, \left( \frac{32 \, n \, k }{3\sigma^2}  + \frac{d \, \pi^2}{B^2} \right)^{-\frac{p}{2}} \right \}.
\end{align}

\subsubsection{Case $1 < p < 2$}

By Theorem \ref{ThVanTreesWass}, if $ 1 < p < 2$, we have that
\begin{align}
&\sup_{\bm{\theta} \in \bm{\Theta}} \mathbb{E}_{\mathbf{M^{(n)}}|\bm{\Theta}} \left[W_p^p(f(\mathbf{x}|\bm{\hat{\theta}}(\mathbf{M^{(n)}})),f(\mathbf{x}|\bm{\theta})) \left. \right \vert \bm{\Theta} \right] \geq \nonumber \\ 
& \sum_{j=1}^d \left( \left \vert \sum_{i=1}^d \mathbb{E}_{\bm{\Theta}} \left[\frac{\partial}{\partial \Theta_i} \left[ \mathbb{E}_{\mathbf{Y} \sim f(\mathbf{y}|\bm{\theta})} [Y_j] \right] \right] \right \vert^p \right) \times \nonumber \\
&\quad \left\{ d^{\frac{1}{p}} (p-1) \left[\sum_{j=1}^n \left(\mathbb{E}_{\bm{\Theta}} \left[ \left( \Omega^{(p)}_{M_j}(\bm{\Theta}) \right)^{\frac{1}{p-1}} \right] \right)^{\frac{2(p-1)}{p}}  \right]^{\frac{1}{2}}  +  \right.  \nonumber \\
&\quad \quad \left. d^{\frac{p-1}{p}} \left(\Omega^{(p)}(\mu) \right)^{\frac{1}{p}} \right \}^{-p} \label{part2WGLM}
\end{align}

For the Gaussian location model, 
\begin{align}
&\mathbb{E}_{\bm{\Theta}} \left[\frac{\partial}{\partial \Theta_i} \left[ \mathbb{E}_{\mathbf{Y} \sim f(\mathbf{y}|\bm{\theta})} [Y_i] \right] \right] = \mathbb{E}_{\bm{\Theta}} \left[\frac{\partial}{\partial \Theta_i} \left[ \Theta_i \right] \right]=1. \label{ThWpart2W2}
\end{align}

We need to compute an upper bound on $\Omega_{M}^{(p)}(\bm{\theta})$. If $1 < p < 2$, then, Theorem \ref{ExtensionTh2RL2} gives us that, for some $r \geq \frac{1}{p-1}$, the upper bound holds
\begin{align}
\Omega_{M}^{(p)}(\bm{\theta}) &\leq \min{\{\Omega_{\mathbf{X}}^{(p)}(\bm{\theta}), d^{\frac{2-p}{2}} \, I_0^{\frac{p}{2}} \, (2^k)^{2-p} \, 2^p \, k^{\frac{p}{r}}\}}. \nonumber
\end{align}
We move on to compute the value of $I_0$. That is, using the same approach as in Corollary $1$ of \cite{Barnes2019}, for $r=2\geq \frac{1}{p-1}$, for $p > 1.5$, we obtain $I_0=\frac{8}{3\sigma^2}$. Then, we obtain
\begin{align}
&\Omega_{M}^{(p)}(\bm{\theta}) \leq \label{LowerBoundonTrMRL2W2} \\
&\min \left \{ \frac{d \sqrt{2}}{\sigma} \left[\frac{\Gamma \left(\frac{1}{2p-2}\right)}{(p-1)\sqrt{2\pi \sigma^2}}  \right]^{p-1}, 2^{p+k(2-p)} d^{\frac{2-p}{2}} \left(\frac{8k}{3\sigma^2} \right)^{\frac{p}{2}} \right \}. \nonumber
\end{align}

Substituting (\ref{ThWpart2W2}) and (\ref{LowerBoundonTrMRL2W2}) in (\ref{part2WGLM}) and by Remark \ref{RemarkDefPrior} $\forall p > 1.5$, which is required for the Beta function $\mathcal{B}(\cdot)$ to exist, we obtain
\begin{align}
&I^{(p)}(\mu_i) = \frac{\pi}{2} \left(\frac{2}{B} \right)^p \left[\mathcal{B}\left(\frac{2p-1}{2p-2}, \frac{2p-3}{2p-2} \right) \right]^{p-1} \text{ and } \nonumber \\
&\nonumber \\
&\sup_{\bm{\theta} \in \bm{\Theta}} \mathbb{E}_{\mathbf{M^{(n)}}|\bm{\Theta}} \left[  W_p^p(f(\mathbf{x}|\bm{\theta}), f(\mathbf{x}|\bm{\hat{\theta}})) \left \vert \right. \bm{\Theta} \right]  \geq d^p \times \nonumber \\
&\max \left \{  \left(\sqrt{n} \, d^{\frac{1}{p}}(p-1) \left(\frac{d \sqrt{2}}{\sigma}\right)^{\frac{1}{p}} \left[\frac{\Gamma \left(\frac{1}{2p-2}\right)}{(p-1)\sqrt{2\pi \sigma^2}}  \right]^{\frac{p-1}{p}} + \right. \right. \nonumber \\
&\quad \left. \left. + \, d \,  \left( \frac{\pi}{2}  \right)^{\frac{1}{p}} \frac{2}{B} \left[\mathcal{B}\left(\frac{2p-1}{2p-2}, \frac{2p-3}{2p-2} \right) \right]^{\frac{p-1}{p}} \right)^{-p}, \right. \nonumber \\
& \left. \left(\sqrt{n} \, (p-1) 2^{\frac{p+k(2-p)}{p}} d^{\frac{4-p}{2p}} \left(\frac{8k}{3\sigma^2} \right)^{\frac{1}{2}} + \right. \right. \nonumber \\
&\quad \left. \left. + \, d \,  \left( \frac{\pi}{2}  \right)^{\frac{1}{p}} \frac{2}{B} \left[\mathcal{B}\left(\frac{2p-1}{2p-2}, \frac{2p-3}{2p-2} \right) \right]^{\frac{p-1}{p}} \right)^{-p} \right \}. \nonumber
\end{align}


\appendix

\subsection{Auxilliary results}

\begin{Lem}[Extension of Lemma $1$ of \cite{Barnes2019} to $p$-norms]
For $p \geq 1$ and the parameter $\bm{\theta}=[\theta_1, \ldots, \theta_d]$, $i=1:d$, the $(i,i)^{th}$ element of the generalized Fisher information matrix is lower bounded by
\begin{align}
[I_M^{(p)}(\bm{\theta})]_{i,i} &= \left( \mathbb{E}_{M|\bm{\theta}} \left[\left \vert \mathbb{E}_{(\mathbf{X}|\bm{\theta},M)} \left[ S_{\theta_i}(\mathbf{X}) |M \right] \right \vert^{\frac{p}{p-1}} \right] \right)^{p-1}. \nonumber
\end{align}
\label{Lemma1RL1}
\end{Lem}

\begin{IEEEproof}
The $(i,i)^{th}$ element of the generalized Fisher information matrix of order $p \geq 1$ associated to $M$ is equal to
\begin{align}
[I_M^{(p)}(\bm{\theta})]_{i,i} &= \left( \mathbb{E}_{M|\bm{\theta}} \left[ \left \vert S_{\theta_i}(M) \right \vert^{\frac{p}{p-1}} \right] \right)^{p-1}. \nonumber
\end{align} 
We start lower bounding the score function as
\begin{align}
&S_{\theta_i}(m) = \frac{\partial }{ \partial \theta_i} \left[\log{p(m|\bm{\theta})} \right] = \frac{1}{p(m|\bm{\theta})} \frac{\partial }{ \partial \theta_i} \left[p(m|\bm{\theta}) \right] \nonumber \\
&\quad = \frac{1}{p(m|\bm{\theta})} \frac{\partial }{ \partial \theta_i} \left[ \int_{\mathbf{x}} f(\mathbf{x}|\bm{\theta}) p(m|\mathbf{x},\bm{\theta}) \, \rm{d} \nu(\mathbf{x}) \right] \nonumber \\
&\quad =  \int_{\mathbf{x}} \frac{p(m|\mathbf{x}) f(\mathbf{x}|\bm{\theta})}{p(m|\bm{\theta})} \frac{1}{f(\mathbf{x}|\bm{\theta})} \frac{\partial f(\mathbf{x}|\bm{\theta})}{ \partial \theta_i} \,  \rm{d} \nu(\mathbf{x}) \nonumber \\
&\quad =  \int_{\mathbf{x}} \frac{f(\mathbf{x},m|\bm{\theta})}{p(m|\bm{\theta})} \frac{\partial }{ \partial \theta_i} \left[\log f(\mathbf{x}|\bm{\theta}) \right] \rm{d} \nu(\mathbf{x}) \nonumber \\
&\quad =  \int_{\mathbf{x}} S_{\theta_i}(\mathbf{x})  f(\mathbf{x}|\bm{\theta},m) \, \rm{d} \nu(\mathbf{x})=\mathbb{E}_{(\mathbf{X}|\bm{\theta},m)} \left[  S_{\theta_i}(\mathbf{X}) |m \right]. \nonumber
\end{align}
Taking the absolute value, raising both sides to the power $\frac{p}{p-1}$, taking the expectation with respect to $M|\bm{\theta}$ and raising again everything to the power $p-1$, we obtain the desired result
\begin{align}
[I_M^{(p)}(\bm{\theta})]_{i,i} &= \left( \mathbb{E}_{M|\bm{\theta}} \left[\left \vert \mathbb{E}_{(\mathbf{X}|\bm{\theta},M)} \left[ S_{\theta_i}(\mathbf{X}) |M \right] \right \vert^{\frac{p}{p-1}} \right] \right)^{p-1}. \nonumber
\end{align}

\end{IEEEproof}

\begin{Lem}[Extension of Lemma $2$ of \cite{Barnes2019} to $p$-norms, $1 \leq p<2$]
For $1 \leq p <2$, $\bm{\theta}=[\theta_1, \ldots, \theta_d]$, the trace of the generalized Fisher information matrix is lower bounded by
\begin{align}
\Omega_M^{(p)}(\bm{\theta})& \leq \sum_{j=1}^{2^k} p^{p-1}(m_j|\bm{\theta}) \left \vert \left \vert \mathbb{E}_{(\mathbf{X}|\bm{\theta},m_j)} \left[ S_{\bm{\theta}}(\mathbf{X}) |m_j \right] \right \vert \right \vert^{p}_{p}. \nonumber
\end{align}
\label{ExtensionLemma2RL2}
\end{Lem}

\begin{IEEEproof}
We begin with the definition of the trace of a matrix and we apply Lemma $1$ yielding
\begin{align}
&\Omega_M^{(p)}(\bm{\theta}) = \sum_{i=1}^{d} [I_M^{(p)}(\bm{\theta})]_{i,i} \nonumber \\
&\quad = \sum_{i=1}^{d} \left( \mathbb{E}_{M|\bm{\theta}} \left[ \left \vert \mathbb{E}_{(\mathbf{X}|\bm{\theta},M)} \left[ S_{\theta_i}(\mathbf{X}) |M \right] \right \vert^{\frac{p}{p-1}} \right] \right)^{p-1} \nonumber \\
&\quad =\sum_{i=1}^{d} \left( \sum_{j=1}^{2^k} p(m_j|\bm{\theta}) \left \vert \mathbb{E}_{(\mathbf{X}|\bm{\theta},M)} \left[ S_{\theta_i}(\mathbf{X}) |M \right] \right \vert^{\frac{p}{p-1}} \right)^{p-1} \nonumber \\
&\quad \leq \sum_{i=1}^{d} \sum_{j=1}^{2^k} p^{p-1}(m_j|\bm{\theta}) \left \vert \mathbb{E}_{(\mathbf{X}|\bm{\theta},M)} \left[ S_{\theta_i}(\mathbf{X}) |M \right] \right \vert^{p} \label{SumIneqpowerRL2} \\
&\quad = \sum_{j=1}^{2^k} p^{p-1}(m_j|\bm{\theta}) \sum_{i=1}^{d} \left \vert \mathbb{E}_{(\mathbf{X}|\bm{\theta},M)} \left[ S_{\theta_i}(\mathbf{X}) |M \right] \right \vert^{p} \nonumber \\
&\quad = \sum_{j=1}^{2^k} p^{p-1}(m_j|\bm{\theta}) \left \vert \left \vert \mathbb{E}_{(\mathbf{X}|\bm{\theta},m_j)} \left[ S_{\bm{\theta}}(\mathbf{X}) |m_j \right] \right \vert \right \vert^{p}_{p} \label{defRnorm}, 
\end{align}
where (\ref{SumIneqpowerRL2}) follows from the inequality $\left(\sum_{j=1}^{2^k} x_j \right)^{p-1} \leq \sum_{j=1}^{2^k} x_j^{p-1}$, for $x_j>0$ and $p-1<1$, and (\ref{defRnorm}) from the definition of the $p$-norm, $\left \vert \left \vert x \right \vert \right \vert_p^p=\sum_{i=1}^{d} \left \vert x_i \right \vert^p$. Then, we obtain the final upper bound
\begin{align}
\Omega_M^{(p)}(\bm{\theta})& \leq \sum_{j=1}^{2^k} p^{p-1}(m_j|\bm{\theta}) \left \vert \left \vert \mathbb{E}_{(\mathbf{X}|\bm{\theta},m_j)} \left[ S_{\bm{\theta}}(\mathbf{X}) |m_j \right] \right \vert \right \vert^{p}_{p}. \nonumber
\end{align}

\end{IEEEproof}

\begin{Th}[Extension of Theorem $2$ of \cite{Barnes2019} to $p$-norms, $1 \leq p<2$]
If for any $\bm{\theta} \in \bm{\Theta}$ and any unit vector $\mathbf{u} \in \mathbb{R}^d$,
\begin{align}
\left \vert \left \vert \langle\mathbf{u}, S_{\bm{\theta}}(\mathbf{X}) \rangle \right \vert \right \vert^{2}_{\Psi_r} &\leq I_0
\end{align}
holds for some $r\geq \frac{p}{2(p-1)}$, then
\begin{align}
\Omega_M^{(p)}(\bm{\theta}) &\leq \min{\{\Omega_{\mathbf{X}}^{(p)}(\bm{\theta}), d^{\frac{2-p}{2}} \, I_0^{\frac{p}{2}} \, (2^k)^{2-p} \, 2^p \, k^{\frac{p}{r}}\}}. \nonumber
\end{align}
\label{ExtensionTh2RL2}
\end{Th}

\begin{IEEEproof}
By the inequality between the generalized Fisher information associated to a random vector and that of its transformation by a measurable function given by \cite{Boekee1977b}, we have that
\begin{align}
\Omega_M^{(p)}(\bm{\theta}) & \leq \Omega_{\mathbf{X}}^{(p)}(\bm{\theta}).
\end{align}

From the proof of Theorem $2$ of \cite{Barnes2019}, with the notation $t=p(m|\bm{\theta})$, we have that
\begin{align}
\left \vert \left \vert \mathbb{E}_{(\mathbf{X}|\bm{\theta},m)} \left[ S_{\bm{\theta}}(\mathbf{X})|m \right] \right \vert \right \vert_2 &\leq I_0^{\frac{1}{2}} \left( \log{\frac{2}{t}} \right)^{\frac{1}{p}},
\end{align}
and, together with the following inequality between different norms, for any vector $\mathbf{x} \in \mathbb{R}^d$, $0 < p < 2$, 
\begin{align}
\left \vert \left \vert \mathbf{x} \right \vert \right \vert_2 \leq  \left \vert \left \vert \mathbf{x} \right \vert \right \vert_p \leq  d^{\frac{1}{p}-\frac{1}{2}}\left \vert \left \vert \mathbf{x} \right \vert \right \vert_2,
\end{align}
 yield the upper bound on the $p$-norm as
\begin{align}
\left \vert \left \vert \mathbb{E}_{(\mathbf{X}|\bm{\theta},m)} \left[ S_{\bm{\theta}}(\mathbf{X})|m \right] \right \vert \right \vert_{p}^{p} &\leq d^{\frac{2-p}{2}} I_0^{\frac{p}{2}} \left( \log{\frac{2}{t}} \right)^{\frac{p}{r}}. \label{useboundrnorm4}
\end{align}
Let $t_j=p(m_j|\bm{\theta})$. Then, Lemma \ref{ExtensionLemma2RL2} upper bounds the trace of the generalized Fisher information matrix
\begin{align}
\Omega_M^{(p)}(\bm{\theta})& \leq \sum_{j=1}^{2^k} t_j^{r-1} \left \vert \left \vert \mathbb{E}_{(\mathbf{X}|\bm{\theta},m_j)} \left[ S_{\bm{\theta}}(\mathbf{X})|m_j \right] \right \vert \right \vert^{p}_{p} \nonumber \\
& \leq d^{\frac{2-p}{2}} I_0^{\frac{p}{2}} \sum_{j=1}^{2^k} t_j^{p-1} \left( \log{\frac{2}{t_j}} \right)^{\frac{p}{r}},
\end{align}
where the last step follows from (\ref{useboundrnorm4}). Using the same argument as in \cite{Barnes2019}, let $\phi(x)$ be the concave envelope of $f:\left( \right. 0,1 \left. \right] \rightarrow \mathbb{R}$, $f(x)=x^{p-1} \left( \log{\frac{2}{x}} \right)^{\frac{p}{r}} $. Then, by this definition and the concavity of $\phi(x)$,
\begin{align}
&\Omega_M^{(p)}(\bm{\theta}) \leq d^{\frac{2-p}{2}} I_0^{\frac{p}{2}} \sum_{j=1}^{2^k} \phi(t_j) \leq d^{\frac{2-p}{2}} I_0^{\frac{p}{2}} 2^k \phi\left(\frac{\sum_{j=1}^{2^k} t_j}{2^k} \right) \nonumber \\
&\quad = d^{\frac{2-p}{2}} I_0^{\frac{p}{2}} 2^k \phi\left(\frac{1}{2^k} \right) \leq d^{\frac{2-p}{2}} I_0^{\frac{p}{2}} (2^k)^{2-p} (k+1)^{\frac{p}{r}} \nonumber \\
&\quad \leq d^{\frac{2-p}{2}} I_0^{\frac{p}{2}} (2^k)^{2-p} 2^p k^{\frac{p}{r}}, \nonumber
\end{align}
where we selected $\phi(x)=x^{p-1} \left( \log{\frac{2}{x}} \right)^{\frac{p}{r}}$, which, for $r \geq \frac{1}{p-1}$ and $1<p<2$, is concave on $x \in \left( \right. 0,\frac{1}{2} \left. \right]$. This is true because the function $f(x)=x \left( \log{\frac{2}{x}} \right)^{\frac{p}{r(p-1)}}$, for $r \geq \frac{p}{2(p-1)}$ and $1<p<2$, is concave on $x \in \left( \right. 0,\frac{1}{2} \left. \right]$ and $\phi(x)=f^{p-1}(x)$ and $x \rightarrow x^{p-1}$ is non-decreasing and concave for $p-1<1$ [page 84 of \cite{BV2004}]. For any $r \geq 1$ and $k \geq 1$, $(k+1)^{\frac{p}{r}} \leq 2^p k^{\frac{p}{r}}$.

\end{IEEEproof}

\begin{Lem}[Extension of Lemma $2$ of \cite{Barnes2019} to $p$-norms, $p \geq 2$]
For $p \geq 2$, $\bm{\theta}=[\theta_1, \ldots, \theta_d]$, the trace of the generalized Fisher information matrix is lower bounded by
\begin{align}
\Omega_M^{(p)}(\bm{\theta})& \leq \sum_{j=1}^{2^k} p(m_j|\bm{\theta}) \left \vert \left \vert \mathbb{E}_{(\mathbf{X}|\bm{\theta},m_j)} \left[ S_{\bm{\theta}}(\mathbf{X}) |m_j \right] \right \vert \right \vert^{p}_{p}. \nonumber
\end{align}
\label{ExtensionLemma2RG2}
\end{Lem}

\begin{IEEEproof}
We begin with the definition of the trace of a matrix and apply Lemma (\ref{Lemma1RL1}), which yields
\begin{align}
\Omega_M^{(p)}(\bm{\theta})&= \sum_{i=1}^{d} \left( \mathbb{E}_{M|\bm{\theta}} \left[ \left \vert \mathbb{E}_{(\mathbf{X}|\bm{\theta},M)} \left[ S_{\theta_i}(\mathbf{X}) |M \right] \right \vert^{\frac{p}{p-1}} \right] \right)^{p-1} \nonumber \\
&\leq \sum_{i=1}^{d}  \mathbb{E}_{M|\bm{\theta}} \left[ \left \vert \mathbb{E}_{(\mathbf{X}|\bm{\theta},M)} \left[ S_{\theta_i}(\mathbf{X}) |M \right] \right \vert^{p} \right]  \label{useJensenconvexRG2} \\
&= \mathbb{E}_{M|\bm{\theta}} \left[ \sum_{i=1}^{d} \left \vert \mathbb{E}_{(\mathbf{X}|\bm{\theta},M)} \left[ S_{\theta_i}(\mathbf{X}) |M \right] \right \vert^{p} \right] \nonumber \\
&= \mathbb{E}_{M|\bm{\theta}} \left[ \left \vert \left \vert \mathbb{E}_{(\mathbf{X}|\bm{\theta},M)} \left[ S_{\bm{\theta}}(\mathbf{X}) |M \right] \right \vert \right \vert^{p}_{p} \right], \label{usedefrnorm2RG2}
\end{align}
where the inequality (\ref{useJensenconvexRG2}) is given by Jensen's inequality for convex functions of expectations, $\phi(x)=x^{p-1}$, $p >2$, $\phi(\mathbb{E}[X])\leq\mathbb{E}[\phi(X)]$, and the last step (\ref{usedefrnorm2RG2}) follows from the definition of the $p$-norm, $\left \vert \left \vert x \right \vert \right \vert_p^p=\sum_{i=1}^{d} \left \vert x_i \right \vert^p$. Writing explicitly the outer expectation from (\ref{usedefrnorm2RG2}), we obtain the final result
\begin{align}
\Omega_M^{(p)}(\bm{\theta})& \leq \sum_{j=1}^{2^k} p(m_j|\bm{\theta}) \left \vert \left \vert \mathbb{E}_{(\mathbf{X}|\bm{\theta},m_j)} \left[ S_{\bm{\theta}}(\mathbf{X}) |m_j \right] \right \vert \right \vert^{p}_{p}. \nonumber
\end{align}

\end{IEEEproof}

\begin{Th}[Extension of Theorem $2$ of \cite{Barnes2019} to $p$-norms, $ p \geq 2 $]
If for any $\bm{\theta} \in \bm{\Theta}$ and any unit vector $\mathbf{u} \in \mathbb{R}^d$,
\begin{align}
\left \vert \left \vert \langle\mathbf{u}, S_{\bm{\theta}}(\mathbf{X}) \rangle \right \vert \right \vert^{2}_{\Psi_r} &\leq I_0
\end{align}
holds for some $r\geq \frac{p}{2}$, then
\begin{align}
\Omega_M^{(p)}(\bm{\theta}) &\leq \min{\{\Omega_{\mathbf{X}}^{(p)}(\bm{\theta}),I_0^{\frac{p}{2}} 2^p k^{\frac{p}{r}}\}}.
\end{align}
\label{ExtensionTh2RG2}
\end{Th}

\begin{IEEEproof}
By the inequality between the generalized Fisher information associated to a random vector and that of its transformation by a measurable function given by \cite{Boekee1977b}, we have that
\begin{align}
\Omega_M^{(p)}(\bm{\theta}) & \leq \Omega_{\mathbf{X}}^{(p)}(\bm{\theta}).
\end{align}

From the proof of Theorem $2$ of \cite{Barnes2019}, with the notation $t=p(m|\bm{\theta})$, we have that
\begin{align}
\left \vert \left \vert \mathbb{E}_{(\mathbf{X}|\bm{\theta},m)} \left[ S_{\bm{\theta}}(\mathbf{X})|m \right] \right \vert \right \vert_2 &\leq I_0^{\frac{1}{2}} \left( \log{\frac{2}{t}} \right)^{\frac{1}{r}},
\end{align}
and, together with the following inequality between different norms, for any vector $\mathbf{x} \in \mathbb{R}^d$, $ p \geq 2$, 
\begin{align}
\left \vert \left \vert \mathbf{x} \right \vert \right \vert_p \leq  \left \vert \left \vert \mathbf{x} \right \vert \right \vert_2 \leq  d^{\frac{1}{2}-\frac{1}{p}}\left \vert \left \vert \mathbf{x} \right \vert \right \vert_p,
\end{align}
 yield the upper bound on the $p$-norm as
\begin{align}
\left \vert \left \vert \mathbb{E}_{(\mathbf{X}|\bm{\theta},m)} \left[ S_{\bm{\theta}}(\mathbf{X})|m \right] \right \vert \right \vert_{p}^{p} &\leq I_0^{\frac{p}{2}} \left( \log{\frac{2}{t}} \right)^{\frac{p}{r}}. \label{useboundrnorm5RG2}
\end{align}
Lemma \ref{ExtensionLemma2RG2} upper bounds the trace of the generalized Fisher information matrix
\begin{align}
\Omega_M^{(p)}(\bm{\theta})& \leq \sum_{j=1}^{2^k} t_j \left \vert \left \vert \mathbb{E}_{(\mathbf{X}|\bm{\theta},m_j)} \left[ S_{\bm{\theta}}(\mathbf{X})|m_j \right] \right \vert \right \vert^{p}_{p} \nonumber \\
&\leq I_0^{\frac{p}{2}} \sum_{j=1}^{2^k} t_j \left( \log{\frac{2}{t_j}} \right)^{\frac{p}{r}},
\end{align}
where the last step follows from (\ref{useboundrnorm5RG2}). Using the same argument as in \cite{Barnes2019}, let $\phi(x)$ be the concave envelope of $f:\left( \right. 0,1 \left. \right] \rightarrow \mathbb{R}$, $f(x)=x \left( \log{\frac{2}{x}} \right)^{\frac{p}{r}} $. Then, by this definition and the concavity of $\phi(x)$,
\begin{align}
\Omega_M^{(p)}(\bm{\theta}) & \leq I_0^{\frac{p}{2}}  \sum_{j=1}^{2^k} \phi(t_j) \leq I_0^{\frac{p}{2}} 2^k \phi\left(\frac{\sum_{j=1}^{2^k} t_j}{2^k} \right)  \nonumber \\
& = I_0^{\frac{p}{2}}  2^k \phi\left(\frac{1}{2^k} \right) \leq I_0^{\frac{p}{2}} (k+1)^{\frac{p}{r}} \leq I_0^{\frac{p}{2}} 2^p k^{\frac{p}{r}}, \nonumber
\end{align}
where we selected $\phi(x)=x \left( \log{\frac{2}{x}} \right)^{\frac{p}{r}}$, which, for $r \geq \frac{p}{2}$ and $p \geq 2$, is concave on $x \in \left( \right. 0,\frac{1}{2} \left. \right]$.

\end{IEEEproof}



\begin{thebibliography}{19}

\bibitem{HanCOLT2018}
Y.~Han, A.~\"{O}zg\"{u}r and T.~Weissman,
"Geometric lower bounds for distributed parameter estimation under communication constraints,"
\emph{Proceedings of the $\mathit{31^{st}}$ Conference On Learning Theory (COLT)}, 75:3163-3188, 2018.

\bibitem{Barnes2019}
L.~P.~Barnes, Y.~Han and A.~\"{O}zg\"{u}r,
"Lower bounds for learning distributions under communication constraints via {F}isher information,"
\emph{arXiv:1902.02890}, 2019.

\bibitem{KOPS15}
S.~Kamath, A.~Orlitsky and V.~Pichapati and A.-T. Suresh,
"On learning distributions from their samples,"
\emph{{JMLR} Workshop and Conference Proceedings}, Vol. 40, pp. 1-35, 2015.


\bibitem{HanISIT2018}
Y.~Han, P.~Mukherjee, A.~\"{O}zg\"{u}r and T.~Weissman,
"Distributed statistical estimation of high-dimensional and nonparametric distributions,"
\emph{Proceedings of the IEEE International Symposium on Information Theory (ISIT)}, 2018.


\bibitem{A11}
S.-I. Amari,
"On optimal data compression in multiterminal statistical inference,"
\emph{IEEE Transactions on Information Theory}, Vol. 57, N0. 9, pp. 5577-5587, 2011.

\bibitem{Villani2009}
C.~Villani,
"Optimal transport old and new",
\emph{Springer-Verlag Berlin Heidelberg}, 2009.

\bibitem{vT68}
H.-L. van Trees,
"Detection, Estimation and Modulation Theory. Part I,"
in \emph{Wiley \& Sons}, 1968.

\bibitem{SZ21}
S. Sarbu and A. Zaidi,
"On learning parametric distributions from quantized samples"
\emph{Draft. Available at \url{http://www-syscom.univ-mlv.fr/~zaidi/publications/proofs-paper-isit2021.pdf}"}, 2021.

\bibitem{Frogner2015}
C.~Frogner, C.~Zhang, H.~Mobahi, M.~Araya-Polo, T.~Poggio,
"Learning with a {W}asserstein loss,"
in \emph{Advances in Neural Information Processing Systems 28}, 2015.

\bibitem{Ambrogioni2018}
L.~Ambrogioni, U.~G\"{u}\c{c}l\"{u}, Y.~G\"{u}\c{c}l\"{u}t\"{u}rk, M.~Hinne, E.~Maris, M.A.J.~van Gerven,
"Wasserstein variational inference,"
\emph{Advances in Neural Information Processing Systems 31}, 2018.

\bibitem{Tolstikhin2018}
I.~Tolstikhin, O.~Bousquet, S.~Gelly, B.~Sch\"{o}lkopf,
"Wasserstein auto-encoders,"
\emph{Proceedings of the $\mathit{6^{th}}$ International Conference on Learning Representations}, 2018.

\bibitem{Arjovski2017}
M.~Arjovski, S.~Chintala, L.~Bottou,
"Wasserstein generative adversarial networks,"
\emph{Proceedings of the $\mathit{34^{th}}$ International Conference on Machine Learning}, 2017.

\bibitem{Gulrajani2017}
I.~Gulrajani, F.~Ahmed, M.~Arjovski, V.~Dumoulin, A.C.~Courville,
"Improved training of {W}asserstein {GAN}s,"
\emph{Advances in Neural Information Processing Systems 30}, 2017.





\bibitem{Boekee1977b}
D.E.~Boekee,
"Generalized {F}isher information with application to estimation problems,"
in \emph{IFAC Workshop on Information and Systems}, 10:75-82, 1977.

\bibitem{Boekee1977a}
D.E.~Boekee,
"A generalization of the {F}isher information measure,"
\emph{Ph.D. Thesis, Dept. of El. Eng., Delft Univ. of Tech., Delft, The Netherlands}, 1977.



\bibitem{Gill1995}
R.D.~Gill and B.Y.~Levit,
"Applications of the van {T}rees inequality: a {B}ayesian {C}ram\'{e}r-{R}ao bound,"
in \emph{Bernoulli}, 1:59-79, 1995.

\bibitem{Yu1997}
B.Yu,
"Assouad, {F}ano, and {L}e {C}am,"
Chapter in D.~Pollard, E.~Torgersen, G.L.~Yang(eds), \emph{Festschrift for Lucien Le Cam}, Springer, New York, NY, 1997.


\bibitem{Ren2001}
Y.-F.~Ren and H.-Y.~Liang,
"On the best constant in {M}arcinkiewicz-{Z}ygmund inequality,"
in \emph{Statistics \& Probability Letters}, 53, 227-233, 2001.

\bibitem{Burkholder1988}
D.L.~Burkholder,
"Sharp inequalities for martingales and stochastic integrals",
in \emph{Ast\'{e}risque}, tome 157-158, p. 75-94, 1988.

\bibitem{V10}
R. Vershynin,
"Introduction to the non-asymptotic analysis of random matrices",
in \emph{arXiv preprint, arxiv.:1011.3027}, 2010.





\bibitem{Goldfeld2020a}
Z.~Goldfeld, K.~Greenewald, J.~Niels-Weed and Y.~Polyanskiy,
"Convergence of smoothed empirical measures with applications to entropy estimation,"
in \emph{IEEE Transactions on Information Theory}, vol. 66, 7:4368-4391, July 2020.


\bibitem{BV2004}
S.~Boyd and L.Vandenberghe,
"Convex Optimization",
\emph{Cambridge University Press}, 2004.



\end{thebibliography}
\end{document}